\documentclass[11pt]{article}
\usepackage{amsfonts}
\usepackage{amsmath}
\usepackage{amssymb}
\usepackage[a4paper,top=3cm,bottom=3cm,left=2cm,right=2.5cm,bindingoffset=5mm]{geometry}
\usepackage{amscd}

\newtheorem{theorem}{Theorem}

\newtheorem{claim}{Claim}

\newtheorem{definition}{Definition}

\newtheorem{lemma}{Lemma}

\newtheorem{proposition}{Proposition}
\newtheorem{remark}{Remark}

\numberwithin{equation}{section}
\input{tcilatex}
\begin{document}

\title{Complete conditional type structures\thanks{%
I thank three anonymous reviewers from TARK 2023, Pierpaolo Battigalli,
Emiliano Catonini and Marciano Siniscalchi for helpful feedback. Financial
support from the Italian Ministry of Education, PRIN 2017, Grant Number
2017K8ANN4,\ is gratefully acknowledged.}}
\author{Nicodemo De Vito\thanks{%
Department of Decision Sciences, Bocconi University, Milan, Italy. E-mail
address: nicodemo.devito@unibocconi.it.}}
\date{November 2023}
\maketitle

\begin{abstract}
Hierarchies of conditional beliefs (Battigalli and Siniscalchi 1999) play a
central role for the epistemic analysis of solution concepts in sequential
games. They are modelled by type structures, which allow the analyst to
represent the players' hierarchies without specifying an infinite sequence
of conditional beliefs. Here, we study type structures that satisfy a
\textquotedblleft richness\textquotedblright\ property, called \textit{%
completeness}. This property is defined on the type structure alone, without
explicit reference to hierarchies of beliefs or other type structures. We
provide conditions under which a complete type structure represents all
hierarchies of conditional beliefs. In particular, we present an extension
of the main result in Friedenberg (2010) to type structures with conditional
beliefs.


\end{abstract}

\section{Introduction}

\textit{Hierarchies of conditional beliefs} (Battigalli and Siniscalchi
1999) are fundamental to the epistemic analysis of sequential games.
Conditional beliefs can be thought of as a form of \textquotedblleft
extended\textquotedblright\ probabilities: every player is endowed with a
collection of conditioning events or \textquotedblleft relevant
hypothesis,\textquotedblright\ and forms beliefs given each hypothesis in a
way that updating is satisfied whenever possible. Such a collection of
probabilities is called conditional probability system (CPS, hereafter), and
it is used to model how the beliefs of a player change as the play unfolds.

Hierarchies of conditional beliefs arise naturally in an interactive
setting, and their explicit description is usually difficult. A player's
first-order conditional beliefs are described by a CPS over the space of
primitive uncertainty, e.g., the set of all strategy profiles; her
second-order conditional beliefs are described by a CPS over the spaces of
primitive uncertainty and of the co-players' first-order conditional
beliefs; and so on.

Battigalli and Siniscalchi (1999) show that hierarchies of CPSs can be
practically represented by\textit{\ type structures}, i.e., compact models
which mimic Harsanyi's representation of hierarchies of probabilistic
beliefs. Namely, for each player there is a \textit{set of types}. Each type
is associated with a CPS over the set of primitive uncertainty and the set
of the co-players' types. As shown by Battigalli and Siniscalchi, it is
always possible to unfold from a type structure the explicit description of
a hierarchy of CPSs of each type.

Here, we study type structures that satisfy a \textquotedblleft
richness\textquotedblright\ property, called \textit{completeness}
(Brandenburger 2003). This property---which plays a crucial role for the
epistemic foundations\footnote{%
See Dekel and Siniscalchi (2015) for a survey.} of some solution
concepts---is defined on the type structure alone, without explicit
reference to hierarchies of CPSs.

Completeness seems to be, prima facie, a natural richness property in terms
of representation of hierarchies: it requires that a type structure must
induce all possible beliefs about types. This leads to the obvious
conjecture that complete type structures represent all hierarchies of CPSs.
Yet, a recent paper by Friedenberg and Keisler (2021) shows that this
conjecture is wrong! Specifically, these authors consider type structures
with beliefs represented by ordinary probabilities (a special case of CPSs).
They show that, absent some conditions, a complete type structure need not
represent all possible hierarchies of beliefs.

In light of the above result, this paper addresses the following question:
When does a complete type structure represent all hierarchies of conditional
beliefs? Our main result (Theorem \ref{Main theorem}) can be summarized as
follows. Suppose that a type structure is complete. Then:

(i) if the structure is Souslin, then it is \textit{finitely terminal},
i.e., it represents all finite order conditional beliefs;

(ii) if the structure is compact and continuous, then it is \textit{terminal}%
, i.e., it represents all hierarchies of conditional beliefs.

Precise definitions are given in the main text. Here we point out that
Theorem \ref{Main theorem} is an extension of the main result in Friedenberg
(2010) to type structures with conditional beliefs. More precisely,
Friedenberg studies complete type structures with beliefs represented by
ordinary probabilities; her main result shows that (i) and (ii) are
sufficient conditions for finite terminality and terminality, respectively.
Friedenberg (2010, Section 5) leaves open the question whether her result
still holds when beliefs are represented by CPSs: Theorem \ref{Main theorem}
in this paper provides an affirmative answer.

Our main result is far from being trivial because, unlike Friedenberg's
framework, the case of CPSs requires a different approach and method of
analysis. In brief, our method of proof is as follows. We first provide a\
construction---based on the set-up in Heifetz (1993)---of the canonical
space of hierarchies of CPSs which allows us to characterize the notion of
(finite) terminality in a convenient way (Proposition \ref{terminality
characterization}). With this, the crucial step relies on Lemma \ref{Lemma:
ontoness lubinlike}, an \textquotedblleft extension\textquotedblright\
result for CPSs whose proof makes use of a (measurable) selection argument.

Although this paper addresses a specific research question, we propose that
the usefulness of the approach and of the results goes well beyond it. As
mentioned, for the proof of our result we construct a \textquotedblleft
canonical\textquotedblright\ space of hierarchies. Such a construction,
which has a \textquotedblleft nice\textquotedblright\ topological structure,
is a generalization of Battigalli and Siniscalchi's (1999) construction
because we allow for weaker assumptions on the family of conditioning
events---namely, we drop the \textquotedblleft clopeness\textquotedblright\
assumption in their framework. The conceptual motivation is that---as we
will elaborate in the Discussion Section---there are situations (i.e.,
applications) whereby the assumptions in Battigalli and Siniscalchi's
framework may not be satisfied.\footnote{%
As Pierpaolo Battigalli put it (private communication), the original main
motivation in Battigalli and Siniscalchi (1999) for assuming
\textquotedblleft clopeness\textquotedblright\ of conditioning events was
essentially technical---although such an assumption is satisfied in games of
interest, like the ones studied in Battigalli and Siniscalchi (2002),
Battigalli and Tebaldi (2019) and others. That said, we are highly indebted
to Battigalli and Siniscalchi (1999) for some of their results that are used
for our construction; see Appendix A.} Thus, our construction provides a
stepping stone for further research on epistemic analyses for a wide class
of sequential games. Moreover, as we will discuss in Section \ref{Section
main result}, it also allows us to shed light on when our result (Theorem %
\ref{Main theorem}) cannot be applied.

Before moving on to the related literature and the formal analysis, two
features of our contribution are worth noting. First, we study (hierarchies
of) conditional beliefs represented by CPSs. But the analysis and the
results still hold if conditional beliefs are represented by \textit{%
consistent conditional probability systems} (Siniscalchi 2022), as we will
elaborate in Section \ref{Section: Discussion}.

Second, because of the nature of our results, some of the terminology is
technical, but the proofs and the conclusions in the main text are simple,
and they can be understood without technical jargon. Most of the literature
on epistemic game theory uses Polish spaces as standard technical
conditions. Here, we consider more general spaces that behave like Polish
spaces, such as Souslin and/or Lusin (again, the details about the
motivation for using such spaces are spelled out in the Discussion Section).
But the essential ingredients of our analysis do not change if a non-expert
reader thinks of them as Polish spaces, modulo some technical details.

\paragraph{\noindent Related literature}

The closest related papers to our work are Friedenberg (2010) and Battigalli
and Siniscalchi (1999). Friedenberg (2010) studies type structures with
beliefs represented by ordinary probabilities; our paper considers instead
the case of beliefs represented by CPSs. Despite the obvious differences in
terms of formalism (ordinary vs. \textquotedblleft
extended\textquotedblright\ probabilities), both papers study complete
structures, but the methods of proof for the main results are distinct.
Friedenberg (2010) uses a proof technique which---roughly speaking---relies
on extending probability measures from a sub-$\sigma $-algebra of events to
the Borel $\sigma $-algebra of a topological space. Such a proof does not
work for CPSs for a simple reason: if we can extend every probability
measure of a CPS from a sub-$\sigma $-algebra to a larger $\sigma $-algebra,
then the corresponding collection of probabilities may not yield a CPS---in
particular, the chain rule of conditional probabilities may not be satisfied.

We briefly explained above how we relate to Battigalli and Siniscalchi
(1999). Besides the aforementioned differences, the starting point of our
analysis is distinct. Battigalli and Siniscalchi (1999) begin by describing
all hierarchies of conditional beliefs, and show that their construction
gives a complete type structure. So, in a certain sense, and like other
papers in the literature on higher-order beliefs, Battigalli and Siniscalchi
address the question: If a type structure represents all hierarchies, is it
complete? But they do not address the converse (i.e., does a complete
structure represent all hierarchies?), which is the main question of this
paper.

In the literature, various kinds of \textquotedblleft
large\textquotedblright\ or \textquotedblleft rich\textquotedblright\ type
structures have been considered. We refer the reader to the survey papers of
Dekel and Siniscalchi (2015) and of Siniscalchi (2008) for an overview. Here
we just mention that the literature has considered a notion for large type
structures based on the idea of \textquotedblleft
embedding\textquotedblright : a type structure represents all hierarchies if
any other type structure can be uniquely \textquotedblleft
embedded\textquotedblright\ into it\ via a map called type morphism. Unlike
completeness, such a notion makes reference to other type structures; yet,
both notions do no make explicit reference to hierarchies of beliefs. We
have already claimed that the \textquotedblleft completeness
test\textquotedblright\ may fail, in the sense that there are situations in
which Theorem \ref{Main theorem} cannot be applied. The same is true for the
alternative \textquotedblleft embedding test\textquotedblright : a type
structure may represent all hierarchies of beliefs, still there could be
distinct type structures that cannot be mapped\ into it via type morphism.
We will further elaborate on this issue in the Discussion Section.

\paragraph{\noindent Structure of the paper}

Section 2 introduces all the relevant notation and auxiliary mathematical
results. Section 3 introduces basic definitions and main properties of type
structures. Section 4 is devoted to the presentation and the proof of the
main results of the paper. Section 5 discusses some conceptual and technical
issues. All omitted proofs are in Appendix.\footnote{%
The Supplementary Appendix (available at https://arxiv.org/abs/2305.08940)
contains additional analyses and extension of results as claimed the main
text.}

\section{Preliminaries\label{Section Mathematical Preliminaries}}

A measurable space is a pair $\left( X,\Sigma _{X}\right) $, where $X$ is a
set and $\Sigma _{X}$ is a $\sigma $-algebra, the elements of which are
called \textbf{events}. Throughout this paper, when it is clear from the
context which $\sigma $-algebra on $X$ we are considering, we suppress
reference to $\Sigma _{X}$ and simply write $X$\ to denote a measurable
space. Furthermore, given a function $f:X\rightarrow Y$\ and a family $%
\mathcal{F}_{Y}$\ of subsets of $Y$, we let $f^{-1}\left( \mathcal{F}%
_{Y}\right) :=\left \{ E\subseteq X:\exists F\in \mathcal{F}%
_{Y},E=f^{-1}\left( F\right) \right \} $. So, if $Y$\ is a measurable space,
then $f^{-1}\left( \Sigma _{Y}\right) $\ is the $\sigma $-algebra on $X$\
generated by $f$.

We write $\Delta \left( X\right) $\ for the set of probability measures on $%
\Sigma _{X}$. Fix measurable spaces $X$ and $Y$. Given a measurable function 
$f:X\rightarrow Y$, we let $\mathcal{L}_{f}:\Delta \left( X\right)
\rightarrow \Delta \left( Y\right) $ denote the pushforward-measure map
induced by $f$; that is, for each $\mu \in \Delta \left( X\right) $, $%
\mathcal{L}_{f}\left( \mu \right) $\ is the image measure of $\mu $\ under $%
f $, and is defined by $\mathcal{L}_{f}\left( \mu \right) \left( E\right)
:=\mu \left( f^{-1}\left( E\right) \right) $\ for every $E\in \Sigma _{Y}$.

If $X$ is a topological space, we keep using $\Sigma _{X}$\ to denote the
Borel $\sigma $-algebra on $X$. All the topological spaces considered in
this paper are assumed to be metrizable. We consider any product, finite or
countable, of metrizable spaces as a metrizable space with the product
topology. Moreover, we endow each subset of a metrizable space with the
subspace topology. A \textbf{Souslin}\ (resp. \textbf{Lusin}) \textbf{space}
is a topological space that is the image of a complete, separable metric
space under a continuous surjection (resp. bijection). Clearly, a Lusin
space is also Souslin. Examples of Souslin (resp. Lusin) spaces include
analytic (resp. Borel) subsets of a complete separable metric space. In
particular, a Polish space (i.e., a topological space which is homeomorphic
to a complete, separable metric space) is a Lusin space. Furthermore, if $X$
is a Lusin space, then $\left( X,\Sigma _{X}\right) $\ is a \textbf{standard
Borel} space, i.e., there is a Polish space $Y$\ such that $\left( X,\Sigma
_{X}\right) $\ is isomorphic to $\left( Y,\Sigma _{Y}\right) $. If $X$ is a
Souslin space, then $\left( X,\Sigma _{X}\right) $\ is an \textbf{analytic
measurable }space, i.e., there is a Polish space $Y$\ and an analytic subset 
$A\subseteq Y$ such that $\left( X,\Sigma _{X}\right) $\ is isomorphic to $%
\left( A,\Sigma _{A}\right) $; see Cohn (2013, Chapter 8).

For a metrizable space $X$, the set $\Delta \left( X\right) $ of (Borel)
probability measures is endowed with the topology of weak convergence. With
this topology, $\Delta \left( X\right) $\ becomes a metrizable space.

\section{Conditional probability systems\label{Section CPS}}

We represent the players' beliefs as conditional probability systems (cf. R%
\'{e}nyi 1955). Fix a measurable space $\left( X,\Sigma _{X}\right) $. A
family of \textbf{conditioning events} of $X$ is a collection $\mathcal{B}%
\subseteq \Sigma _{X}$ that does not include the empty set. A possible
interpretation is that an individual is uncertain about the realization of
the \textquotedblleft state\textquotedblright \ $x\in X$, and $\mathcal{B}$\
represents a family of observable events or \textquotedblleft relevant
hypotheses.\textquotedblright \ If $X$ is a metrizable space, then each
conditioning event $B\in \mathcal{B}$\ is a Borel subset of $X$. In this
case, we say that $\mathcal{B}$ is \textbf{clopen} if each element $B\in 
\mathcal{B}$ is both closed and open. For instance, $\mathcal{B}$ is clopen
if $X$ is a (finite) set endowed with the discrete topology; if $\mathcal{B=}%
\left \{ X\right \} $, then $\mathcal{B}$\ is trivially clopen.

\begin{definition}
\label{Definition CPS}Let $\left( X,\Sigma _{X}\right) $\ be a measurable
space and $\mathcal{B}\subseteq \Sigma _{X}$ be a family\ of conditioning
events. A \textbf{conditional probability system} (\textbf{CPS}) on $\left(
X,\Sigma _{X},\mathcal{B}\right) $\ is an array of probability measures $\mu
:=\left( \mu \left( \cdot |B\right) \right) _{B\in \mathcal{B}}$ such that:%
\newline
\ \ (i) for all $B\in \mathcal{B}$, $\mu \left( B|B\right) =1$;\newline
\ \ (ii) for all $A\in \Sigma _{X}$ and $B,C\in \mathcal{B}$, if $A\subseteq
B\subseteq C$ then $\mu \left( A|B\right) \mu \left( B|C\right) =\mu \left(
A|C\right) $.
\end{definition}

Definition \ref{Definition CPS} says that a CPS $\mu $ is an element of the
set $\Delta \left( X\right) ^{\mathcal{B}}$, i.e., $\mu $\ is a function
from $\mathcal{B}$\ to $\Delta \left( X\right) $.\footnote{%
For every pair of sets $X$ and $Y$, we let $Y^{X}$ denote the set of
functions with domain $X\,$\ and codomain $Y$.} We write $\mu \left( \cdot
|B\right) $ to stress the interpretation as a conditional probability given
event $B\in \mathcal{B}$. Condition (ii) is the \textbf{chain rule} of
conditional probabilities and it can be written as follows: if $A\subseteq
B\subseteq C$, then%
\begin{equation*}
\mu \left( B|C\right) >0\Rightarrow \mu \left( A|B\right) =\frac{\mu \left(
A|C\right) }{\mu \left( B|C\right) }\text{.}
\end{equation*}

We let $\Delta ^{\mathcal{B}}\left( X\right) $\ denote the set of CPSs on $%
\left( X,\Sigma _{X},\mathcal{B}\right) $. The following result (whose proof
can be found in Appendix A) records some topological properties of $\Delta ^{%
\mathcal{B}}\left( X\right) $\ when $X$ is a metrizable space and $\mathcal{B%
}$\ is countable.\footnote{%
Lemma \ref{Lemma souslin space of CPSs}\ is a generalization of analogous
results (for the case when $X$ is Polish) in Battigalli and Siniscalchi
(1999, Lemma 1).}

\begin{lemma}
\label{Lemma souslin space of CPSs}Fix a metrizable space $X$\ and a
countable family $\mathcal{B}\subseteq \Sigma _{X}$ of conditioning events.%
\newline
\ \ (i) The space $\Delta ^{\mathcal{B}}\left( X\right) $\ is metrizable.%
\newline
\ \ (ii) If $X$ is Souslin or Lusin, so is $\Delta ^{\mathcal{B}}\left(
X\right) $. \newline
\ \ (iii) Suppose that $\mathcal{B}$\ is clopen. Then $\Delta ^{\mathcal{B}%
}\left( X\right) $ is compact if and only if $X$ is compact.
\end{lemma}

Note that if $X$ is a Polish space, then $\Delta ^{\mathcal{B}}\left(
X\right) $ may fail to be Polish. But, by Lemma \ref{Lemma souslin space of
CPSs}.(ii), $\Delta ^{\mathcal{B}}\left( X\right) $\ is a Lusin space. We
can conclude that $\Delta ^{\mathcal{B}}\left( X\right) $\ is a Polish
space\ provided that $\mathcal{B}$\ is clopen (cf. Battigalli and
Siniscalchi 1999, Lemma 1).

Fix measurable spaces $\left( X,\Sigma _{X}\right) $\ and $\left( Y,\Sigma
_{Y}\right) $, and families $\mathcal{B}_{X}\subseteq \Sigma _{X}$\ and $%
\mathcal{B}_{Y}\subseteq \Sigma _{Y}$ of conditioning events. Suppose that $%
f:X\rightarrow Y$\ is a measurable function such that%
\begin{equation*}
f^{-1}\left( \mathcal{B}_{Y}\right) =\mathcal{B}_{X}\text{.}
\end{equation*}%
The function $\overline{\mathcal{L}}_{f}:\Delta ^{\mathcal{B}_{X}}\left(
X\right) \rightarrow \Delta ^{\mathcal{B}_{Y}}\left( Y\right) $ defined by%
\begin{equation*}
\overline{\mathcal{L}}_{f}\left( \mu \right) \left( E|B\right) :=\mu \left(
f^{-1}\left( E\right) |f^{-1}\left( B\right) \right) \text{,}
\end{equation*}%
where $E\in \Sigma _{Y}$\ and $B\in \mathcal{B}_{Y}$, is the \textbf{%
pushforward-CPS map} induced by $f$. We call $\overline{\mathcal{L}}%
_{f}\left( \mu \right) $ the \textbf{image CPS} of $\mu $\ under $f$. Note
that, for any $\mu \in \Delta ^{\mathcal{B}_{X}}\left( X\right) $, we can
write $\overline{\mathcal{L}}_{f}\left( \mu \right) $\ as%
\begin{equation*}
\overline{\mathcal{L}}_{f}\left( \mu \right) =\left( \mathcal{L}_{f}\left(
\mu \left( \cdot |f^{-1}\left( B\right) \right) \right) \right) _{B\in 
\mathcal{B}_{Y}}\text{,}
\end{equation*}%
where $\mathcal{L}_{f}:\Delta \left( X\right) \rightarrow \Delta \left(
Y\right) $\ is the pushforward-measure map induced by $f$.

We record some basic results on the pushforward-CPS map that will be used
extensively throughout the paper (for the proofs, see Appendix A). In
particular, Lemma \ref{Lemma on general measurability of image CPS}.(i)
ensures that $\overline{\mathcal{L}}_{f}$\ is well-defined and justifies the
terminology: if $\mu \in \Delta ^{\mathcal{B}_{X}}\left( X\right) $, then $%
\overline{\mathcal{L}}_{f}\left( \mu \right) $\ is a CPS on $\left( Y,\Sigma
_{Y},\mathcal{B}_{Y}\right) $.

\begin{lemma}
\label{Lemma on general measurability of image CPS}Fix measurable spaces $%
\left( X,\Sigma _{X}\right) $\ and $\left( Y,\Sigma _{Y}\right) $, and
families $\mathcal{B}_{X}\subseteq \Sigma _{X}$ and $\mathcal{B}%
_{Y}\subseteq \Sigma _{Y}$\ of conditioning events. Suppose that $%
f:X\rightarrow Y$\ is a measurable function such that $f^{-1}\left( \mathcal{%
B}_{Y}\right) =\mathcal{B}_{X}$. The following statements hold.\newline
\ \ (i) The map $\overline{\mathcal{L}}_{f}:\Delta ^{\mathcal{B}_{X}}\left(
X\right) \rightarrow \Delta ^{\mathcal{B}_{Y}}\left( Y\right) $\ is
well-defined.\newline
\ \ (ii) Suppose that $\mathcal{B}_{X}$\ and $\mathcal{B}_{Y}$\ are
countable, $X$\ is a metrizable space and $Y$ is a Souslin space. If $f$ is
Borel measurable (resp. continuous), then $\overline{\mathcal{L}}_{f}$\ is
Borel measurable (resp. continuous).
\end{lemma}

A special case of pushforward-CPS map induced by a function is of particular
interest---namely, the marginalization of a CPS on a product space. Consider
measurable spaces $X$ and $Y$, and denote by $\pi _{X}$\ the coordinate
projection from $X\times Y$\ onto $X$. Fix a family $\mathcal{B}\subseteq
\Sigma _{X}$ of conditioning events, and define $\mathcal{B}_{X\times Y}$\ as%
\begin{equation}
\mathcal{B}_{X\times Y}:=\left( \pi _{X}\right) ^{-1}\left( \mathcal{B}%
\right) =\left \{ C\subseteq X\times Y:\exists B\in \mathcal{B},C=B\times
Y\right \} \text{,}  \label{conditioning events cylinders}
\end{equation}%
that is, $\mathcal{B}_{X\times Y}$\ is the set of all cylinders $B\times Y$\
with $B\in \mathcal{B}$. The function $\overline{\mathcal{L}}_{\pi
_{X}}:\Delta ^{\mathcal{B}_{X\times Y}}\left( X\times Y\right) \rightarrow
\Delta ^{\mathcal{B}}\left( X\right) $ defined by%
\begin{equation*}
\overline{\mathcal{L}}_{\pi _{X}}\left( \mu \right) :=\left( \mathcal{L}%
_{\pi _{X}}\left( \mu \left( \cdot |B\right) \right) \right) _{B\in \mathcal{%
B}}
\end{equation*}%
is called \textbf{marginal-CPS map}, and $\overline{\mathcal{L}}_{\pi
_{X}}\left( \mu \right) $ is called the \textbf{marginal}\ on $X$ of $\mu
\in \Delta ^{\mathcal{B}_{X\times Y}}\left( X\times Y\right) $.

\section{Type structures and hierarchies of conditional beliefs \label%
{Section of type structures and belief hierarchies}}

Throughout, we fix a two-player set $I$;\footnote{%
The assumption of a two-player set is merely for notational convenience. The
analysis can be equivalently carried out with any finite set $I$ with
cardinality greater than two.} given a player $i\in I$, we let $j$ denote
the other player in $I$. We assume that both players share a common
measurable space $\left( S,\Sigma _{S}\right) $, called \textbf{space of
primitive uncertainty}\textit{. }For each $i\in I$, there is a family $%
\mathcal{B}_{i}\subseteq \Sigma _{S}$ of conditioning events. One
interpretation (which is borrowed from Battigalli and De Vito 2021) is the
following: $S$ is a product set, viz. $S:=\tprod_{i\in I}S_{i}$, and each
element $s:=\left( s_{i}\right) _{i\in I}$\ is an objective description of
players' behavior in a game with complete information and without chance
moves---technically, $\left( s_{i}\right) _{i\in I}$\ is a strategy profile.
Each player is uncertain about the \textquotedblleft true\textquotedblright
\ behavior $s\in S$, including his own. If the game has sequential moves,
then each $\mathcal{B}_{i}$\ is a collection of \textit{observable events};
that is, each $B\in \mathcal{B}_{i}$\ is the set of strategy profiles
inducing an information set of player $i$. Other interpretations of $S$ and $%
\left( \mathcal{B}_{i}\right) _{i\in I}$ are also possible; a more thorough
discussion can be found in Battigalli and Siniscalchi (1999, pp. 191-192).
The results in this paper do not hinge on a specific interpretation.

From now on, we maintain the following technical assumptions on $S$\ and $%
\left( \mathcal{B}_{i}\right) _{i\in I}$:

\begin{itemize}
\item $S$ is a Souslin space, and

\item $\mathcal{B}_{i}\subseteq \Sigma _{S}$\ is countable for every $i\in I$%
.
\end{itemize}

Following Battigalli and Siniscalchi (1999), we adopt the following
notational convention.

\bigskip

\noindent \textbf{Convention 1}. Given a product space $X\times Y$ and a
family $\mathcal{B}\subseteq \Sigma _{X}$ of conditioning events of $X$, the
family of conditioning events of $X\times Y$ is $\mathcal{B}_{X\times Y}$ as
defined in (\ref{conditioning events cylinders}). Accordingly, we let $%
\Delta ^{\mathcal{B}}\left( X\times Y\right) $\ denote the set of CPSs on $%
\left( X\times Y,\Sigma _{X\times Y},\mathcal{B}_{X\times Y}\right) $.

\subsection{Type structures}

We use the framework of type structures (or \textquotedblleft type
spaces\textquotedblright ) to model players' hierarchies of conditional
beliefs. We adopt the following definition of type structure (cf. Battigalli
and Siniscalchi 1999).

\begin{definition}
\label{Definition of type structure}An $\left( S,\left( \mathcal{B}%
_{i}\right) _{i\in I}\right) $-\textbf{based type structure} is a tuple 
\begin{equation*}
\mathcal{T}:=\left( S,\left( \mathcal{B}_{i},T_{i},\beta _{i}\right) _{i\in
I}\right)
\end{equation*}%
such that, for every $i\in I$,\newline
\ \ (i) the \textbf{type set} $T_{i}$\ is a metrizable space;\newline
\ \ (ii) the \textbf{belief map} $\beta _{i}:T_{i}\rightarrow \Delta ^{%
\mathcal{B}_{i}}\left( S\times T_{j}\right) $ is Borel measurable.

Each element of $T_{i}$, viz. $t_{i}$, is called (player $i$'s) \textbf{type}%
.
\end{definition}

Definition \ref{Definition of type structure}\ says that, for any $i\in I$, $%
T_{i}$ represents the set of player $i$'s possible \textquotedblleft ways to
think.\textquotedblright \ Each type $t_{i}\in T_{i}$ is associated with a
CPS on the set of primitive uncertainty as well as on the possible
\textquotedblleft ways to think\textquotedblright \ (types)\ of player $j$.
Each conditioning event for $\beta _{i}\left( t_{i}\right) $ has the form $%
B\times T_{j}$, with $B\in \mathcal{B}_{i}$.

Note that in Definition \ref{Definition of type structure}, and like in
Battigalli and Siniscalchi (1999), it is implicitly assumed that players
have introspective beliefs---that is, beliefs about their own types. We
maintain the implicit assumption that players are introspective, hence they
know their own way to think, and there is common certainty of this
conditional on each relevant hypothesis $B\in \mathcal{B}_{i}$ ($i\in I$).
This assumption is innocuous: an explicit formalization of introspective
beliefs would not change the substance of our results (cf. Heifetz and Samet
1998, p. 330).

If $\mathcal{B}_{i}\mathcal{=}\left \{ S\right \} $ for every player $i\in I$%
, then each set $\Delta ^{\mathcal{B}_{i}}\left( S\times T_{j}\right) $\ can
be naturally identified with $\Delta \left( S\times T_{j}\right) $. In this
case, Definition \ref{Definition of type structure}\ coincides essentially
with the definition in Friedenberg (2010),\footnote{%
The only difference is that $S$ is assumed to be a Polish space in
Friedenberg (2010). Such difference is immaterial for the remainder of the
analysis.} and we say that $\mathcal{T}$\ is an \textbf{ordinary type
structure}. Moreover, we will sometimes refer to type structures via
Definition \ref{Definition of type structure}\ as \textbf{conditional type
structures}.

For the statement of our main results, we also need the following
definitions.

\begin{definition}
An $\left( S,\left( \mathcal{B}_{i}\right) _{i\in I}\right) $-based type
structure $\mathcal{T}:=\left( S,\left( \mathcal{B}_{i},T_{i},\beta
_{i}\right) _{i\in I}\right) $ is\newline
\ \ (i) \textbf{Souslin} (resp. \textbf{Lusin}, \textbf{compact}) if, for
every $i\in I$, the type set $T_{i}$ is a Souslin (resp. Lusin, compact)
space;\footnote{%
In Friedenberg (2010), a(n ordinary) type structure is called analytic if
each type set is an analytic subset of a Polish space---hence, a metrizable
Souslin (sub)space. We adopt the definition of Souslin type structure
because we can extend our analysis (as we do in the Supplementary Appendix)
without assuming metrizability of the topological spaces.}\newline
\ \ (ii) \textbf{continuous} if, for every $i\in I$, the belief map $\beta
_{i}$\ is continuous.
\end{definition}

Next, we introduce the notion of completeness for a type structure. This
notion is due to Brandenburger (2003).

\begin{definition}
\label{Definition complete type structure}An $\left( S,\left( \mathcal{B}%
_{i}\right) _{i\in I}\right) $-based type structure $\mathcal{T}:=\left(
S,\left( \mathcal{B}_{i},T_{i},\beta _{i}\right) _{i\in I}\right) $ is 
\textbf{complete} if, for every $i\in I$, the belief map $\beta _{i}$ is
surjective.
\end{definition}

In words, completeness says that, for every player $i\in I$ and for every
conditional belief $\mu \in \Delta ^{\mathcal{B}_{i}}\left( S\times
T_{j}\right) $\ that player $i$ can hold, there is a type of player $i$\
which induces that belief. Thus, it is a \textquotedblleft
richness\textquotedblright \ requirement which may not be satisfied by some
type structures. For instance, suppose that $S$\ is not a singleton.\ Then a
type structure\ where the type set of some player\ has finite cardinality is
not complete.

A type structure provides an implicit representation of the hierarchies of
beliefs. To address the question whether a complete type structure
represents all hierarchies of beliefs, we need to formally clarify \textit{%
how} type structures generate a collection of hierarchies of beliefs for
each player. This is illustrated in the following section.

\subsection{The canonical space of hierarchies\label{Section: Canonical
space of hierarchies}}

In this section we first offer a construction of the set of all hierarchies
of conditional beliefs satisfying a \textit{coherence} condition. Loosely
speaking, coherence means that lower-order beliefs are the marginals of
higher-order beliefs. The construction---which is based on the set-up in
Heifetz (1993)---shows that this set of hierarchies identifies in a natural
way a type structure, which we call it \textquotedblleft
canonical.\textquotedblright \ Next, we show how each profile of types in a
type structure can be associated with an element of the constructed set of
hierarchies. This part is standard (cf. Heifetz and Samet 1998, 1999).

Our construction of the \textquotedblleft canonical\textquotedblright \
space of hierarchies is formally different from the one in Battigalli and
Siniscalchi (1999), as it relies on a costruction scheme based on the notion
of projective limit. Here, we report the essential ingredients for the main
results of the paper. A detailed analysis is deferred to Appendix B where we
briefly review the background notions and results on projective limit
theory. In Section \ref{Section: Discussion}\ we will compare our\
construction of the canonical space with the\ construction \textit{\`{a} la}
Battigalli and Siniscalchi (1999).

Unless otherwise stated, the proofs of remarks and propositions in the
following subsections can be found in Appendix B.

\subsubsection{From hierarchies to types\label{Section from hierarchies to
types}}

To construct the set of hierarchies of conditional beliefs we define
recursively, for each player, two sequences of sets as well as a sequence of
conditioning events. The first sequence, $\left( \Theta _{i}^{n}\right)
_{n\geq 0}$, represents player $i$'s $\left( n+1\right) $-order domain of
uncertainty, for each $n\geq 0$. The second sequence, $\left(
H_{i}^{n}\right) _{n\geq 1}$, consists of player $i$'s $n$-tuples of \textit{%
coherent} conditional beliefs over these space. The notion of coherence,
formally defined below, says that, conditional on any relevant hypothesis,
beliefs at different order do not contradict one another.

Formally, for each player $i\in I$,\ let%
\begin{equation*}
\Theta _{i}^{0}:=S\text{,}
\end{equation*}%
\begin{equation*}
\mathcal{B}_{i}^{0}:=\mathcal{B}_{i}\text{,}
\end{equation*}%
\begin{equation*}
H_{i}^{1}:=\Delta ^{\mathcal{B}_{i}^{0}}\left( \Theta _{i}^{0}\right) \text{.%
}
\end{equation*}%
The set $\Theta _{i}^{0}$ is player $i$'s $1$-order (primitive) domain of
uncertainty, and a first-order belief, viz. $\mu _{i}^{1}$, is an element of
the set $H_{i}^{1}$.

For $n\geq 1$ assume that $\left( \Theta _{i}^{m}\right) _{m=0,...,n-1}$, $%
\left( \mathcal{B}_{i}^{m}\right) _{m=0,...,n-1}$ and $\left(
H_{i}^{m}\right) _{m=1,...,n}$\ have been defined for each player $i\in I$.
Then, for each $i\in I$,\ let%
\begin{equation*}
\Theta _{i}^{n}:=\Theta _{i}^{0}\times H_{j}^{n}\text{.}
\end{equation*}%
That is, $\Theta _{i}^{n}$\ is player $i$'s $\left( n+1\right) $-order
domain of uncertainty: it consists of the space of primitive uncertainty and
what player $j\neq i$ believes about the space of primitive uncertainty,
what player $j$ believes about what player $i$ believes about the space of
primitive uncertainty,..., and so on, up to level $n$. For each $i\in I$ and 
$n\geq 1$, let $\pi _{i}^{n,n+1}:H_{i}^{n+1}\rightarrow H_{i}^{n}$\ and $%
\rho _{i}^{n-1,n}:\Theta _{i}^{n}\rightarrow \Theta _{i}^{n-1}$ denote the
coordinate projections. By construction, these maps satisfy the following
property:%
\begin{equation*}
\forall i\in I,\forall n\geq 2,\rho _{i}^{n-1,n}=\left( \mathrm{Id}_{\Theta
_{i}^{0}},\pi _{j}^{n-1,n}\right) \text{,}
\end{equation*}%
where $\mathrm{Id}_{\Theta _{i}^{0}}$\ is the identity on $\Theta _{i}^{0}$.

To define players' conditional beliefs on the $\left( n+1\right) $-th order
domain of uncertainty, for each player $i\in I$, let%
\begin{equation*}
\mathcal{B}_{i}^{n}:=\left( \rho _{i}^{n-1,n}\right) ^{-1}\left( \mathcal{B}%
_{i}^{n-1}\right) =\left \{ C\subseteq \Theta _{i}^{n}:\exists B\in \mathcal{%
B}_{i}^{n-1},C=\left( \rho _{i}^{n-1,n}\right) ^{-1}\left( B\right) \right
\} \text{,}
\end{equation*}%
\begin{equation*}
H_{i}^{n+1}:=\left \{ \left( \left( \mu _{i}^{1},...,\mu _{i}^{n}\right)
,\mu _{i}^{n+1}\right) \in H_{i}^{n}\times \Delta ^{\mathcal{B}%
_{i}^{n}}\left( \Theta _{i}^{n}\right) :\overline{\mathcal{L}}_{\rho
_{i}^{n-1,n}}\left( \mu _{i}^{n+1}\right) =\mu _{i}^{n}\right \} \text{.}
\end{equation*}%
Specifically, $\mathcal{B}_{i}^{n}$\ represents the set of relevant
hypotheses upon which player $i$'s $\left( n+1\right) $-th order conditional
beliefs are defined. That is, $\mu _{i}^{n+1}\in \Delta ^{\mathcal{B}%
_{i}^{n}}\left( \Theta _{i}^{n}\right) $ is player $i$'s $\left( n+1\right) $%
-th order CPS with $\mu _{i}^{n+1}\left( \cdot |B\right) \in \Delta \left(
\Theta _{i}^{n}\right) $, $B\in \mathcal{B}_{i}^{n}$. Recursively, it can be
checked that, for all $n\geq 1$,%
\begin{equation*}
\mathcal{B}_{i}^{n}=\left \{ C\subseteq \Theta _{i}^{n}:\exists B\in 
\mathcal{B}_{i},C=B\times H_{j}^{n}\right \} \text{,}
\end{equation*}%
i.e.,\ $\mathcal{B}_{i}^{n}$\ is a set of cylinders in $\Theta _{i}^{n}$\
generated by $\mathcal{B}_{i}$. If $\mathcal{B}_{i}$\ is clopen, then every $%
B\in \mathcal{B}_{i}^{n}$\ is clopen in $\Theta _{i}^{n}$ since each
coordinate projection $\rho _{i}^{n-1,n}$\ is a continuous function. By
definition of each $\Theta _{i}^{n}$, we write, according to Convention 1,%
\begin{equation*}
\Delta ^{\mathcal{B}_{i}^{n}}\left( \Theta _{i}^{n}\right) =\Delta ^{%
\mathcal{B}_{i}}\left( \Theta _{i}^{n}\right) \text{.}
\end{equation*}

The set $H_{i}^{n+1}$\ is the set of player $i$'s $\left( n+1\right) $%
-tuples of CPSs on $\Theta _{i}^{0}$, $\Theta _{i}^{1}$,..., $\Theta
_{i}^{n} $. The condition on $\mu _{i}^{n+1}$\ in the definition of $%
H_{i}^{n+1}$\ is the \textbf{coherence} condition mentioned above. Given the
recursive construction of the sets, CPSs $\mu _{i}^{n+1}$\ and $\mu _{i}^{n}$%
\ both specify a (countable) array of conditional beliefs on the domain of
uncertainty $\Theta _{i}^{n-1}$, and those beliefs cannot be contradictory.
Formally, for all $B\in \mathcal{B}_{i}^{n-1}$ and for every event $%
E\subseteq \Theta _{i}^{n-1}$,%
\begin{equation*}
\mu _{i}^{n+1}\left( \left( \rho _{i}^{n-1,n}\right) ^{-1}\left( E\right)
\left \vert \left( \rho _{i}^{n-1,n}\right) ^{-1}\left( B\right) \right.
\right) =\mu _{i}^{n}\left( E|B\right) \text{.}
\end{equation*}%
That is, the conditional belief $\mu _{i}^{n+1}(\cdot |(\rho
_{i}^{n-1,n})^{-1}\left( B\right) )$\ must assign to event $\left( \rho
_{i}^{n-1,n}\right) ^{-1}\left( E\right) $\ the same number as $\mu
_{i}^{n}\left( \cdot |B\right) $\ assigns to event $E$.

\begin{remark}
\label{Remark souslin compact}For each $i\in I$\ and $n\geq 1$, $H_{i}^{n+1}$%
\ is a Souslin (resp. Lusin) space provided $S$ is a Souslin (resp. Lusin)
space. If $\mathcal{B}_{i}$ is clopen for each $i\in I$, then $H_{i}^{n+1}$
is compact if and only if $S$ is compact.
\end{remark}

In the limit, for each $i\in I$, let%
\begin{align*}
H_{i}& :=\left \{ \left( \mu _{i}^{1},\mu _{i}^{2},...\right) \in
\tprod_{m=0}^{\infty }\Delta ^{\mathcal{B}_{i}}\left( \Theta _{i}^{m}\right)
:\forall n\geq 1,\left( \mu _{i}^{1},...,\mu _{i}^{n}\right) \in
H_{i}^{n}\right \} \text{,} \\
\Theta _{i}& :=S\times H_{j}\text{.}
\end{align*}

\begin{remark}
\label{Remark on closure "projective limit"}For each $i\in I$, $H_{i}$ is a
Souslin (resp. Lusin) space provided $S$ is a Souslin (resp. Lusin) space.
If $\mathcal{B}_{i}$ is clopen for each $i\in I$, then $H_{i}$ is compact if
and only if $S$ is compact.
\end{remark}

As we elaborate in Appendix B, the spaces $H_{i}$\ and $\Theta _{i}$\ ($i\in
I$) are identified with (i.e., they are homeomorphic to) the non-empty
projective limits of the sequences $\left( H_{n}^{i}\right) _{n\geq 1}$\ and 
$\left( \Theta _{n}^{i}\right) _{n\geq 0}$. The following result corresponds
to Proposition 2 in Battigalli and Siniscalchi (1999).

\begin{proposition}
\label{Proposition on the canonical homeomorphism}For each $i\in I$, the
spaces $H_{i}$\ and $\Delta ^{\mathcal{B}_{i}}\left( S\times H_{j}\right) $\
are homeomorphic.
\end{proposition}

The set $H:=\tprod_{i\in I}H_{i}$ is the set of all pairs of \textit{%
collectively coherent} hierarchies of conditional beliefs; that is, $H$\ is
the set of pairs of coherent hierarchies satisfying common full belief of
coherence.\footnote{%
An event $E$ is fully believed under a CPS $\left( \mu \left( \cdot
|B\right) \right) _{B\in \mathcal{B}}$ \ if $\mu \left( E|B\right) =1$ for
every $B\in \mathcal{B}$. The notion of \textquotedblleft common full belief
of coherence\textquotedblright \ is made explicit in the alternative,
\textquotedblleft top-down\textquotedblright \ construction of the canonical
space \textit{\`{a} la} Battigalli and Siniscalchi (1999), as we discuss in
Section \ref{Section: Discussion}. A note on terminology: in Battigalli and
Siniscalchi (1999) the expression \textquotedblleft
certainty\textquotedblright \ is used in place of \textquotedblleft full
belief.\textquotedblright}

The homeomorphisms in Proposition \ref{Proposition on the canonical
homeomorphism} are \textquotedblleft canonical\textquotedblright \ in the
following sense. Every coherent hierarchy $\left( \mu _{i}^{1},\mu
_{i}^{2},...\right) $\ of player $i$ is associated with a unique CPS $\mu
_{i}$\ on the space of primitive uncertainty and the coherent hierarchies of
the co-player, i.e., $S\times H_{j}$. Then, for all $n\geq 0$, the marginal
of $\mu _{i}$\ on player $i$'s $\left( n+1\right) $-order domain of
uncertainty, viz. $\Theta _{i}^{n}$, is precisely what it should be, namely $%
\mu _{i}^{n+1}$. A formal definition of such homeomorphisms is not needed
for the statements and proofs of the results in this paper---the reader can
consult Appendix B, in particular the proof of Proposition \ref{Proposition
on the canonical homeomorphism} for details. Instead, we will make use of
the following implication of Proposition \ref{Proposition on the canonical
homeomorphism}: we can define an $\left( S,\left( \mathcal{B}_{i}\right)
_{i\in I}\right) $-based type structure $\mathcal{T}^{\mathrm{c}}:=\left(
S,\left( \mathcal{B}_{i},T_{i}^{\mathrm{c}},\beta _{i}^{\mathrm{c}}\right)
_{i\in I}\right) $\ by letting, for each $i\in I$,%
\begin{equation*}
T_{i}^{\mathrm{c}}:=H_{i}\text{,}
\end{equation*}%
and $\beta _{i}^{\mathrm{c}}:T_{i}^{\mathrm{c}}\rightarrow \Delta ^{\mathcal{%
B}_{i}}\left( S\times T_{j}^{\mathrm{c}}\right) $\ is the \textquotedblleft
canonical\textquotedblright \ homeomorphism. Following the terminology in
the literature, we call $\mathcal{T}^{\mathrm{c}}$\ the \textbf{canonical} 
\textbf{type structure}.

The following properties of $\mathcal{T}^{\mathrm{c}}$\ are immediate from
Remark \ref{Remark on closure "projective limit"}\ and Proposition \ref%
{Proposition on the canonical homeomorphism}.

\begin{remark}
\label{Remark topology canonical type structure}Structure $\mathcal{T}^{%
\mathrm{c}}$\ is Souslin, continuous and complete. If $S$\ is Lusin, then $%
\mathcal{T}^{\mathrm{c}}$\ is Lusin. If $\mathcal{B}_{i}$\ is clopen for
each $i\in I$, then $\mathcal{T}^{\mathrm{c}}$\ is a compact type structure
if and only if $S$\ is compact.
\end{remark}

\subsubsection{From types to hierarchies\label{Section from types to
hierarchies of beliefs}}

The next step is to consider the relationship between the set of hierarchies
constructed in the previous section and any other type structure. In so
doing, we specify how types generate (collectively) coherent hierarchies of
conditional beliefs. As we did in the previous section, given a set $X$, we
let $\mathrm{Id}_{X}$\ denote\ the identity map.

Fix an $\left( S,\left( \mathcal{B}_{i}\right) _{i\in I}\right) $-based type
structure $\mathcal{T}:=\left( S,\left( \mathcal{B}_{i},T_{i},\beta
_{i}\right) _{i\in I}\right) $.\ We construct a natural (Borel) measurable
map, called \textbf{hierarchy map}, which unfolds the higher-order beliefs
of each player $i\in I$. This map assigns to each $t_{i}\in T_{i}$ a
hierarchy of beliefs in $H_{i}$.

For each $i\in I$, let $h_{-i}^{0}:\Theta _{i}^{0}\times T_{j}\rightarrow
\Theta _{i}^{0}$ be the projection map (recall that $\Theta _{i}^{0}:=S$\
for each $i\in I$). The \textquotedblleft first-order map\textquotedblright
\ for each player $i$, viz. $h_{i}^{1}:T_{i}\rightarrow H_{i}^{1}$, is
defined by%
\begin{equation*}
h_{i}^{1}\left( t_{i}\right) :=\overline{\mathcal{L}}_{h_{-i}^{0}}\left(
\beta _{i}\left( t_{i}\right) \right) \text{.}
\end{equation*}%
In words, $h_{i}^{1}\left( t_{i}\right) $\ is the marginal on $S$\ of CPS $%
\beta _{i}\left( t_{i}\right) $. Measurability of each map $h_{i}^{1}$ holds
by Lemma \ref{Lemma on general measurability of image CPS} and measurability
of belief maps.

With this, for each $i\in I$, let $h_{-i}^{1}:\Theta _{i}^{0}\times
T_{j}\rightarrow \Theta _{i}^{0}\times H_{j}^{1}=\Theta _{i}^{1}$ be the map
defined as $h_{-i}^{1}:=\left( \mathrm{Id}_{S},h_{j}^{1}\right) $. That is,
for each pair $\left( s,t_{j}\right) \in \Theta _{i}^{0}\times T_{j}$, the
expression $h_{-i}^{1}\left( s,t_{j}\right) =\left( s,h_{j}^{1}\left(
t_{j}\right) \right) \in \Theta _{i}^{1}$\ describes the profile\ $s$\ and
the first-order beliefs for type $t_{j}\in T_{j}$. Standard arguments show
that, for each $i\in I$, $h_{-i}^{1}$\ is a measurable map;\footnote{%
Since $\Theta _{i}^{0}$ and $H_{j}^{1}$\ are Souslin spaces, they are
separable; hence, the Borel $\sigma $-algebra generated by the product
topology coincides with the product $\sigma $-algebra (Bogachev 2007, Lemma
6.4.2). With this, Lemma 4.49 in Aliprantis and Border (2006) yields the
result.} furthermore, it can be checked that $h_{-i}^{1}$ satisfies%
\begin{equation*}
\mathcal{B}_{\Theta _{i}^{0}\times T_{j}}=\left( h_{-i}^{1}\right)
^{-1}\left( \mathcal{B}_{i}^{1}\right) \text{ \ and \ }h_{-i}^{0}=\rho
_{i}^{0,1}\circ h_{-i}^{1}\text{.}
\end{equation*}

Recursively, we define the \textquotedblleft $\left( n+1\right) $%
th-orders\textquotedblright \ maps. For $n\geq 1$, assume that measurable
maps $h_{i}^{n}:T_{i}\rightarrow H_{i}^{n}$ have been defined for each
player $i\in I$. Moreover, for each $i\in I$, assume that $h_{-i}^{n}:\Theta
_{i}^{0}\times T_{j}\rightarrow \Theta _{i}^{0}\times H_{j}^{n}=\Theta
_{i}^{n}$\ is the unique measurable function, defined as $h_{-i}^{n}:=\left( 
\mathrm{Id}_{S},h_{j}^{n}\right) $, which satisfies%
\begin{equation}
\mathcal{B}_{\Theta _{i}^{0}\times T_{j}}=\left( h_{-i}^{n}\right)
^{-1}\left( \mathcal{B}_{i}^{n}\right)
\label{hierarchy description conditioning events}
\end{equation}%
and%
\begin{equation}
h_{-i}^{n-1}=\rho _{i}^{n-1,n}\circ h_{-i}^{n}\text{.}
\label{Hierarchy description condition}
\end{equation}%
Fix a player $i\in I$. Note that, since (\ref{hierarchy description
conditioning events})\ holds, $\overline{\mathcal{L}}_{h_{-i}^{n}}:\Delta ^{%
\mathcal{B}_{i}}\left( S\times T_{j}\right) \rightarrow \Delta ^{\mathcal{B}%
_{i}^{n}}\left( \Theta _{i}^{n}\right) $ is a well-defined measurable map by
Lemma \ref{Lemma on general measurability of image CPS}. With this, define $%
h_{i}^{n+1}:T_{i}\rightarrow H_{i}^{n}\times \Delta ^{\mathcal{B}%
_{i}^{n}}\left( \Theta _{i}^{n}\right) $ by%
\begin{equation*}
h_{i}^{n+1}\left( t_{i}\right) :=\left( h_{i}^{n}\left( t_{i}\right) ,%
\overline{\mathcal{L}}_{h_{-i}^{n}}\left( \beta _{i}\left( t_{i}\right)
\right) \right) \text{.}
\end{equation*}%
Using the same arguments as above, it can be verified that $h_{i}^{n+1}$ is
a measurable map. Moreover, the image of $h_{i}^{n+1}$\ is contained in the
set $H_{i}^{n+1}\subseteq H_{i}^{n}\times \Delta ^{\mathcal{B}%
_{i}^{n}}\left( \Theta _{i}^{n}\right) $.

\begin{claim}
\label{Claim recursive step hierarchy map}$h_{i}^{n+1}\left( T_{i}\right)
\subseteq H_{i}^{n+1}$\ for each $i\in I$.
\end{claim}

\noindent \textbf{Proof}. We need to show that, for all $t_{i}\in T_{i}$,%
\begin{equation*}
\overline{\mathcal{L}}_{\rho _{i}^{n-1,n}}\left( \overline{\mathcal{L}}%
_{h_{-i}^{n}}\left( \beta _{i}\left( t_{i}\right) \right) \right) =\overline{%
\mathcal{L}}_{h_{-i}^{n-1}}\left( \beta _{i}\left( t_{i}\right) \right) 
\text{.}
\end{equation*}%
Pick any $t_{i}\in T_{i}$. Then, for all events $E\subseteq \Theta
_{i}^{n-1} $\ and all $B\in \mathcal{B}_{i}^{n-1}$,%
\begin{eqnarray*}
\overline{\mathcal{L}}_{\rho _{i}^{n-1,n}}\left( \overline{\mathcal{L}}%
_{h_{-i}^{n}}\left( \beta _{i}\left( t_{i}\right) \right) \right) \left(
E\left \vert B\right. \right) &=&\overline{\mathcal{L}}_{h_{-i}^{n}}\left(
\beta _{i}\left( t_{i}\right) \right) \left( \left( \rho _{i}^{n-1,n}\right)
^{-1}\left( E\right) \left \vert \left( \rho _{i}^{n-1,n}\right) ^{-1}\left(
B\right) \right. \right) \\
&=&\beta _{i}\left( t_{i}\right) \left( \left( \rho _{i}^{n-1,n}\circ
h_{-i}^{n}\right) ^{-1}\left( E\right) \left \vert \left( \rho
_{i}^{n-1,n}\circ h_{-i}^{n}\right) ^{-1}\left( B\right) \right. \right) \\
&=&\overline{\mathcal{L}}_{h_{-i}^{n-1}}\left( \beta _{i}\left( t_{i}\right)
\right) \left( E\left \vert B\right. \right) \text{,}
\end{eqnarray*}%
where the last equality follows from (\ref{Hierarchy description condition}%
).\hfill $\blacksquare $

\bigskip

It is easily seen that, for each $t_{i}\in T_{i}$,%
\begin{equation*}
h_{i}^{n+1}\left( t_{i}\right) =\left( \overline{\mathcal{L}}%
_{h_{-i}^{0}}\left( \beta _{i}\left( t_{i}\right) \right) ,...,\overline{%
\mathcal{L}}_{h_{-i}^{n-1}}\left( \beta _{i}\left( t_{i}\right) \right) ,%
\overline{\mathcal{L}}_{h_{-i}^{n}}\left( \beta _{i}\left( t_{i}\right)
\right) \right) \text{.}
\end{equation*}%
Finally, for each $i\in I$, the map $h_{i}:T_{i}\rightarrow
\tprod_{n=0}^{\infty }\Delta ^{\mathcal{B}_{i}}\left( S\times
H_{j}^{n}\right) $\ is defined by%
\begin{equation*}
h_{i}\left( t_{i}\right) :=\left( \overline{\mathcal{L}}_{h_{-i}^{n}}\left(
\beta _{i}\left( t_{i}\right) \right) \right) _{n\geq 0}\text{.}
\end{equation*}%
Thus, $h_{i}\left( t_{i}\right) $\ is the hierarchy generated by type $%
t_{i}\in T_{i}$. Each type generates a (collectively) coherent hierarchy of
beliefs, i.e., $h_{i}\left( T_{i}\right) \subseteq H_{i}$.

\begin{remark}
\label{Remark on measurability and continuity of description map}For each $%
i\in I$, the map $h_{i}:T_{i}\rightarrow H_{i}$\ is well-defined and Borel
measurable. If $\mathcal{T}$\ is continuous, then $h_{i}$\ is continuous for
each $i\in I$.
\end{remark}

For future reference, we record the following property of the hierarchy map.
We omit the proof because it is an adaptation of the proofs in Heifetz and
Samet (1998) and Battigalli and Siniscalchi (1999).\footnote{%
The Supplementary Appendix contains omitted proofs of some results/claims in
this paper.}

\begin{remark}
\label{Remark commutative diagram}The map $h:=\left( h_{i}\right) _{i\in
I}:\tprod_{i\in I}T_{i}\rightarrow \tprod_{i\in I}H_{i}$ is a type morphism
from $\mathcal{T}$\ to $\mathcal{T}^{\mathrm{c}}$. That is, for each $i\in I$
and for each $t_{i}\in T_{i}$,%
\begin{equation}
\beta _{i}^{\mathrm{c}}\left( h_{i}\left( t_{i}\right) \right) =\overline{%
\mathcal{L}}_{\left( \mathrm{Id}_{S},h_{j}\right) }\left( \beta _{i}\left(
t_{i}\right) \right) \text{.}  \label{commutative diagram}
\end{equation}
\end{remark}

\section{Terminal type structures\label{Section: terminal type structures}}

Having specified how types generate hierarchies of beliefs, we can formally
introduce the notion of (finite) terminality for conditional type
structures. The following definitions are extensions to conditional type
structures of the definitions put forward by Friedenberg (2010, Section 2)
for ordinary type structures.

\begin{definition}
\label{Finite terminality definition}An $\left( S,\left( \mathcal{B}%
_{i}\right) _{i\in I}\right) $-based type structure $\mathcal{T}:=\left(
S,\left( \mathcal{B}_{i},T_{i},\beta _{i}\right) _{i\in I}\right) $\ is 
\textbf{finitely} \textbf{terminal} if, for each $\left( S,\left( \mathcal{B}%
_{i}\right) _{i\in I}\right) $-based type structure $\mathcal{T}^{\ast
}:=\left( S,\left( \mathcal{B}_{i},T_{i}^{\ast },\beta _{i}^{\ast }\right)
_{i\in I}\right) $, each type $t_{i}^{\ast }\in T_{i}^{\ast }$\ and each $%
n\in \mathbb{N}$, there is a type $t_{i}\in T_{i}$\ such that $h_{i}^{\ast
,n}\left( t_{i}^{\ast }\right) =h_{i}^{n}\left( t_{i}\right) $.
\end{definition}

\begin{definition}
\label{Terminality definition}An $\left( S,\left( \mathcal{B}_{i}\right)
_{i\in I}\right) $-based type structure $\mathcal{T}:=\left( S,\left( 
\mathcal{B}_{i},T_{i},\beta _{i}\right) _{i\in I}\right) $\ is \textbf{%
terminal} if, for each $\left( S,\left( \mathcal{B}_{i}\right) _{i\in
I}\right) $-based type structure $\mathcal{T}^{\ast }:=\left( S,\left( 
\mathcal{B}_{i},T_{i}^{\ast },\beta _{i}^{\ast }\right) _{i\in I}\right) $\
and each type $t_{i}^{\ast }\in T_{i}^{\ast }$, there is a type $t_{i}\in
T_{i}$\ such that $h_{i}^{\ast }\left( t_{i}^{\ast }\right) =h_{i}\left(
t_{i}\right) $.
\end{definition}

Definition \ref{Finite terminality definition} says that $\mathcal{T}$ is
finitely terminal if, for every type $t_{i}^{\ast }$ that occurs in some
structure $\mathcal{T}^{\ast }$ and every $n\in \mathbb{N}$, there exists a
type $t_{i}$ in $\mathcal{T}$ whose hierarchy agrees with the hierarchy
generated by $t_{i}^{\ast }$\ up to level $n$. Definition \ref{Terminality
definition} says that $\mathcal{T}$ is terminal if, for every type $%
t_{i}^{\ast }$ that occurs in some structure $\mathcal{T}^{\ast }$, there
exists a type $t_{i}$ in $\mathcal{T}$ which generates the same hierarchy as 
$t_{i}^{\ast }$.

The notion of terminality in Definition \ref{Terminality definition} can be
equivalently expressed as follows: $\mathcal{T}$\ is terminal if, for every
structure $\mathcal{T}^{\ast }$, there exists a \textbf{hierarchy morphism}
from $\mathcal{T}^{\ast }$ to $\mathcal{T}$, i.e., a map that preserves the
hierarchies of beliefs. In Section \ref{Section: Discussion}\ we will
compare this notion of terminality with related notions in the literature.
Here we show that: (a) Definition \ref{Finite terminality definition} is
equivalent to the requirement that a type structure generates all
finite-order beliefs consistent with coherence and common full belief of
coherence; and (b) Definition \ref{Terminality definition} is equivalent to
the requirement that a type structure generates all collectively coherent
hierarchies of beliefs.

To this end, let us reformulate Definitions \ref{Finite terminality
definition}\ and \ref{Terminality definition}\ in a convenient way. The
statements in the following remark hold by inspection of the definitions.

\begin{remark}
\label{Remark on terminality definition}An $\left( S,\left( \mathcal{B}%
_{i}\right) _{i\in I}\right) $-based type structure $\mathcal{T}$\ is
finitely terminal if and only if, for each $\left( S,\left( \mathcal{B}%
_{i}\right) _{i\in I}\right) $-based type structure $\mathcal{T}^{\ast }$,
each player $i\in I$\ and each $n\in \mathbb{N}$,%
\begin{equation*}
h_{i}^{\ast ,n}\left( T_{i}^{\ast }\right) \subseteq h_{i}^{n}\left(
T_{i}\right) \text{.}
\end{equation*}%
An $\left( S,\left( \mathcal{B}_{i}\right) _{i\in I}\right) $-based type
structure $\mathcal{T}$\ is terminal if and only if, for each $\left(
S,\left( \mathcal{B}_{i}\right) _{i\in I}\right) $-based type structure $%
\mathcal{T}^{\ast }$, and for each player $i\in I$,%
\begin{equation*}
h_{i}^{\ast }\left( T_{i}^{\ast }\right) \subseteq h_{i}\left( T_{i}\right) 
\text{.}
\end{equation*}
\end{remark}

With this, the following result establishes the relationship between any
(finitely) terminal type structure and the canonical space of hierarchies.

\begin{proposition}
\label{terminality characterization}Fix an $\left( S,\left( \mathcal{B}%
_{i}\right) _{i\in I}\right) $-based type structure $\mathcal{T}:=\left(
S,\left( \mathcal{B}_{i},T_{i},\beta _{i}\right) _{i\in I}\right) $.\newline
\ \ (i) $\mathcal{T}$ is finitely terminal if and only if $h_{i}^{n}\left(
T_{i}\right) =H_{i}^{n}$ for each $i\in I$\ and each $n\in \mathbb{N}$.%
\newline
\ \ (ii) $\mathcal{T}$\ is terminal if and only if $h_{i}\left( T_{i}\right)
=H_{i}$ for each $i\in I$.
\end{proposition}

\noindent \textbf{Proof}. We show only part (ii). (The proof of part (i) is
virtually identical.) Suppose first that $\mathcal{T}$\ is terminal. Fix a
player $i\in I$. Since\ $h_{i}\left( T_{i}\right) \subseteq H_{i}$ holds by
Remark \ref{Remark on measurability and continuity of description map}, we
only need to show that\ $H_{i}\subseteq h_{i}\left( T_{i}\right) $. Consider
structure $\mathcal{T}^{\mathrm{c}}$. Then $h_{i}^{\mathrm{c}}\left(
H_{i}\right) =H_{i}\subseteq h_{i}\left( T_{i}\right) $, where the equality
holds because $h_{i}^{\mathrm{c}}:T_{i}^{\mathrm{c}}\rightarrow H_{i}$ is
the identity map, and the inclusion follows from Remark \ref{Remark on
terminality definition}.

Suppose now that $\mathcal{T}$\ satisfies $h_{i}\left( T_{i}\right) =H_{i}$
for each $i\in I$. For any structure $\mathcal{T}^{\ast }$, we have $%
h_{i}^{\ast }\left( T_{i}^{\ast }\right) \subseteq H_{i}=h_{i}\left(
T_{i}\right) $ for each $i\in I$, where the inclusion follows from Remark %
\ref{Remark on measurability and continuity of description map}. By Remark %
\ref{Remark on terminality definition}, $\mathcal{T}$\ is terminal.\hfill $%
\blacksquare $

\bigskip

Proposition \ref{terminality characterization}\ provides a characterization
of (finite) terminality which turns out to be useful for the proof of our
results. It is an analogue of Result 2.1 (and Proposition B1.(ii)) in
Friedenberg (2010).

In Section \ref{Section main result}\ we state\ and prove the main result
(Theorem \ref{Main theorem}) of this paper, that is, conditions under which
complete type structures are (finitely) terminal. In Section \ref{Section
second result}\ we address the converse question, i.e., if terminal type
structures are complete, and we prove a simple result (Theorem \ref{Theorem
terminality implies completeness}). We believe that the comparison of both
results provides a more complete understanding of the issues that arise.

\subsection{When complete structures are terminal\label{Section main result}}

The main result of this paper is the following theorem.

\begin{theorem}
\label{Main theorem}Fix an $\left( S,\left( \mathcal{B}_{i}\right) _{i\in
I}\right) $-based type structure $\mathcal{T}:=\left( S,\left( \mathcal{B}%
_{i},T_{i},\beta _{i}\right) _{i\in I}\right) $.\newline
\ \ (i) If $\mathcal{T}$\ is\ Souslin and complete, then $\mathcal{T}$ is
finitely terminal.\newline
\ \ (ii) If $\mathcal{T}$\ is complete, compact and continuous, then $%
\mathcal{T}$ is terminal.
\end{theorem}

If $\mathcal{B}_{i}\mathcal{=}\left \{ S\right \} $ for every $i\in I$, then
Theorem \ref{Main theorem}\ corresponds to Theorem 3.1\ in Friedenberg
(2010). The proof of Theorem \ref{Main theorem} relies on the following
result, whose proof---which\ makes use of Von Neumann Selection
Theorem---can be found in Appendix C.

\begin{lemma}
\label{Lemma: ontoness lubinlike}\textit{Fix Souslin spaces }$X$\textit{, }$%
Y $\textit{\ and }$Z$\textit{, and a countable family }$\mathcal{B}\subseteq
\Sigma _{X}$\textit{\ of conditioning events}. \textit{Let }$%
f_{1}:Y\rightarrow Z$\textit{\ be Borel measurable, and define }$%
f_{2}:X\times Y\rightarrow X\times Z$\textit{\ as }$f_{2}:=\left( \mathrm{Id}%
_{X},f_{1}\right) $\textit{. Then:\newline
\ \ (i) }$f_{2}$\textit{\ is Borel measurable, and the map }$\overline{%
\mathcal{L}}_{f_{2}}:\Delta ^{\mathcal{B}}\left( X\times Y\right)
\rightarrow \Delta ^{\mathcal{B}}\left( X\times Z\right) $\textit{\ is
well-defined;\newline
\ \ (ii) if }$f_{1}$\textit{\ is surjective, then }$\overline{\mathcal{L}}%
_{f_{2}}$\textit{\ is surjective.}
\end{lemma}

The proof of part (i) of Theorem \ref{Main theorem} is by induction on $n\in 
\mathbb{N}$. The proof of the base step does not rely on the hypothesis that 
$\mathcal{T}$ is Souslin. Lemma \ref{Lemma: ontoness lubinlike}\ is used 
\textit{only} in the inductive (and crucial) step. The proof of part (ii) of
Theorem \ref{Main theorem} uses the same arguments as in Friedenberg (2010).

\bigskip

\noindent \textbf{Proof of Theorem \ref{Main theorem}}. Part (i): We show
that $h_{i}^{n}\left( T_{i}\right) =H_{i}^{n}$\ for each $n\in \mathbb{N}$\
and each $i\in I$. With this, the result follows from Proposition \ref%
{terminality characterization}.(i). The proof is by induction on $n\in 
\mathbb{N}$.

(Base step: $n=1$) Fix an arbitrary type $\bar{t}_{i}\in T_{i}$ for each
player $i\in I$. Next, for each $i\in I$, let $\varphi _{i}:S\rightarrow
T_{i}$\ be the map such that $\varphi _{i}\left( s\right) =\bar{t}_{i}$\ for
every $s\in S$. Each map $\varphi _{i}$ is continuous (so measurable), since
it is a constant function. For each $i\in I$, define the function $\psi
_{i}:S\rightarrow S\times T_{j}$\ as $\psi _{i}:=\left( \mathrm{Id}%
_{S},\varphi _{j}\right) $. This function is a topological embedding, and it
satisfies $h_{-i}^{0}\circ \psi _{i}=\mathrm{Id}_{S}$ and $\mathcal{B}%
_{i}=\psi _{i}^{-1}\left( \mathcal{B}_{S\times T_{j}}\right) $. Fix a player 
$i\in I$. We show that $h_{i}^{1}\left( T_{i}\right) =H_{i}^{1}$. To this
end, pick any $\mu _{i}^{1}\in H_{i}^{1}$. By Lemma \ref{Lemma souslin space
of CPSs}.(i), $\overline{\mathcal{L}}_{\psi _{i}}\left( \mu _{i}^{1}\right)
\in \Delta ^{\mathcal{B}_{i}}\left( S\times T_{j}\right) $. Moreover, the
marginal of $\overline{\mathcal{L}}_{\psi _{i}}\left( \mu _{i}^{1}\right) $
on $S$ is $\mu _{i}^{1}$: for all events $E\subseteq S$\ and all $B\in 
\mathcal{B}_{i}$,%
\begin{eqnarray*}
\overline{\mathcal{L}}_{h_{-i}^{0}}\left( \overline{\mathcal{L}}_{\psi
_{i}}\left( \mu _{i}^{1}\right) \right) \left( E\left\vert B\right. \right)
&=&\overline{\mathcal{L}}_{\psi _{i}}\left( \mu _{i}^{1}\right) \left(
\left( h_{-i}^{0}\right) ^{-1}\left( E\right) \left\vert \left(
h_{-i}^{0}\right) ^{-1}\left( B\right) \right. \right) \\
&=&\mu _{i}^{1}\left( \left( h_{-i}^{0}\circ \psi _{i}\right) ^{-1}\left(
E\right) \left\vert \left( h_{-i}^{0}\circ \psi _{i}\right) ^{-1}\left(
B\right) \right. \right) \\
&=&\mu _{i}^{1}\left( E\left\vert B\right. \right) \text{,}
\end{eqnarray*}%
where the last equality follows from the fact that $h_{-i}^{0}\circ \psi
_{i}=\mathrm{Id}_{S}$. By completeness, there exists $t_{i}\in T_{i}$\ such
that $\beta _{i}\left( t_{i}\right) =\overline{\mathcal{L}}_{\psi
_{i}}\left( \mu _{i}^{1}\right) $; thus, $h_{i}^{1}\left( t_{i}\right) :=%
\overline{\mathcal{L}}_{h_{-i}^{0}}\left( \beta _{i}\left( t_{i}\right)
\right) =\mu _{i}^{1}$.

(Inductive step: $n\geq 1$) Suppose that the result is true for $n\geq 1$.
We show that it is also true for $n+1$. Fix a player $i\in I$ and $\left(
\mu _{i}^{1},...,\mu _{i}^{n},\mu _{i}^{n+1}\right) \in H_{i}^{n+1}$. We
need to show the existence of $t_{i}\in T_{i}$\ such that $h_{i}^{n+1}\left(
t_{i}\right) =\left( \mu _{i}^{1},...,\mu _{i}^{n},\mu _{i}^{n+1}\right) $.
First note that, by the inductive hypothesis, the function $%
h_{-i}^{n}:S\times T_{j}\rightarrow S\times H_{j}^{n}$ defined as $%
h_{-i}^{n}:=\left( \mathrm{Id}_{S},h_{j}^{n}\right) $\ is a Borel measurable
surjection. By Lemma \ref{Lemma: ontoness lubinlike}, the induced function $%
\overline{\mathcal{L}}_{h_{-i}^{n}}:\Delta ^{\mathcal{B}_{i}}\left( S\times
T_{j}\right) \rightarrow \Delta ^{\mathcal{B}_{i}}\left( S\times
H_{j}^{n}\right) $\ is surjective as well. With this, we can claim that,
since $\mu _{i}^{n+1}\in \Delta ^{\mathcal{B}_{i}^{n}}\left( \Theta
_{i}^{n}\right) =\Delta ^{\mathcal{B}_{i}}\left( S\times H_{j}^{n}\right) $,
there is $v_{i}\in \Delta ^{\mathcal{B}_{i}}\left( S\times T_{j}\right) $\
such that $\overline{\mathcal{L}}_{h_{-i}^{n}}\left( v_{i}\right) =\mu
_{i}^{n+1}$.\ Completeness yields the existence of a type $t_{i}\in T_{i}$\
such that $\beta _{i}\left( t_{i}\right) =v_{i}$. Thus, $\overline{\mathcal{L%
}}_{h_{-i}^{n}}\left( \beta _{i}\left( t_{i}\right) \right) =\mu _{i}^{n+1}$%
. Now note that, by coherence of $\mu _{i}^{n+1}$ with respect to $\mu
_{i}^{m}$ ($m\leq n$),%
\begin{equation*}
\mu _{i}^{m}=\overline{\mathcal{L}}_{h_{-i}^{m-1}}\left( \beta _{i}\left(
t_{i}\right) \right)
\end{equation*}%
for all $m\in \mathbb{N}$\ such that $1\leq m\leq n$ (this can be verified
using arguments analogous to those in Claim \ref{Claim recursive step
hierarchy map}). Thus,%
\begin{eqnarray*}
h_{i}^{n+1}\left( t_{i}\right) &=&\left( \overline{\mathcal{L}}%
_{h_{-i}^{0}}\left( \beta _{i}\left( t_{i}\right) \right) ,...,\overline{%
\mathcal{L}}_{h_{-i}^{n-1}}\left( \beta _{i}\left( t_{i}\right) \right) ,%
\overline{\mathcal{L}}_{h_{-i}^{n}}\left( \beta _{i}\left( t_{i}\right)
\right) \right) \\
&=&\left( \mu _{i}^{1},...,\mu _{i}^{n},\mu _{i}^{n+1}\right) \text{,}
\end{eqnarray*}%
as required. Since $i\in I$ is arbitrary, this concludes the proof of the
inductive step.

Part (ii): We show that $h_{i}\left( T_{i}\right) =H_{i}$ for each $i\in I$.
With this, Proposition \ref{terminality characterization}.(ii) yields the
result. Fix a player $i\in I$. Fix also an arbitrary $\bar{h}:=\left( \mu
_{i}^{1},\mu _{i}^{2},...\right) \in H_{i}$. We show the existence of $%
t_{i}\in T_{i}$\ such that $h_{i}\left( t_{i}\right) =\bar{h}$. Since $%
\mathcal{T}$\ is a continuous type structure, Remark \ref{Remark on
measurability and continuity of description map} entails that the maps $%
h_{i}^{n}$ ($n\in \mathbb{N}$) and $h_{i}$\ are continuous. For each $n\in 
\mathbb{N}$, let%
\begin{equation*}
T_{i}\left[ \bar{h}\left \vert n\right. \right] :=\left \{ t_{i}\in
T_{i}:h_{i}^{n}\left( t_{i}\right) =\left( \mu _{i}^{1},...,\mu
_{i}^{n}\right) \right \}
\end{equation*}%
be the set of types whose generated hierarchy of beliefs agree with $\bar{h}$%
\ up to level $n$. Note that $T_{i}\left[ \bar{h}\left \vert n\right. \right]
$\ is\ the inverse image under $h_{i}^{n}$ of the singleton $\left \{ \left(
\mu _{i}^{1},...,\mu _{i}^{n}\right) \right \} \subseteq H^{n}$. Also note
that $\mathcal{T}$\ is a Souslin type structure, because each type space $%
T_{i}$\ is compact metrizable. Thus, by part (i), $T_{i}\left[ \bar{h}%
\left
\vert n\right. \right] \neq \emptyset $. Moreover, by continuity of $%
h_{i}^{n}$, $T_{i}\left[ \bar{h}\left \vert n\right. \right] $ is a closed
subset of $T_{i}$. Since $T_{i}$\ is compact, so is $T_{i}\left[ \bar{h}%
\left \vert n\right. \right] $. By inspection of the definition of $%
h_{i}^{n} $, it readily follows that $T_{i}\left[ \bar{h}\left \vert
n\right. \right] \subseteq T_{i}\left[ \bar{h}\left \vert m\right. \right] $
for every $m\leq n $. Hence, $\left( T_{i}\left[ \bar{h}\left \vert n\right. %
\right] \right) _{n\in \mathbb{N}}$\ is decreasing sequence of non-empty
compact sets. By the finite intersection property, $\cap _{n\in \mathbb{N}%
}T_{i}\left[ \bar{h}\left \vert n\right. \right] \neq \emptyset $. Thus,
there exists $t_{i}\in T_{i}$\ such that $h_{i}\left( t_{i}\right) =\bar{h}$%
, as desired.\hfill $\blacksquare $

\bigskip

Some comments on Theorem \ref{Main theorem} are in order. First, complete
type structures that are finitely terminal can be easily constructed. A
simple example---which uses the ideas in Brandenburger et al. (2008, proof
of Proposition 7.2)---is the following. For each $i\in I$, let $T_{i}$ be
the Baire space, i.e., the (non-compact) Polish space $\mathbb{N}^{\mathbb{N}%
}$. Every Souslin space is the image of $\mathbb{N}^{\mathbb{N}}$\ under a
continuous function.\footnote{%
It is well-known (see, e.g., Cohn 2003, Proposition 8.2.7) that every
non-empty Polish space is the image of $\mathbb{N}^{\mathbb{N}}$\ under a
continuous function. Using this result, it is easy to check---by inspection
of definitions---that an analogous conclusion also holds for Souslin spaces
(cf. Cohn 2003, Corollary 8.2.8).} Since $\Delta ^{\mathcal{B}_{i}}\left(
S\times T_{j}\right) $\ is a Souslin space by Lemma \ref{Lemma souslin space
of CPSs}, there exists a continuous surjection $\beta _{i}:T_{i}\rightarrow
\Delta ^{\mathcal{B}_{i}}\left( S\times T_{j}\right) $. These maps give us a
Souslin and complete type structure $\mathcal{T}$ that is finitely terminal,
but not necessarily terminal. A more complex example of a(n ordinary)
complete, finitely terminal type structure which is \textit{not} terminal
can be found in Friedenberg and Keisler (2021, Section 6).

Second, we point out that complete, compact and continuous type structures
may not exist. This is because the structural hypotheses on the families of
conditioning events $\mathcal{B}_{i}$ ($i\in I$) are quite weak---each
element of $\mathcal{B}_{i}$\ is a Borel subset of $S$. To elaborate,
suppose that $\mathcal{T}$\ is complete and compact. Completeness yields $%
\beta _{i}\left( T_{i}\right) =\Delta ^{\mathcal{B}_{i}}\left( S\times
T_{j}\right) $\ for each player $i\in I$. If $\mathcal{T}$\ were continuous,
then compactness of $T_{i}$\ would imply compactness of $\Delta ^{\mathcal{B}%
_{i}}\left( S\times T_{j}\right) $\ as well, because the continuous image of
a compact set is compact. But, in general, $\Delta ^{\mathcal{B}_{i}}\left(
S\times T_{j}\right) $\ is not a compact space regardless of whether the
underlying space $S\times T_{j}$\ is compact (cf. Lemma \ref{Lemma souslin
space of CPSs}).\footnote{%
Let $X$ be a compact space. The set $\Delta ^{\mathcal{B}}\left( X\right) $\
may fail to be a closed subset of $\Delta \left( X\right) ^{\mathcal{B}}$,
which is a compact metrizable space. Yet, by Lemma \ref{Lemma souslin space
of CPSs}.(ii), $\Delta ^{\mathcal{B}}\left( X\right) $\ is a Lusin subspace
of $\Delta \left( X\right) ^{\mathcal{B}}$. In particular, $\Delta ^{%
\mathcal{B}}\left( X\right) $ is a Borel subset of $\Delta \left( X\right) ^{%
\mathcal{B}}$; \ see Remark A.1 in Appendix A.} In other words, unless each
family $\mathcal{B}_{i}$ ($i\in I$) satisfies some specific assumptions,
there is no guarantee that a complete, compact and continuous type structure
exists---in particular, complete and continuous structures may fail the
compactness requirement.\footnote{%
Analogously, complete and compact type structures may fail the continuity
requirement.} If this is the case, Theorem \ref{Main theorem}.(ii) still
holds, but \textit{vacuously} because the antecedent of the conditional is
false. An immediate implication of this fact is that the completeness
test---as formalized by completeness, compactness and continuity---cannot be
applied.

With this in mind, suppose now that each family $\mathcal{B}_{i}$ ($i\in I$%
)\ is clopen. If $S$ is a compact space, then complete, compact and
continuous type structures do exist. The canonical structure is an example,
but there are also complete, compact and continuous structures which can be
distinct from the canonical one. For instance, one can take each $T_{i}$\ to
be the Cantor space $\left \{ 0,1\right \} ^{\mathbb{N}}$, a compact
metrizable space. Lemma \ref{Lemma souslin space of CPSs}.(iii) yields that
each $\Delta ^{\mathcal{B}_{i}}\left( S\times T_{j}\right) $\ is compact
metrizable, so there exists a continuous surjection $\beta
_{i}:T_{i}\rightarrow \Delta ^{\mathcal{B}_{i}}\left( S\times T_{j}\right) $
(Aliprantis and Border 2006, Theorem 3.60). The resulting structure $%
\mathcal{T}$\ is complete, compact and continuous, so terminal.

Finally, note that if each $\mathcal{B}_{i}$\ is clopen, then compactness of 
$S$ is a necessary condition for the existence of a complete, compact and
continuous structure $\mathcal{T}$. Indeed, continuity and surjectivity of
the belief maps entail that each set $\Delta ^{\mathcal{B}_{i}}\left(
S\times T_{j}\right) $\ is compact; by Lemma \ref{Lemma souslin space of
CPSs}.(iii)\ and Tychonoff's theorem, $S$\ is a compact space.

\subsection{When terminal structures are complete\label{Section second
result}}

This section addresses the following question: If $\mathcal{T}$\ is a
terminal type structure, is $\mathcal{T}$ complete? As shown by Friedenberg
(2010), the answer is No. Specifically, Section 4.1 in Friedenberg (2010)
provides an example of an ordinary type structure which is terminal but not
complete. The key feature of Friedenberg's example is that the type
structure is redundant, i.e., there are types associated with the same
hierarchy of beliefs.\footnote{%
For ease of reference, we briefly report Friedenberg's example. Suppose that
the two-player set is $I=\left \{ a,b\right \} $\ and $S$ is a singleton,
viz. $S=\left \{ s\right \} $. Hence, there is a unique hierarchy for each
player, consisting of a sequence of Dirac point masses. Consider the type
structure $\mathcal{T}$ in which $T_{a}=\left \{ t_{a}^{\prime
},t_{a}^{\prime \prime }\right \} $, $T_{b}=\left \{ t_{b}\right \} $, $%
\beta _{a}\left( t_{a}^{\prime }\right) \left( \left \{ \left(
s,t_{b}\right) \right \} \right) =\beta _{a}\left( t_{a}^{\prime \prime
}\right) \left( \left \{ \left( s,t_{b}\right) \right \} \right) =1$, and $%
\beta _{b}\left( t_{b}\right) \left( \left \{ \left( s,t_{a}^{\prime
}\right) \right \} \right) =1$. Structure $\mathcal{T}$ generates the unique
belief hierarchy, and it is terminal. But $\mathcal{T}$ is not complete
because $\beta _{b}$\ is not surjective. It should be noted that types $%
t_{a}^{\prime }$\ and $t_{a}^{\prime \prime }$\ generate the same hierarchy.}
Proposition 4.1\ in Friedenberg (2010) shows that if an ordinary, terminal
type structure satisfies a certain\ measurability condition on the type
sets, then it is complete.

That said, we now ask: Is there an analogue of Friedenberg's result for
conditional type structures? The answer is Yes and the proof is short and
simple. To state and prove the desired result, we need a formal definition
of non-redundancy.

\begin{definition}
An $\left( S,\left( \mathcal{B}_{i}\right) _{i\in I}\right) $-based type
structure $\mathcal{T}:=\left( S,\left( \mathcal{B}_{i},T_{i},\beta
_{i}\right) _{i\in I}\right) $ is \textbf{non-redundant} if, for every $i\in
I$, the map $h_{i}:T_{i}\rightarrow H_{i}$\ is injective.
\end{definition}

In words, a structure is non-redundant if distinct types generate distinct
hierarchies of beliefs. The canonical structure $\mathcal{T}^{\mathrm{c}}$\
is non-redundant.

\begin{theorem}
\label{Theorem terminality implies completeness}Let $\mathcal{T}:=\left(
S,\left( \mathcal{B}_{i},T_{i},\beta _{i}\right) _{i\in I}\right) $\ be a
Souslin $\left( S,\left( \mathcal{B}_{i}\right) _{i\in I}\right) $-based
type structure. If $\mathcal{T}$\ is terminal and non-redundant, then $%
\mathcal{T}$ is complete.
\end{theorem}

\noindent \textbf{Proof}. If $\mathcal{T}$\ is non-redundant and terminal,
then, by Proposition \ref{terminality characterization}.(ii), each map $%
h_{i}:T_{i}\rightarrow H_{i}$\ is a Borel measurable bijection. Since $T_{i}$%
\ and $H_{i}$ are Souslin spaces (hence, $\left( T_{i},\Sigma
_{T_{i}}\right) $\ and $\left( H_{i},\Sigma _{H_{i}}\right) $ are analytic
measurable spaces), it follows from Proposition 8.6.2 in Cohn (2013) (see
also Bogachev 2007, Theorem 6.7.3, or Schwartz 1973, Lemma 16, pp. 107-108)
that each hierarchy map $h_{i}$\ is a Borel isomorphism. Fix a player $i\in
I $. Since $\beta _{i}^{\mathrm{c}}\circ h_{i}$ and $\overline{\mathcal{L}}%
_{\left( \mathrm{Id}_{S},h_{j}\right) }$\ are Borel isomorphisms, it follows
from (\ref{commutative diagram}) in Remark \ref{Remark commutative diagram}
that $\beta _{i}$\ is a Borel isomorphism. Thus, $\mathcal{T}$\ is
complete.\hfill $\blacksquare $

\bigskip

We conclude this section with three remarks on Theorem \ref{Theorem
terminality implies completeness}. First, the hypothesis that $\mathcal{T}$\
is Souslin is essential for the proof of Theorem \ref{Theorem terminality
implies completeness}: it allows us to say that each map $%
h_{i}:T_{i}\rightarrow H_{i}$\ is a Borel isomorphism, and to conclude that
also $\overline{\mathcal{L}}_{\left( \mathrm{Id}_{S},h_{j}\right) }$\ is a
Borel isomorphism. Suppose now that\ $\mathcal{T}$\ is non-redundant and
terminal, but \textit{not} Souslin. In this case, $h_{i}$\ is a Borel
measurable bijection but not necessarily an isomorphism. Is this enough to
conclude that $\overline{\mathcal{L}}_{\left( \mathrm{Id}_{S},h_{j}\right) }$%
\ is a bijection? The answer is No because $\overline{\mathcal{L}}_{\left( 
\mathrm{Id}_{S},h_{j}\right) }$ may fail to be a surjection! This issue is
discussed with an example in Appendix C.

Second, suppose that $S$\ is a Polish space as in Friedenberg's framework.
In this case, every Souslin, non-redundant and terminal type structure $%
\mathcal{T}$ has a \textquotedblleft nice\textquotedblright \ measurability
property: for every $i\in I$, $\left( T_{i},\Sigma _{T_{i}}\right) $ is
standard Borel. Indeed, note that\ $\mathcal{T}^{\mathrm{c}}$\ is a Lusin
type structure (Remark \ref{Remark topology canonical type structure}), so
that each measurable space $\left( H_{i},\Sigma _{H_{i}}\right) $ is
standard Borel. Since each hierarchy map $h_{i}:T_{i}\rightarrow H_{i}$\ is
a Borel isomorphism, it readily follows that $\left( T_{i},\Sigma
_{T_{i}}\right) $\ is standard Borel as well.\footnote{%
Put differently, we can conclude that if $S$ is a Polish space, then every
Souslin, terminal and non-redundant type structure is Lusin.} In the context
of ordinary type structures, Proposition 4.1\ in Friedenberg (2010) states
that every non-redundant, terminal structure with standard Borel type spaces
is complete. Theorem \ref{Theorem terminality implies completeness} shows
that this result remains true also for conditional type structures.

Finally, the reverse implication of Theorem \ref{Theorem terminality implies
completeness} is not true. In particular, there are complete and terminal
type structures with distinct types generating the same hierarchy of
beliefs: an example can be found in Friedenberg (2010, Section 5).

\section{Discussion\label{Section: Discussion}}

In this section we discuss some conceptual aspects of the paper, and we
compare our work with the closest related literature.

\paragraph{\noindent On the canonical type structure}

The method of the proof of the main result (Theorem \ref{Main theorem})
takes the canonical structure constructed in Section \ref{Section: Canonical
space of hierarchies}, viz. $\mathcal{T}^{\mathrm{c}}$, as a natural
benchmark for the characterization of terminality of (possibly different)
type structures. The modelling strategy used in Section \ref{Section:
Canonical space of hierarchies} to obtain $\mathcal{T}^{\mathrm{c}}$ is
complementary to the one following the approch of Brandenburger and Dekel
(1993)\ to coherence, and adopted by Battigalli and Siniscalchi (1999). This
alternative\ strategy first constructs the space of all possible (i.e., not
necessarily coherent) belief hierarchies, and then it imposes coherence and
common full belief of coherence. Such a construction, unlike the\
construction of Section \ref{Section: Canonical space of hierarchies}, is
more explicit about what is assumed to be common full belief.

Both approaches are equivalent---they yield two homeomorphic structures. We
have chosen the\ construction \textit{\`{a} la} Heifetz (1993) because it
allows to characterize the notion of (finite) terminality in a convenient
way. For instance, to formalize a statement like \textquotedblleft structure 
$\mathcal{T}$\ generates all finite order beliefs,\textquotedblright \ we
simply require that each map $h_{i}^{n}:T_{i}\rightarrow H_{i}^{n}$ be onto
for all $n\in \mathbb{N}$ (cf. Proposition \ref{terminality characterization}%
.(i)).

In terms of technical assumptions, there are differences between our
construction of $\mathcal{T}^{\mathrm{c}}$\ and the one in Battigalli and
Siniscalchi (1999). Actually, our construction is a generalization of theirs
since we allow for weaker conditions on (\textit{i}) the family of
conditioning events, and (\textit{ii}) the topological property of the
primitive uncertainty space $S$. As for (\textit{i}), Battigalli and
Siniscalchi (1999) restrict attention to clopen families of conditioning
events, while our approach requires that each conditioning event be a Borel
set. We will discuss in the next subsection why, from a conceptual
standpoint, clopeness of conditioning events could be restrictive in some
applications of interest.

As for (\textit{ii}), Battigalli and Siniscalchi (1999) consider the case,
typically adopted in the literature, of a Polish space of primitive
uncertainty. Here we considered the weaker requirement of Souslin or Lusin
spaces. There are reasons to find such a requirement useful. The main reason
is that the Lusin/Souslin properties devolve from $X$ to $\Delta ^{\mathcal{B%
}}\left( X\right) $ (see Lemma \ref{Lemma souslin space of CPSs}). An
analogous conclusion does not hold if $X$ is a Polish space---unless
clopeness of conditioning events is assumed. Yet, $\Delta ^{\mathcal{B}%
}\left( X\right) $ has a \textquotedblleft nice\textquotedblright\
topological property: it is a Lusin space, so standard Borel as a measurable
space. This is all that matters because the Borel structure on $\Delta ^{%
\mathcal{B}}\left( X\right) $\ generated by a Souslin topology on a set $X$
coincides with the \textquotedblleft natural\textquotedblright\
measure-theoretic structure on $\Delta ^{\mathcal{B}}\left( X\right) $. Let $%
\mathcal{A}^{+}$\ be the $\sigma $-algebra on $\Delta ^{\mathcal{B}}\left(
X\right) $ generated by the sets%
\begin{equation*}
b_{p}\left( E|B\right) :=\left\{ v\in \Delta ^{\mathcal{B}}\left( X\right)
:v\left( E|B\right) \geq p\right\} \text{,}
\end{equation*}%
where $E\subseteq X$ is an event, $B\in \mathcal{B}$, and $p\in \left[ 0,1%
\right] $. A statement like \textquotedblleft conditional on $B$, player $j$
assigns probability at least $p$ to $E$\textquotedblright\ corresponds
exactly to the set $b_{p}\left( E|B\right) $. As discussed by Battigalli and
Siniscalchi (1999, pp. 192-193), $\mathcal{A}^{+}$ is the conceptually
appropriate structure which allows to formalize complex statements
concerning higher-order beliefs. Appendix A shows that when $X$\ is
separable and metrizable, the Borel $\sigma $-algebra on $\Delta ^{\mathcal{B%
}}\left( X\right) $\ coincides with $\mathcal{A}^{+}$.\footnote{%
Formally, this follows from Fact A.4 in Appendix A. An analogous conclusion
holds whenever $X$ is not metrizable but it is a completely regular Souslin
space; see the Supplementary Appendix.}

To sum up, one message of this paper is the following. If we consider the
case of a Polish space of primitive uncertainty and a \textit{non-clopen}
family of conditioning events, we should not worry if \textquotedblleft
Polishness\textquotedblright \ of the higher-order spaces is lost: the
Souslin/Lusin property is always preserved, and it is well-suited for the
formal construction of a canonical structure.

\paragraph{\noindent The (non-)clopeness property of conditioning events}

In many game-theoretic applications, the conditioning events of a CPS
correspond---roughly speaking---to information sets; they can be thought of
as representing the possible evidence that a player can obtain in a game. To
illustrate, consider a sequential game with finite sets of strategies,
complete information and imperfect monitoring of past moves. If we
assume---contrary to what we did in Section \ref{Section of type structures
and belief hierarchies}---that each player $i$ is certain about his own
behavior/strategy, then his space of primitive uncertainty is the
co-player's strategy set $S_{j}$. In this environment, a conditioning event
for player $i$'s first-order beliefs is the set $S_{j}\left( h_{i}\right) $
of co-player's strategies that do no prevent an information set $h_{i}$ to
be reached;\footnote{%
If $j$ plays strategy $s_{j}\in S_{j}\left( h_{i}\right) $, then $h_{i}$ can
be reached; whether it is reached depends on $i$'s play.} that is, if $h_{i}$%
\ occurs, then $i$ is informed that the co-player is playing strategies in $%
S_{j}\left( h_{i}\right) $. Due to finiteness of the game, these
conditioning events are clopen subsets of $S_{j}$. Any type structure
associated with this game has the property that each conditioning event has
the form $S_{j}\left( h_{i}\right) \times T_{j}$,\ which is a clopen set.

Yet, there are situations under which clopeness is not satisfied. Consider
again a sequential game with a finite set of strategy profiles, imperfect
monitoring of past actions, and with \textit{incomplete} \textit{information}
about payoff functions. This incomplete-information scenario can be modelled
by the introduction of sets $\Theta _{i}$ of \textquotedblleft
information-types\textquotedblright \ for each player $i\in I$. An
information-type of player $i$, viz. $\theta _{i}\in \Theta _{i}$, describes
the private information of $i$ about payoffs: for example, $\theta _{i}$ may
be a preference or a productivity parameter. All uncertainty and private
information about payoffs is captured by the profile $\left( \theta
_{i}\right) _{i\in I}$ with the understanding that each $i$ knows only his
information-type $\theta _{i}$. With this, player $i$'s space of primitive
uncertainty is $\Theta _{j}\times S_{j}$, and a conditioning event for
player $i$'s first-order beliefs is the set $S_{j}\left( h_{i}\right) $ of
profiles $\left( \theta _{j},s_{j}\right) $ consistent with an information
set $h_{i}$.

The requirement that each conditioning event $S_{j}\left( h_{i}\right) $\ is
clopen could be restrictive in this case. One concrete example is when $%
\Theta _{j}$\ is an uncountable set, say $\Theta _{j}=\left[ 0,1\right] $,
and at the beginning of the game $i$ gets a private signal about the
realization of $\theta _{j}$. Even if the signal is discrete, the set $%
S_{j}\left( h_{i}\right) $\ may not be clopen. For instance, player $i$'s
signal reveals that $j$'s information type $\theta _{j}$\ is greater or
equal than a threshold $\hat{\theta}_{j}\in \left[ 0,1\right] $.

Other examples of non-clopen conditioning events are discussed by Battigalli
and Tebaldi (2019). These authors extend the epistemic analysis of
Battigalli and Siniscalchi (2002) from finite sequential games to a special
class of infinite sequential games---called \textit{simple dynamic games}%
---where the clopeness assumption is satisfied and conceptually
well-motivated. Battigalli and Tebaldi (2002, Section 5.6) address the
question whether their results still hold in a more general set-up (games
and type structures) whereby clopeness is not satisfied. We think that the
canonical structure constructed in this paper may be used to extend their
results to a wide class of infinite sequential games.

\paragraph{\noindent Hierarchy vs. type morphisms}

The notion of terminality used in this paper reflects the following
requirement on a type structure: $\mathcal{T}$\ is terminal if there exists
a hierarchy morphism---i.e., a map which preserves the hierarchies of
beliefs---from every type structure $\mathcal{T}^{\ast }$ to $\mathcal{T}$.
A distinct notion of terminality, often called universality (e.g., Heifetz
and Samet 1998, Battigalli and Siniscalchi 1999), requires the existence of
a \textit{unique} type morphism---i.e., a map which preserves the belief
maps---from every type structure $\mathcal{T}^{\ast }$ to $\mathcal{T}$. The
latter notion is stronger than the former: by a straightforward extension of
the proofs in Heifetz and Samet (1998)\ to conditional type structures, it
can be shown that a type morphism is a hierarchy morphism; the converse,
however, does not hold. The canonical structure $\mathcal{T}^{\mathrm{c}}$\
is terminal/universal because the hierarchy map is a type morphism (cf.
Remark \ref{Remark commutative diagram}) and, as it can be easily proved by
standard techniques, it is also unique.

What is the relationship between terminal structures and the
canonical/universal type structure $\mathcal{T}^{\mathrm{c}}$? Fix any
terminal type structure $\mathcal{T}$. In light of the above discussion,
there is a unique type\ morphism from $\mathcal{T}$ to $\mathcal{T}^{\mathrm{%
c}}$. But there is also a hierarchy morphism from $\mathcal{T}^{\mathrm{c}}$
to $\mathcal{T}$. This hierarchy morphism need not be unique because $%
\mathcal{T}$\ may be redundant (cf. Theorem \ref{Theorem terminality implies
completeness}). If this is the case, there could be many
hierarchy-preserving maps from $\mathcal{T}^{\mathrm{c}}$ to $\mathcal{T}$
that do not preserve the beliefs maps; that is, the commutativity property
of a type morphism---see (\ref{commutative diagram}) in Remark \ref{Remark
commutative diagram}---need not be satisfied. In particular, there could be 
\textit{no} type morphism from $\mathcal{T}^{\mathrm{c}}$ to $\mathcal{T}$.

To understand this issue, we briefly review a known result in the
literature. In the context of ordinary type structures, Friedenberg and
Meier (2011) provide a necessary and sufficient condition, called \textit{%
strong measurability}, under which hierarchy and type morphisms coincide: a
hierarchy morphism from $\mathcal{T}^{\ast }$ to $\mathcal{T}$\ is a type
morphism if and only if $\mathcal{T}$\ is strongly measurable. They also
provide sufficient conditions on structure $\mathcal{T}$\ which guarantee
strong measurability. For instance, if both the space of primitive
uncertainty and the type spaces are standard Borel spaces, then strong
measurability is equivalent to non-redundancy of $\mathcal{T}$.

With this in mind, consider a Polish space of primitive uncertainty, and
assume that: (1) $\mathcal{T}^{\mathrm{c}}$ to $\mathcal{T}$ are ordinary
Lusin type structures, and (2) $\mathcal{T}$ is terminal and redundant.
Then, Friedenberg and Meier (2011)'s results entail that there is no type
morphism from $\mathcal{T}$ to $\mathcal{T}^{\mathrm{c}}$. We suspect (but
have not proved) that similar results are also valid for conditional type
structures, although Friedenberg and Meier (2011)'s proof cannot directly
extended to our case. This extension requires additional investigation which
is beyond the scope of this paper.

\paragraph{\noindent Complete type structures and first-order beliefs}

As observed in Section \ref{Section main result}, a careful reading of the
base step in the proof of Theorem \ref{Main theorem}.(i) reveals that the
hypothesis that $\mathcal{T}$\ is a Souslin structure is not used. Put
differently, the proof of the base step shows that every complete type
structure generates all first-order beliefs, and the result remains true
under the assumption that $S$\ is a metrizable space. This observation
suggests that, in the topology-free framework \textit{\`{a} la} Heifetz and
Samet (1998), a complete type structure generates all first-order
conditional beliefs. The intuition is correct, and the proof---with obvious
modifications---is similar.\footnote{%
To provide a topology-free definition of conditional type structures, the
following changes should be made in Definition \ref{Definition of type
structure}.\ Every type set $T_{i}$\ is endowed with a $\sigma $-algebra $%
\mathcal{A}_{i}$, and every belief map $\beta _{i}$\ is $\mathcal{A}_{i}$%
-measurable. For any measurable space $\left( X,\Sigma _{X}\right) $, the
set $\Delta ^{\mathcal{B}}\left( X\right) $\ is endowed with the $\sigma $%
-algebra $\mathcal{A}^{+}$\ described above. Any product of measurable
spaces is endowed with the product $\sigma $-algebra.} This conclusion is
the only possible one if the topological assumptions in this paper are
relaxed. Indeed, Theorem 5 in Friedenberg and Keisler (2021) shows a
powerful negative result: Within Zermelo-Fraenkel set theory with the axiom
of choice, one cannot prove that a complete (ordinary) type structure
generates all second-order beliefs.

\paragraph{\noindent Conditional beliefs as consistent CPSs}

Siniscalchi (2020, 2022) put forward a notion of rationality in sequential
games, called \textit{structural rationality}, that allows the elicitation
of players' (higher-order) beliefs both on and off the predicted path of
play. A key aspect of Siniscalchi's analysis is that the players' beliefs
are represented by consistent conditional probability systems (CCPSs). A
CCPS is a CPS satisfying a requirement that is stronger than the chain rule,
since it imposes a more stringent consistency condition on beliefs across
different conditioning events. With appropriate changes in some proofs, it
can be shown (see the Supplementary Appendix) that \textit{all the results
in this paper still hold if beliefs are represented by CCPSs}. Consequently,
we propose that our results can be fruitfully applied to epistemic analyses
of solution concepts based on the notion of structural rationality (cf.
Siniscalchi 2023).

\paragraph{\noindent Non-metrizable spaces}

Technically, we have restricted our analysis to metrizable topological
spaces as this simplifies some of the proofs. Actually, all of our results
are valid in a more general setup where each topological space is assumed to
be a perfectly normal Hausdorff space. Such a generalization can be found in
the Supplementary Appendix.

\section*{Appendix A. Proofs for Section \protect\ref{Section CPS}}

We first provide the proof of Lemma \ref{Lemma souslin space of CPSs}, which
relies on three mathematical facts.

\bigskip

\noindent \textbf{Fact A.1} \textit{Fix a metrizable space }$X$\textit{.
Then }$\Delta \left( X\right) $\textit{\ is metrizable. Furthermore, }$%
\Delta \left( X\right) $\textit{\ is Souslin (resp. Lusin, compact) if and
only if }$X$\textit{\ is Souslin (resp. Lusin, compact).}

\bigskip

\noindent \textbf{Fact A.2} \textit{Fix a metrizable space }$X$\textit{\ and
a countable family }$\mathcal{B}\subseteq \Sigma _{X}$\textit{\ of
conditioning events. If }$\mathcal{B}$\textit{\ is clopen, then }$\Delta ^{%
\mathcal{B}}\left( X\right) $\textit{\ is a closed subset of }$\Delta \left(
X\right) ^{\mathcal{B}}$\textit{. }

\bigskip

\noindent \textbf{Fact A.3 }\textit{Suppose }$X$\textit{\ is a metrizable
Souslin (resp. Lusin) space. Then for every countable family }$\mathcal{C}$%
\textit{\ of Borel subsets of }$X$\textit{\ there exists a finer Souslin
(resp. Lusin), metrizable topology on }$X$\textit{\ generating the same
Borel }$\sigma $\textit{-algebra and making each }$C\in \mathcal{C}$\textit{%
\ clopen.}

\bigskip

Fact A1 is well-known; see Bogachev (2007, Chapter 8.9). The proof of Fact
A.2 can be found in Battigalli and Siniscalchi (1999, Lemma 1): indeed,
their proof for the statement in Fact A.2 remains valid if $X$ is only
assumed to be a metrizable space, not necessarily Polish.\footnote{%
Specifically, metrizability of $X$ implies that $\Delta \left( X\right) $\
is metrizable. Since $\mathcal{B}$\ is countable, the product topology on $%
\Delta \left( X\right) ^{\mathcal{B}}$ is metrizable as well. The proof in
Battigalli and Siniscalchi (1999) relies on two relevant properties implied
by metrizability of the space $X$. First, every Borel probability measure is
both outer regular and inner regular---see Aliprantis and Border (2006,
Theorem 12.5). Second, every metrizable space is a normal space, and this
yields the existence of a Urysohn function satisfying the required
properties for some steps of the proof. Finally, the assumption that $%
\mathcal{B}$\ is clopen allows to apply the Portmanteau Theorem for
continuity sets (Bogachev 2007, Corollary 8.2.10).} The proof of Fact A.3
follows from simple modifications of the results in Srivastava (1998,
Section 3.2).\footnote{%
The results in Srivastava (1998) are stated for Polish spaces. But the
arguments in the proofs can be easily adapted to show that the results
continue to hold with the weaker requirement of Souslin or Lusin topology.
Details can be found in the Supplementary Appendix.}

\bigskip

\noindent \textbf{Proof of Lemma \ref{Lemma souslin space of CPSs}}. Since $%
\mathcal{B}$\ is countable and $\Delta \left( X\right) $\ is metrizable
(Fact A.1), the product topology on $\Delta \left( X\right) ^{\mathcal{B}}$\
is metrizable. By definition of subspace topology, $\Delta ^{\mathcal{B}%
}\left( X\right) $\ is a metrizable space, and this proves part (i). Next,
recall that the class of Souslin (Lusin, compact) spaces is closed under
countable products. With this, Fact A.1 yields that $\Delta \left( X\right)
^{\mathcal{B}}$\ is Souslin (resp. Lusin, compact) provided that $X$ is
Souslin (resp. Lusin, compact). Moreover, closed subsets of a Souslin (resp.
Lusin) spaces are Souslin (resp. Lusin) spaces when endowed with the
subspace topology. It follows from Fact A.2 that, if $\mathcal{B}$\ is
clopen and $X$ is Souslin (Lusin),\ so is $\Delta ^{\mathcal{B}}\left(
X\right) $. We make use of this result to show part (ii), that is, $\Delta ^{%
\mathcal{B}}\left( X\right) $ is Souslin (Lusin) subspace of $\Delta \left(
X\right) ^{\mathcal{B}}$ even if $\mathcal{B}$\ is not clopen.

By Fact A.3, there exists a metrizable Souslin (Lusin) topology on the
Souslin (Lusin) space $X$ which is finer than the original one, and such
that: (a) it generates the same Borel $\sigma $-algebra, and (b) it makes
each element of $\mathcal{B}$\ clopen. Let $X^{\ast }$ denote the
topological space endowed with this topology. Note that $\Delta \left(
X\right) =\Delta \left( X^{\ast }\right) $ because the generated Borel $%
\sigma $-algebra is the same for both spaces. The topology on $\Delta \left(
X^{\ast }\right) $\ is finer than the topology on $\Delta \left( X\right) $.
To see this, consider a sequence $\left( \mu _{n}\right) _{n\in \mathbb{N}}$
in $\Delta \left( X^{\ast }\right) $\ that converges weakly to $\mu $. By
definition, $\tint fd\mu _{n}\rightarrow \tint fd\mu $ for every continuous
function $f:X^{\ast }\rightarrow \mathbb{R}$. Since every continuous
function $f:X\rightarrow \mathbb{R}$\ is also continuous on $X^{\ast }$, it
follows that $\left( \mu _{n}\right) _{n\in \mathbb{N}}$\ converges weakly\
to $\mu $\ in $\Delta \left( X\right) $.

With this, it follows that the product topology on $\Delta \left( X^{\ast
}\right) ^{\mathcal{B}}$\ is finer than the product topology on $\Delta
\left( X\right) ^{\mathcal{B}}$. This entails that the identity map from $%
\Delta \left( X^{\ast }\right) ^{\mathcal{B}}$\ to $\Delta \left( X\right) ^{%
\mathcal{B}}$ is a continuous bijection. Thus, $\Delta ^{\mathcal{B}}\left(
X\right) $\ is the image of $\Delta ^{\mathcal{B}}\left( X^{\ast }\right) $
(a Souslin space) under a injective continuous map. Since $\Delta ^{\mathcal{%
B}}\left( X^{\ast }\right) $\ is the image of a complete, separable metric
space under a continuous map, it follows that $\Delta ^{\mathcal{B}}\left(
X\right) $\ is a Souslin space. In particular, if $X$ is Lusin, then $\Delta
^{\mathcal{B}}\left( X^{\ast }\right) $\ is the image of a complete,
separable metric space under a continuous injective map; hence, $\Delta ^{%
\mathcal{B}}\left( X\right) $ is Lusin. This concludes the proof of part
(ii).

The proof of part (iii) is immediate. By Fact A.1 and Tychonoff's theorem, $%
X $ is compact if and only if $\Delta \left( X\right) ^{\mathcal{B}}$\ is
compact. If $\mathcal{B}$\ is clopen, then the result follows from Fact
A.2.\hfill $\blacksquare $

\bigskip

\noindent \textbf{Remark A.1 }\textit{If }$X$\textit{\ is a Polish space and 
}$\mathcal{B}$\textit{\ is countable and clopen, then }$\Delta ^{\mathcal{B}%
}\left( X\right) $\textit{\ is a Polish space (cf. Battigalli and
Siniscalchi 1999). An analogous conclusion does not hold if }$\mathcal{B}$%
\textit{\ is assumed only countable. Yet, by Lemma \ref{Lemma souslin space
of CPSs}, }$\Delta ^{\mathcal{B}}\left( X\right) $\textit{\ is a Lusin
space. Moreover, it can be deduced from the above proof that }$\Delta ^{%
\mathcal{B}}\left( X\right) $\textit{\ is a Borel subset of }$\Delta \left(
X\right) ^{\mathcal{B}}$\textit{. To see this, recall that }$X^{\ast }$%
\textit{\ is the space obtained by endowing the set }$X$\textit{\ with a
Souslin topology satisfying the same properties as in Fact A.3. Both }$%
\Delta \left( X^{\ast }\right) ^{\mathcal{B}}$\textit{\ and }$\Delta \left(
X\right) ^{\mathcal{B}}$\textit{\ are Souslin spaces, and the identity map
from }$\Delta \left( X^{\ast }\right) ^{\mathcal{B}}$\textit{\ to }$\Delta
\left( X\right) ^{\mathcal{B}}$\textit{\ is a continuous bijection. Since
every Borel measurable bijection between Souslin spaces is a Borel
isomorphism (Cohn 2013, Proposition 8.6.2; Bogachev 2007, Theorem 6.7.3),
and }$\Delta ^{\mathcal{B}}\left( X^{\ast }\right) $\textit{\ is a closed
(hence, Borel) subset of }$\Delta \left( X^{\ast }\right) ^{\mathcal{B}}$%
\textit{, the result follows.}

\bigskip

Next, we provide the proof of Lemma \ref{Lemma on general measurability of
image CPS}. To this end, we need some additional definitions and a result.

Fix a metrizable space $X$. Let $\mathcal{F}_{\Delta \left( X\right) }$\ be
the $\sigma $-algebra\ on $\Delta \left( X\right) $ generated by all sets of
the form%
\begin{equation*}
\left \{ \mu \in \Delta \left( X\right) :\mu \left( E\right) \geq p\right \} 
\text{,}
\end{equation*}%
where $E\in \Sigma _{X}$ and $p\in \mathbb{Q\cap }\left[ 0,1\right] $. We
let $\Sigma _{\Delta \left( X\right) }$\ denote the Borel $\sigma $-algebra
on $\Delta \left( X\right) $\ generated by the topology of weak convergence.
With this, fix a countable family $\mathcal{B}\subseteq \Sigma _{X}$\ of
conditioning events, and, for each $B\in \mathcal{B}$, let $e_{B}:\Delta
\left( X\right) ^{\mathcal{B}}\rightarrow \Delta \left( X\right) $ denote
the coordinate projection. We use the following notation for three $\sigma $%
-algebras on the set $\Delta \left( X\right) ^{\mathcal{B}}$:

\begin{itemize}
\item $\mathcal{F}_{\Delta \left( X\right) ^{\mathcal{B}}}$ is the $\sigma $%
-algebra generated by the cylinder algebra $\cup _{B\in \mathcal{B}%
}e_{B}^{-1}\left( \mathcal{F}_{\Delta \left( X\right) }\right) $;

\item $\mathcal{G}_{\Delta \left( X\right) ^{\mathcal{B}}}$ is\ the $\sigma $%
-algebra generated by the cylinder algebra $\cup _{B\in \mathcal{B}%
}e_{B}^{-1}\left( \Sigma _{\Delta \left( X\right) }\right) $;

\item $\Sigma _{\Delta \left( X\right) ^{\mathcal{B}}}$\ is the Borel $%
\sigma $-algebra generated by the product topology on $\Delta \left(
X\right) ^{\mathcal{B}}$.
\end{itemize}

\bigskip

\noindent \textbf{Fact A.4 }\textit{Fix\ a metrizable space }$X$\textit{\
and a countable family }$\mathcal{B}\subseteq \Sigma _{X}$\textit{\ of
conditioning events. The following statements hold.}\newline
\textit{\ \ (i)} $\mathcal{F}_{\Delta \left( X\right) ^{\mathcal{B}}}$\ 
\textit{is generated by all sets of the form}%
\begin{equation*}
\left \{ \mu \in \Delta \left( X\right) ^{\mathcal{B}}:\mu \left( E|B\right)
\geq p\right \} \text{,}
\end{equation*}%
\textit{where }$E\in \Sigma _{Y}$\textit{, }$B\in B$\textit{\ and }$p\in 
\mathbb{Q}\cap \left[ 0,1\right] $\textit{.}\newline
\textit{\ \ (ii) }$\mathcal{F}_{\Delta \left( X\right) ^{\mathcal{B}%
}}\subseteq \mathcal{G}_{\Delta \left( X\right) ^{\mathcal{B}}}\subseteq
\Sigma _{\Delta \left( X\right) ^{\mathcal{B}}}$\textit{.}\newline
\textit{\ \ (iii) If }$X$\textit{\ is separable, then }$\mathcal{F}_{\Delta
\left( X\right) ^{\mathcal{B}}}=\mathcal{G}_{\Delta \left( X\right) ^{%
\mathcal{B}}}=\Sigma _{\Delta \left( X\right) ^{\mathcal{B}}}$\textit{.}

\bigskip

\noindent \textbf{Proof}. Part (i) follows by inspection of the definitions
of $\mathcal{F}_{\Delta \left( X\right) }$ and $\mathcal{F}_{\Delta \left(
X\right) ^{\mathcal{B}}}$. To show part (ii), first note that, by Theorem
2.2 in Gaudard and Handwin (1989), $\mathcal{F}_{\Delta \left( X\right)
}\subseteq \Sigma _{\Delta \left( X\right) }$. This entails that $\mathcal{F}%
_{\Delta \left( X\right) ^{\mathcal{B}}}\subseteq \mathcal{G}_{\Delta \left(
X\right) ^{\mathcal{B}}}$. Moreover, by Lemma 6.4.1 in Bogachev (2007), $%
\mathcal{G}_{\Delta \left( X\right) ^{\mathcal{B}}}\subseteq \Sigma _{\Delta
\left( X\right) ^{\mathcal{B}}}$. As for part (iii), assume that $X$\ is
separable. By Theorem 2.3 in Gaudard and Handwin (1989), $\mathcal{F}%
_{\Delta \left( X\right) }=\Sigma _{\Delta \left( X\right) }$. This yields $%
\mathcal{F}_{\Delta \left( X\right) ^{\mathcal{B}}}=\mathcal{G}_{\Delta
\left( X\right) ^{\mathcal{B}}}$. Furthermore, separability of $X$ implies
separability of $\Delta \left( X\right) $ (Bogachev 2007, Theorem 8.9.4).
Hence, by Lemma 6.4.2 in Bogachev (2007), $\mathcal{G}_{\Delta \left(
X\right) ^{\mathcal{B}}}=\Sigma _{\Delta \left( X\right) ^{\mathcal{B}}}$%
.\hfill $\blacksquare $

\bigskip

\noindent \textbf{Proof of Lemma \ref{Lemma on general measurability of
image CPS}}. We begin with part (i) of the lemma. Fix some $\mu \in \Delta ^{%
\mathcal{B}_{X}}\left( X\right) $. We show that $\overline{\mathcal{L}}%
_{f}\left( \mu \right) $\ is a CPS on $\left( Y,\Sigma _{Y},\mathcal{B}%
_{Y}\right) $. Pick any $B\in \mathcal{B}_{Y}$. Since $f^{-1}\left( \mathcal{%
B}_{Y}\right) =\mathcal{B}_{X}$, we have $f^{-1}\left( B\right) \in \mathcal{%
B}_{X}$ and%
\begin{equation*}
\overline{\mathcal{L}}_{f}\left( \mu \right) \left( B|B\right) =\mu \left(
f^{-1}\left( B\right) |f^{-1}\left( B\right) \right) =1\text{,}
\end{equation*}%
because $\mu $\ is a CPS. Thus, Condition (i) in Definition \ref{Definition
CPS} holds.To show that the chain rule is satisfied, consider any $A\in
\Sigma _{Y}$\ and $B,C\in \mathcal{B}_{Y}$\ such that $A\subseteq B\subseteq
C$. We have $f^{-1}\left( B\right) $, $f^{-1}\left( C\right) \in \mathcal{B}%
_{X}$. Furthermore, $f^{-1}\left( A\right) \in \Sigma _{X}$\ and $%
f^{-1}\left( A\right) \subseteq f^{-1}\left( B\right) \subseteq f^{-1}\left(
C\right) $. Using the fact that $\mu $\ is a CPS, we get%
\begin{eqnarray*}
\left( \overline{\mathcal{L}}_{f}\left( \mu \right) \left( A|B\right)
\right) \left( \overline{\mathcal{L}}_{f}\left( \mu \right) \left(
B|C\right) \right) &=&\mu \left( f^{-1}\left( A\right) |f^{-1}\left(
B\right) \right) \mu \left( f^{-1}\left( B\right) |f^{-1}\left( C\right)
\right) \\
&=&\mu \left( f^{-1}\left( A\right) |f^{-1}\left( C\right) \right) \\
&=&\overline{\mathcal{L}}_{f}\left( \mu \right) \left( A|C\right) \text{,}
\end{eqnarray*}%
as required.

Now we turn to part (ii) of the lemma. Suppose that $f$\ is Borel
measurable. To show that $\overline{\mathcal{L}}_{f}$\ is Borel measurable,
first note the following facts. Consider the set $\Delta \left( Y\right) ^{%
\mathcal{B}_{Y}}$. Since every Souslin space is separable, it follows from
Fact A.4.(iii) that $\mathcal{F}_{\Delta \left( Y\right) ^{\mathcal{B}%
_{Y}}}=\Sigma _{\Delta \left( Y\right) ^{\mathcal{B}_{Y}}}$. Hence, by Fact
A.4.(i), the Borel $\sigma $-algebra on $\Delta ^{\mathcal{B}_{Y}}\left(
Y\right) $\ is the relative $\sigma $-algebra of $\mathcal{F}_{\Delta \left(
Y\right) ^{\mathcal{B}_{Y}}}$ on $\Delta ^{\mathcal{B}_{Y}}\left( Y\right) $%
, and it is generated by all sets of the form%
\begin{equation*}
b_{p}\left( E|B\right) :=\left \{ v\in \Delta ^{\mathcal{B}_{Y}}\left(
Y\right) :v\left( E|B\right) \geq p\right \} \text{,}
\end{equation*}%
where $E\in \Sigma _{Y}$, $B\in \mathcal{B}_{Y}$\ and $p\in \mathbb{Q\cap }%
\left[ 0,1\right] $. We now claim that $\overline{\mathcal{L}}%
_{f}^{-1}\left( b_{p}\left( E|B\right) \right) $ belongs to the relative $%
\sigma $-algebra of $\mathcal{F}_{\Delta \left( X\right) ^{\mathcal{B}_{X}}}$
on $\Delta ^{\mathcal{B}_{X}}\left( X\right) $. Since $\mathcal{F}_{\Delta
\left( X\right) ^{\mathcal{B}_{X}}}\subseteq \Sigma _{\Delta \left( X\right)
^{\mathcal{B}_{X}}}$\ by Fact A.4.(ii), this will entail that $\overline{%
\mathcal{L}}_{f}^{-1}\left( b_{p}\left( E|B\right) \right) $\ is a Borel
subset of $\Delta ^{\mathcal{B}_{X}}\left( X\right) $ (i.e., it belongs to
the relative $\sigma $-algebra of $\Sigma _{\Delta \left( X\right) ^{%
\mathcal{B}_{X}}}$ on $\Delta ^{\mathcal{B}_{X}}\left( X\right) $). Note that%
\begin{eqnarray*}
\overline{\mathcal{L}}_{f}^{-1}\left( b_{p}\left( E|B\right) \right)
&=&\left \{ \mu \in \Delta ^{\mathcal{B}_{X}}\left( X\right) :\overline{%
\mathcal{L}}_{f}\left( \mu \right) \in b_{p}\left( E|B\right) \right \} \\
&=&\left \{ \mu \in \Delta ^{\mathcal{B}_{X}}\left( X\right) :\mu \left(
f^{-1}\left( E\right) |f^{-1}\left( B\right) \right) \geq p\right \} \text{.}
\end{eqnarray*}%
Since $f$ is measurable and $f^{-1}\left( \mathcal{B}_{Y}\right) =\mathcal{B}%
_{X}$, we have $f^{-1}\left( E\right) \in \Sigma _{X}$\ and $f^{-1}\left(
B\right) \in \mathcal{B}_{X}$. The claim\ follows from the definition of
relative $\sigma $-algebra of $\mathcal{F}_{\Delta \left( X\right) ^{%
\mathcal{B}_{X}}}$ on $\Delta ^{\mathcal{B}_{X}}\left( X\right) $. Finally,
as $b_{p}\left( E|B\right) $\ is arbitrary and it belongs to the system of
generators of the Borel $\sigma $-algebra on $\Delta ^{\mathcal{B}%
_{Y}}\left( Y\right) $, standard results (e.g., Aliprantis and Border 2006,
Corollary 4.24) yield measurability of $\overline{\mathcal{L}}_{f}$.

Suppose now that $f$ is continuous. Consider any sequence $\left( \mu
_{n}\right) _{n\in \mathbb{N}}$ in $\Delta ^{\mathcal{B}_{X}}\left( X\right) 
$ converging to $\mu $. It must be shown that $\overline{\mathcal{L}}%
_{f}\left( \mu _{n}\right) $ converges to $\overline{\mathcal{L}}_{f}\left(
\mu \right) $\ as $n\rightarrow \infty $. That is, $\mathcal{L}_{f}\left(
\mu _{n}\left( .|f^{-1}\left( C\right) \right) \right) $ converges to $%
\mathcal{L}_{f}\left( \mu \left( .|f^{-1}\left( C\right) \right) \right) $\
for every $C\in \mathcal{B}_{Y}$. This readily follows from the fact that
the map $\mathcal{L}_{f}:\Delta \left( X\right) \rightarrow \Delta \left(
Y\right) $\ is continuous (Bogachev 2007, Theorem 8.4.1.(i)).\hfill $%
\blacksquare $

\section*{Appendix B. Proofs for Section \protect\ref{Section of type
structures and belief hierarchies}}

We first provide the proofs of Remarks \ref{Remark souslin compact}-\ref%
{Remark on closure "projective limit"} and Remark \ref{Remark on
measurability and continuity of description map}.

\bigskip

\noindent \textbf{Proof of Remark \ref{Remark souslin compact}}. Consider
first the case when $S$ is a Souslin space. We show, by induction on $n\geq
1 $, that the following statements hold for every $i\in I$:

(i) $\tprod_{m=0}^{n-1}\Delta ^{\mathcal{B}_{i}}\left( \Theta
_{i}^{m}\right) $\ is a Souslin space, and

(ii) $H_{i}^{n}$ is a closed subset of $\tprod_{m=0}^{n-1}\Delta ^{\mathcal{B%
}_{i}}\left( \Theta _{i}^{m}\right) $.

The proof for $n=1$ is immediate from the definition of $H_{i}^{1}$\ and
Lemma \ref{Lemma souslin space of CPSs}.(ii). Thus, suppose that the result
is true for $n\geq 1$. We show that it is true for $n+1$. Fix a player $i\in
I$. By the inductive hypothesis, $H_{j}^{n}$ is a closed subset of $%
\tprod_{m=0}^{n-1}\Delta ^{\mathcal{B}_{j}}\left( \Theta _{j}^{m}\right) $,
so a Souslin space. With this, $\Theta _{i}^{n}=\Theta _{i}^{0}\times
H_{j}^{n}$ is Souslin as well. By Lemma \ref{Lemma souslin space of CPSs}%
.(ii), $\Delta ^{\mathcal{B}_{i}}\left( \Theta _{i}^{n}\right) $\ is a
Souslin space. Using again the inductive hypothesis, we conclude that $%
\tprod_{m=0}^{n}\Delta ^{\mathcal{B}_{i}}\left( \Theta _{i}^{m}\right) $\ is
a Souslin space, establishing (i). To show (ii), pick any sequence $\left(
\left( \mu _{i,m}^{1},...,\mu _{i,m}^{n}\right) ,\mu _{i,m}^{n+1}\right)
_{m\in \mathbb{N}}$ in $H_{i}^{n+1}$ converging to $\left( \left( \mu
_{i}^{1},...,\mu _{i}^{n}\right) ,\mu _{i}^{n+1}\right) \in
\tprod_{m=0}^{n}\Delta ^{\mathcal{B}_{i}}\left( \Theta _{i}^{m}\right) $.
Thus, $\left( \mu _{i}^{1},...,\mu _{i}^{n}\right) \in H_{i}^{n}$ because,
by the inductive hypothesis, $H_{i}^{n}$\ is a closed subset of $%
\tprod_{m=0}^{n-1}\Delta ^{\mathcal{B}_{i}}\left( \Theta _{i}^{m}\right) $.
Note that $\overline{\mathcal{L}}_{\rho _{i}^{n-1,n}}\left( \mu
_{i,m}^{n+1}\right) =\mu _{i,m}^{n}$\ for each $m\in \mathbb{N}$. Continuity
of the map $\rho _{i}^{n-1,n}$ yields---by Lemma \ref{Lemma on general
measurability of image CPS}.(ii)---continuity of $\overline{\mathcal{L}}%
_{\rho _{i}^{n-1,n}}$. Hence,%
\begin{equation*}
\mu _{i}^{n}=\lim_{m\rightarrow \infty }\mu _{i,m}^{n}=\lim_{m\rightarrow
\infty }\overline{\mathcal{L}}_{\rho _{i}^{n-1,n}}\left( \mu
_{i,m}^{n+1}\right) =\overline{\mathcal{L}}_{\rho _{i}^{n-1,n}}\left( \mu
_{i}^{n+1}\right) \text{.}
\end{equation*}%
This shows that $\left( \left( \mu _{i}^{1},...,\mu _{i}^{n}\right) ,\mu
_{i}^{n+1}\right) \in H_{i}^{n+1}$, establishing (ii).

The Lusin case is identical. Finally, if each $\mathcal{B}_{i}$\ is clopen,
the compact case follows from Lemma \ref{Lemma souslin space of CPSs}.(iii)
and Tychonoff's theorem.\hfill $\blacksquare $

\bigskip

\noindent \textbf{Proof of Remark \ref{Remark on closure "projective limit"}}%
. Note that, for each $i\in I$,%
\begin{equation*}
H_{i}:=\tbigcap_{n\geq 1}\left\{ \left( \mu _{i}^{1},\mu _{i}^{2},...\right)
\in \tprod_{m=0}^{\infty }\Delta ^{\mathcal{B}_{i}}\left( \Theta
_{i}^{m}\right) :\left( \mu _{i}^{1},...,\mu _{i}^{n}\right) \in
H_{i}^{n}\right\} \text{.}
\end{equation*}%
As shown in Remark \ref{Remark souslin compact}, each $H_{i}^{n}$ is a
closed subset of $\tprod_{m=0}^{n-1}\Delta ^{\mathcal{B}_{i}}\left( \Theta
_{i}^{m}\right) $. Thus, $H_{i}$\ is a countable intersection of closed
cylinders in $\tprod_{m=0}^{\infty }\Delta ^{\mathcal{B}_{i}}\left( \Theta
_{i}^{m}\right) $. Recall that $\Theta _{i}^{m}=\Theta _{i}^{0}\times
H_{j}^{n}$ for each $m\geq 1$. Thus, the results readily follow from Remark %
\ref{Remark souslin compact}.\hfill $\blacksquare $

\bigskip

\noindent \textbf{Proof of Remark \ref{Remark on measurability and
continuity of description map}}. The induction argument on $n\geq 1$
provided in the main text shows that each map $h_{i}^{n}$ ($i\in I$)\ is
measurable.\ Moreover, for every player $i\in I$, $h_{i}:T_{i}\rightarrow
H_{i}$ is a well-defined map because, by Claim \ref{Claim recursive step
hierarchy map}, $h_{i}^{n}\left( T_{i}\right) \subseteq H_{i}^{n}$\ for all $%
n\geq 1$. To prove measurability of the maps $h_{i}$ ($i\in I$) notice that,
for every $n\geq 0$, $\overline{\mathcal{L}}_{h_{-i}^{n}}\circ \beta
_{i}:T_{i}\rightarrow \Delta ^{\mathcal{B}_{i}}\left( S\times
H_{j}^{n}\right) =\Delta ^{\mathcal{B}_{i}^{n}}\left( \Theta _{i}^{n}\right) 
$\ is a Borel measurable function from a metrizable space to a Souslin space
(cf. Lemma \ref{Lemma souslin space of CPSs}\ and Remark \ref{Remark souslin
compact}). It follows from Schwartz (1973, Lemma 11, p. 106) that each map $%
h_{i}$\ is Borel measurable. Finally, if $\mathcal{T}$\ is continuous, then
an induction argument on $n\geq 0$\ shows---by virtue of Lemma \ref{Lemma on
general measurability of image CPS}---that the maps $\overline{\mathcal{L}}%
_{h_{-i}^{n}}\circ \beta _{i}$ ($i\in I$)\ are continuous.\ The conclusion
that also the maps $h_{i}$\ ($i\in I$) are continuous\ follows from
Engelking (1989, Proposition 2.3.6).\hfill $\blacksquare $

\bigskip

As claimed in the main text, $H_{i}$\ and $\Theta _{i}$\ ($i\in I$) can be
identified as the (non-empty) projective limits of the sequences $\left(
H_{n}^{i}\right) _{n\geq 1}$\ and $\left( \Theta _{n}^{i}\right) _{n\geq 0}$%
, respectively. We now make this claim precise. To this end, we first review
some notions from the theory of projective (or inverse) systems that are
necessary for the proof of Proposition \ref{Proposition on the canonical
homeomorphism}. For a more thorough treatment see Engelking (1989) or Rao
(1981). Here we restrict attention to projective systems of spaces indexed
by $\mathbb{N}$, the set of natural numbers.

A \textbf{projective sequence} is a sequence $\left( Y_{n},f_{m,n}\right)
_{m,n\in \mathbb{N}}$\ of spaces $Y_{n}$ and functions $f_{m,n}:Y_{n}%
\rightarrow Y_{m}$ such that:

\begin{itemize}
\item for all $n\in \mathbb{N}$, $Y_{n}$\ is non-empty topological space;

\item $f_{m,n}$\ is continuous for all $m,n\in \mathbb{N}$ such that $m\leq
n $;

\item $f_{m,p}=f_{m,n}\circ f_{n,p}$ for all $m,n,p\in \mathbb{N}$ such that 
$m\leq n\leq p$, and $f_{n,n}=\mathrm{Id}_{Y_{n}}$ for each $n\in \mathbb{N}$%
.
\end{itemize}

Each space $Y_{n}$\ ($n\in \mathbb{N}$) is called \textbf{coordinate} (or 
\textbf{factor})\textit{\ }\textbf{space} and the functions $f_{m,n}$\ are
called \textbf{bonding maps}. The \textbf{projective limit}\textit{\ }of $%
\left( Y_{n},f_{m,n}\right) _{m,n\in \mathbb{N}}$ is defined as follows:%
\begin{equation*}
\underleftarrow{\lim }Y_{n}:=\left \{ \left( y_{n}\right) _{n\in \mathbb{N}%
}\in \tprod_{n\in \mathbb{N}}Y_{n}:\forall m,n\in \mathbb{N},\text{ }m\leq
n,y_{m}=f_{m,n}\left( y_{n}\right) \right \} \text{.}
\end{equation*}%
For each $m\in \mathbb{N}$, the function $\overline{f}_{m}:\underleftarrow{%
\lim }Y_{n}\rightarrow Y_{m}$\ is defined as the restriction of the
projection map $\mathrm{P}$\textrm{ro}$\mathrm{j}_{Y_{m}}:\tprod_{n\in 
\mathbb{N}}Y_{n}\rightarrow Y_{m}$\ to $\underleftarrow{\lim }Y_{n}$. For
all $m,n\in \mathbb{N}$ such that $m\leq n$,\ the functions $\overline{f}%
_{n} $\ and $\overline{f}_{m}$\ satisfy the equality $\overline{f}%
_{m}=f_{m,n}\circ \overline{f}_{n}$.

The set $\underleftarrow{\lim }Y_{n}$\ inherits the subspace topology as a
subset of the product $\prod_{n\in \mathbb{N}}Y_{n}$. It is known (see
Engelking 1989, Proposition 2.5.1) that\ if $\left( Y_{n},f_{m,n}\right)
_{m,n\in \mathbb{N}}$ is a projective sequence\ of Hausdorff spaces $Y_{n}$,
then $\underleftarrow{\lim }Y_{n}$ is a closed subset of the Cartesian
product $\prod_{n\in \mathbb{N}}Y_{n}$. Moreover, $\underleftarrow{\lim }%
Y_{n}$\ is a metrizable space if each $Y_{n}$\ is a metrizable space
(Engelking 1989, Corollary 4.2.5).

\bigskip

\noindent \textbf{Lemma B.1 }\textit{Let }$\left( Y_{n},f_{m,n}\right)
_{m,n\in \mathbb{N}}$\textit{\ be a projective sequence of Hausdorff spaces }%
$Y_{n}$\textit{\ and surjective bonding maps }$f_{m,n}$\textit{. The
following statements hold.\newline
\ \ (i) }$Y:=\underleftarrow{\lim }Y_{n}$\textit{\ is non-empty, and, for
each }$n\in \mathbb{N}$\textit{, }$\overline{f}_{n}:Y\rightarrow Y_{n}$%
\textit{\ is continuous and surjective.\newline
\ \ (ii) Suppose that each }$Y_{n}$\textit{\ is a Souslin space. Then }$Y$%
\textit{\ is a Souslin space and the Borel }$\sigma $\textit{-algebra }$%
\Sigma _{Y}$\textit{\ is generated by the algebra }$\mathcal{A}_{Y}:=\cup
_{n\in \mathbb{N}}\overline{f}_{n}^{-1}\left( \Sigma _{n}\right) $\textit{.}

\bigskip

\noindent \textbf{Proof}. For part (i), see Engelking (1989, Exercises 2.5.A
and 2.5.B). To show part (ii), suppose that each $Y_{n}$\ is a Souslin
space. Then $\prod_{n\in \mathbb{N}}Y_{n}$\ is a Souslin space, and since $%
Y:=\underleftarrow{\lim }Y_{n}$\ is a closed subset of $\prod_{n\in \mathbb{N%
}}Y_{n}$, it follows that $Y$ is Souslin in the subspace topology (see
Schwartz 1973, Part I, Chapter II). Hence, $\Sigma _{Y}$ is the relative $%
\sigma $-algebra induced by the Borel $\sigma $-algebra generated by the
product topology on $\prod_{n\in \mathbb{N}}Y_{n}$ (Bogachev 2007, Lemma
6.2.4). By Lemma 6.4.2 and Lemma 6.6.4 in Bogachev (2007), the Borel $\sigma 
$-algebra generated by the product topology equals the Borel $\sigma $%
-algebra generated by the cylinder algebra $\cup _{n\in \mathbb{N}}\mathrm{%
Proj}_{Y_{n}}^{-1}\left( \Sigma _{n}\right) $. The conclusion follows from
the fact that each $\overline{f}_{n}$\ is the restriction to $Y$ of the
projection map $\mathrm{P}$\textrm{ro}$\mathrm{j}_{Y_{n}}$.\hfill $%
\blacksquare $

\bigskip

Consider the sequences $\left( H_{i}^{n}\right) _{n\geq 1}$\ and $\left(
\Theta _{i}^{n}\right) _{n\geq 0}$, respectively. Recall that, for each $%
i\in I$ and $n\geq 1$, we defined the projections $\pi
_{i}^{n,n+1}:H_{i}^{n+1}\rightarrow H_{i}^{n}$ and $\rho _{i}^{n-1,n}:\Theta
_{i}^{n}\rightarrow \Theta _{i}^{n-1}$ which satisfy, by construction, $\rho
_{i}^{n-1,n}=\left( \mathrm{Id}_{\Theta _{i}^{0}},\pi _{j}^{n-1,n}\right) $\
for each $n\geq 2$ (cf. Mertens et al. 2015, Theorem III.1.1.(3)).

Clearly, $\left( H_{i}^{n},\pi _{i}^{n,n+1}\right) _{n\in \mathbb{N}}$\ and $%
\left( \Theta _{i}^{n-1},\rho _{i}^{n-1,n}\right) _{n\in \mathbb{N}}$\ are
projective sequences of Souslin spaces. Proposition B.1 below, which is an
analogue of Theorem 6 in Heifetz (1993), ensures that each bonding map $\pi
_{i}^{n,n+1}$\ is a surjective open map---hence, it is a well-defined
projection. Moreover, since each set $H_{i}^{1}$\ is non-empty by
construction, it will follow that for each $n\in \mathbb{N}$\ the set $%
H_{i}^{n+1}$ is non-empty as well. (This fact ensures that the results in
Remark \ref{Remark souslin compact}\ do not hold vacuously.) The proof is
similar to Heifetz's proof and thus omitted.\footnote{%
The proof can be found in the Supplementary Appendix. Differently from
Theorem 6 of Heifetz (1993), the proof additionally shows that each bonding
map is an open function, i.e., the image of any open subset of the domain is
an open subset of the codomain.}

\bigskip

\noindent \textbf{Proposition B.1 }\textit{Consider the projective sequences 
}$\left( H_{i}^{n},\pi _{i}^{n,n+1}\right) _{n\in \mathbb{N}}$\textit{\ and }%
$\left( \Theta _{i}^{n-1},\rho _{i}^{n-1,n}\right) _{n\in \mathbb{N}}$%
\textit{. Then, for each }$i\in I$\textit{\ and each }$n\in \mathbb{N}$%
\textit{, the bonding maps }$\pi _{i}^{n,n+1}:H_{i}^{n+1}\rightarrow
H_{i}^{n}$\textit{\ and }$\rho _{i}^{n-1,n}:\Theta _{i}^{n}\rightarrow
\Theta _{i}^{n-1}$\textit{\ are surjective and open.}

\bigskip

Let $\underleftarrow{\lim }H_{i}^{n}$\ and $\underleftarrow{\lim }\Theta
_{i}^{n-1}$ be the projective limits of $\left( H_{i}^{n},\pi
_{i}^{n,n+1}\right) _{n\in \mathbb{N}}$\ and $\left( \Theta _{i}^{n-1},\rho
_{i}^{n-1,n}\right) _{n\in \mathbb{N}}$. By Proposition B.1\ and Lemma B.1, $%
\underleftarrow{\lim }H_{i}^{n}$\ and $\underleftarrow{\lim }\Theta
_{i}^{n-1}$\ are non-empty Souslin spaces. Furthermore, let $\overline{\pi }%
_{i}^{n}:\underleftarrow{\lim }H_{i}^{n}\rightarrow H_{i}^{n}$\ and $%
\overline{\rho }_{i}^{n}:\underleftarrow{\lim }\Theta _{i}^{n-1}\rightarrow
\Theta _{i}^{n}$\ denote the corresponding projections. By Lemma B.1, both $%
\overline{\pi }_{i}^{n}$\ and $\overline{\rho }_{i}^{n}$\ are continuous and
surjective. Moreover, they are also open functions (Engelking 1989, Exercise
2.7.19.(b)).

Next, refer back to the spaces $H_{i}$\ and $\Theta _{i}$. For each $i\in I$%
\ and each $n\in \mathbb{N}$, we let $\pi _{i}^{n}:H_{i}\rightarrow
H_{i}^{n} $ denote the projection map. For all $m,n\in \mathbb{N}$ such that 
$m\leq n$,\ the functions $\pi _{i}^{m}$\ and $\pi _{i}^{n}$\ satisfy the
equality $\pi _{i}^{m}=\pi _{i}^{m,n}\circ \pi _{i}^{n}$. Furthermore, for
each $i\in I $ and each $n\geq 0$, we let $\rho _{i}^{n}:\Theta
_{i}\rightarrow \Theta _{i}^{n}$\ denote the projection map, which satisfies%
\begin{equation*}
\rho _{i}^{n}=\left( \mathrm{Id}_{\Theta _{i}^{0}},\pi _{j}^{n}\right)
\end{equation*}%
for all $n\geq 1$. It should be noted that the sets $H_{i}$\ and $\Theta
_{i} $\ are formally distinct from $\underleftarrow{\lim }H_{i}^{n}$\ and $%
\underleftarrow{\lim }\Theta _{i}^{n-1}$. In particular, the elements of $%
H_{i}$\ are sequences of CPSs, while---according to the definition of
projective limit given here---the elements of $\underleftarrow{\lim }%
H_{i}^{n}$\ are sequences of \textit{finite sequences} of CPSs. Yet, the
sets $H_{i}$\ and $\Theta _{i}$\ are naturally identified as $%
\underleftarrow{\lim }H_{i}^{n}$\ and $\underleftarrow{\lim }\Theta
_{i}^{n-1}$. Specifically, Proposition B.2 below states the existence of
homeomorphisms $\gamma _{i}:H_{i}\rightarrow \underleftarrow{\lim }H_{i}^{n}$
and $\varphi _{i}:\Theta _{i}\rightarrow \underleftarrow{\lim }\Theta
_{i}^{n}$\ that preserve the projections: for each $n\in \mathbb{N}$,%
\begin{equation*}
\pi _{i}^{n}=\overline{\pi }_{i}^{n}\circ \gamma _{i}\text{ \ and \ }\rho
_{i}^{n-1}=\overline{\rho }_{i}^{n-1}\circ \varphi _{i}\text{.}
\end{equation*}

\bigskip

\noindent \textbf{Proposition B.2 }\textit{For each }$i\in I$\textit{, }$%
H_{i}$\textit{\ and }$\Theta _{i}$\textit{\ are homeomorphic to }$%
\underleftarrow{\lim }H_{i}^{n}$\textit{\ and }$\underleftarrow{\lim }\Theta
_{i}^{n-1}$\textit{, respectively.}

\bigskip

The proof (which can be found in the Supplementary Appendix) is omitted,
since it uses standard techniques in the theory of projective systems; see,
e.g., Rao (1981, pp. 117-118) and Engelking (1989, Examples 2.5.3-2.5.4).

The next step is to provide the proof of Proposition \ref{Proposition on the
canonical homeomorphism}. To this end, we need the following definition and
result.

\bigskip

\noindent \textbf{Definition B.1} \textit{We call }$\left( \left(
Y_{n},\Sigma _{Y_{n}},\mathcal{B}_{n}\right) ,f_{m,n},\mu _{n}\right)
_{m,n\in \mathbb{N}}$\textit{\ a }\textbf{projective sequence of CPSs}%
\textit{\ if:}\newline
\textit{\ \ (i) }$\left( Y_{n},f_{m,n}\right) _{m,n\in \mathbb{N}}$\textit{\
is a projective sequence of Hausdorff spaces }$Y_{n}$\textit{\ and
surjective bonding maps }$f_{m,n}$\textit{;}\newline
\textit{\ \ (ii) for all }$n\in \mathbb{N}$\textit{, the family }$\mathcal{B}%
_{n}\subseteq \Sigma _{Y_{n}}$\textit{\ of conditioning events satisfies the
following property:} $\mathcal{B}_{n}:=f_{m,n}^{-1}\left( \mathcal{B}%
_{m}\right) $\textit{\ for all\ }$m,n\in \mathbb{N}$\textit{\ such that }$%
m\leq n$\textit{;\newline
\ \ (iii) }$\mu _{n}$\textit{\ is a CPS on }$\left( Y_{n},\Sigma _{Y_{n}},%
\mathcal{B}_{n}\right) $\textit{\ for all }$n\in \mathbb{N}$\textit{;}%
\newline
\textit{\ \ (iv) }$\mu _{m}=\overline{\mathcal{L}}_{f_{m,n}}\left( \mu
_{n}\right) $\textit{\ for all }$m,n\in \mathbb{N}$\textit{\ such that }$%
m\leq n$\textit{.}

\textit{We call }$\left( \left( Y,\Sigma _{Y},\mathcal{B}_{Y}\right) ,%
\overline{f}_{n},\mu \right) _{n\in \mathbb{N}}$\textit{\ the }\textbf{CPS
projective limit}\textit{\ of the projective sequence of CPSs }$\left(
\left( Y_{n},\Sigma _{Y_{n}},\mathcal{B}_{n}\right) ,f_{m,n},\mu _{n}\right)
_{m,n\in \mathbb{N}}$\textit{\ if:}\newline
\textit{\ \ (v) }$Y=\underleftarrow{\lim }Y_{n}$,\textit{\ and }$\overline{f}%
_{m}:Y\rightarrow Y_{m}$\textit{\ satisfies }$\overline{f}_{m}=f_{m,n}\circ 
\overline{f}_{n}$\textit{\ for all\ }$m,n\in \mathbb{N}$\textit{\ such that }%
$m\leq n$\textit{;\newline
\ \ (vi) }$\mathcal{B}_{Y}\subseteq \Sigma _{Y}$ \textit{is such that }$%
B_{Y}=\overline{f}_{n}^{-1}\left( \mathcal{B}_{n}\right) $\textit{\ for all }%
$n\in \mathbb{N}$\textit{, so that }$B_{Y}=\overline{f}_{1}^{-1}\left( 
\mathcal{B}_{1}\right) $\textit{;\newline
\ \ (vii) }$\mu $\textit{\ is a CPS (called }\textbf{limit CPS}\textit{) on }%
$\left( Y,\Sigma _{Y},\mathcal{B}_{Y}\right) $\textit{\ such that, for all }$%
n\in \mathbb{N}$\textit{,}%
\begin{equation*}
\overline{\mathcal{L}}_{\overline{f}_{n}}\left( \mu \right) =\mu _{n}\text{%
\textit{.}}
\end{equation*}

\bigskip

Note that a CPS projective limit may fail to exist, that is, there could be
no limit CPS $\mu $\ which \textquotedblleft extends\textquotedblright \ the
sequence $\left( \mu _{n}\right) _{n\in \mathbb{N}}$. The following result
(a version of Kolmogorov's extension theorem for CPSs) gives a sufficient
condition for the existence and uniqueness of a limit CPS.

\bigskip

\noindent \textbf{Theorem B.1} \textit{Let }$\left( \left( Y_{n},\Sigma
_{Y_{n}},\mathcal{B}_{n}\right) ,f_{m,n},\mu _{n}\right) _{m,n\in \mathbb{N}%
} $\textit{\ be a projective sequence of CPSs such that, for each }$n\in 
\mathbb{N}$\textit{, }$Y_{n}$\textit{\ is a Souslin space. Then the CPS
projective limit }$\left( \left( Y,\Sigma _{Y},\mathcal{B}_{Y}\right) ,%
\overline{f}_{n},\mu \right) _{n\in \mathbb{N}}$\textit{\ exists and }$\mu $%
\textit{\ is the unique limit CPS.}

\bigskip

\noindent \textbf{Proof}. Each CPS $\mu _{n}$\ on $\left( Y_{n},\Sigma
_{Y_{n}},\mathcal{B}_{n}\right) $\ is an array%
\begin{equation*}
\mu _{n}=\left( \mu _{n}\left( \cdot |B\right) \right) _{B\in \mathcal{B}%
_{n}}
\end{equation*}%
such that, for all $B\in \mathcal{B}_{n}$, $\mu _{n}\left( \cdot |B\right) $%
\ is a Radon probability measure on $\left( Y_{n},\Sigma _{Y_{n}}\right) $,
i.e., for each $E\in \Sigma _{Y_{n}}$ and each $\varepsilon >0$, there
exists a compact set $K\subseteq E$\ such that $\mu \left( E|B_{n}\right)
-\mu \left( K|B\right) \leq \varepsilon $. This follows from the fact that
every Borel probability measure on a Souslin space is Radon (Bogachev 2007,
Theorem 7.4.3). With this, we can apply Prokhorov's Theorem for Radon
probability measures (Schwartz 1974, Theorem 21 and Corollary, p. 81) to
claim the existence of a unique array $\left( \mu \left( \cdot |B\right)
\right) _{B\in \mathcal{B}_{Y}}$\ of (Radon) probability measures on the
Borel $\sigma $-algebra $\Sigma _{Y}$ of the Souslin space $Y$\ (see Lemma
B1.(ii)) such that%
\begin{equation*}
\mu \left( \overline{f}_{n}^{-1}\left( \cdot \right) \left \vert \overline{f}%
_{n}^{-1}\left( B_{n}\right) \right. \right) =\mu _{n}\left( \cdot
|B_{n}\right)
\end{equation*}%
for all $n\in \mathbb{N}$ and $B_{n}\in \mathcal{B}_{n}$. It remains to
check that the array $\left( \mu \left( \cdot |B\right) \right) _{B\in 
\mathcal{B}_{Y}}$\ is a CPS. Condition (i) of Definition \ref{Definition CPS}
clearly holds. To show that Condition (ii) of Definition \ref{Definition CPS}
is satisfied, pick any $B,C\in \mathcal{B}_{Y}$ such that $B\subseteq C$. We
show that the chain rule is satisfied for every $A\in \Sigma _{Y}$. To this
end, we first consider the case when $A$\ belongs to the algebra $\mathcal{A}%
_{Y}:=\cup _{n\in \mathbb{N}}\overline{f}_{n}^{-1}\left( \Sigma _{n}\right) $%
, which generates the Borel $\sigma $-algebra $\Sigma _{Y}$ (Lemma B1.(ii)).
If $A\in \mathcal{A}_{Y}$ and $A\subseteq B$, then there exist $n\in \mathbb{%
N}$\ and $A_{n}\in \Sigma _{Y_{n}}$\ such that $A=\overline{f}%
_{n}^{-1}\left( A_{n}\right) $. Since $\mathcal{B}_{Y}=\overline{f}%
_{n}^{-1}\left( \mathcal{B}_{n}\right) $, there are $B_{n},C_{n}\in \mathcal{%
B}_{n}$\ such that $B=\overline{f}_{n}^{-1}\left( B_{n}\right) $ and\ $C=%
\overline{f}_{n}^{-1}\left( C_{n}\right) $. By surjectivity of $\overline{f}%
_{n}$ (Lemma B.1.(i)), $A_{n}\subseteq B_{n}\subseteq C_{n}$. Using the fact
that $\mu _{n}$\ is a CPS, we obtain%
\begin{eqnarray*}
\mu \left( A|B\right) \mu \left( B|C\right) &=&\mu \left( \overline{f}%
_{n}^{-1}\left( A_{n}\right) |\overline{f}_{n}^{-1}\left( B_{n}\right)
\right) \mu \left( \overline{f}_{n}^{-1}\left( B_{n}\right) |\overline{f}%
_{n}^{-1}\left( C_{n}\right) \right) \\
&=&\mu _{n}\left( A_{n}|B_{n}\right) \mu _{n}\left( B_{n}|C_{n}\right) =\mu
_{n}\left( A_{n}|C_{n}\right) \\
&=&\mu \left( \overline{f}_{n}^{-1}\left( A_{n}\right) |\overline{f}%
_{n}^{-1}\left( C_{n}\right) \right) =\mu \left( A|C\right) \text{.}
\end{eqnarray*}%
Thus, the chain rule holds whenever $A\in \mathcal{A}_{Y}$. Next, let $%
\mathcal{M}\left( B,C\right) \subseteq \Sigma _{Y}$\ be the collection of
all Borel subsets of $Y$ such that Condition (ii) of Definition \ref%
{Definition CPS} holds for those fixed $B,C\in \mathcal{B}_{Y}$. By $\sigma $%
-additivity of probability measures, $\mathcal{M}\left( B,C\right) $\ is a
monotone class. As seen, $\mathcal{A}_{Y}\subseteq \mathcal{M}\left(
B,C\right) $. It follows that $\Sigma _{Y}\subseteq \mathcal{M}\left(
B,C\right) $ because, by the Monotone Class Theorem, the smallest monotone
class containing $\mathcal{A}_{Y}$ is $\Sigma _{Y}$. Thus, $\mathcal{M}%
\left( B,C\right) =\Sigma _{Y}$. This shows that the chain rule is
satisfied, and concludes the proof.\hfill $\blacksquare $

\bigskip

With this, we prove Proposition \ref{Proposition on the canonical
homeomorphism}.

\bigskip

\noindent \textbf{Proof of Proposition \ref{Proposition on the canonical
homeomorphism}}. Fix $i\in I$. By Proposition B.2\ and Theorem B.1, to every 
$\left( \mu _{i}^{1},\mu _{i}^{2},...\right) \in H_{i}$\ there corresponds a
unique $\mu _{i}\in \Delta ^{\mathcal{B}_{i}}\left( \Theta _{i}\right) $\
such that $\overline{\mathcal{L}}_{\rho _{i}^{n-1}}\left( \mu _{i}\right)
=\mu _{i}^{n}$ for all $n\in \mathbb{N}$.\ Hence, there exists a bijection
between $H_{i}$\ and $\Delta ^{\mathcal{B}_{i}}\left( \Theta _{i}\right) $.
Let $g_{i}:\Delta ^{\mathcal{B}_{i}}\left( \Theta _{i}\right) \rightarrow
H_{i}$\ be the function defined by%
\begin{equation*}
g_{i}\left( \mu _{i}\right) :=\left( \overline{\mathcal{L}}_{\rho
_{i}^{n-1}}\left( \mu _{i}\right) \right) _{n\in \mathbb{N}}\text{.}
\end{equation*}%
Note that, for all $n\in \mathbb{N}$,%
\begin{equation*}
\left( \overline{\mathcal{L}}_{\rho _{i}^{m-1}}\left( \mu _{i}\right)
\right) _{m\leq n}\in H_{i}^{n}\text{,}
\end{equation*}%
so that $g_{i}$\ is a well-defined bijection. Since each coordinate
projection\ is a continuous function, it follows from Lemma \ref{Lemma on
general measurability of image CPS} that $\overline{\mathcal{L}}_{\rho
_{i}^{n-1}}$\ is continuous for all $n\in \mathbb{N}$. With this, continuity
of $g_{i}$\ follows from Engelking (1989, Proposition 2.3.6). We now show
that the inverse of $g_{i}$, denoted by $f_{i}$, is also continuous.
Consider a sequence in $H_{i}$, viz. $\left( h_{m}\right) _{m\in \mathbb{N}%
}:=\left( \left( \mu _{i,m}^{1},\mu _{i,m}^{2},...\right) \right) _{m\in 
\mathbb{N}}$, converging to $h:=\left( \mu _{i}^{1},\mu _{i}^{2},...\right)
\in H_{i}$. That is, for all $n\in \mathbb{N}$,\ it is the case that $\mu
_{i,m}^{n}\rightarrow \mu _{i}^{n}$ as $m\rightarrow +\infty $. For
convenience, let $\mu _{i,m}:=f_{i}\left( h_{m}\right) $\ for all $m\in 
\mathbb{N}$\ and $\mu _{i}:=f_{i}\left( h\right) $. We need to prove that $%
\mu _{i,m}\rightarrow \mu _{i}$ as $m\rightarrow +\infty $. We will prove
the following claim:%
\begin{equation*}
\forall B\in \mathcal{B}_{\Theta _{i}},\mu _{i,m}\left( \cdot \left \vert
B\right. \right) \rightarrow \mu _{i}\left( \cdot \left \vert B\right.
\right) \text{.}
\end{equation*}%
Fix some $B\in \mathcal{B}_{\Theta _{i}}$, so that, for all $n\geq 0$, $%
B=\left( \rho _{i}^{n}\right) ^{-1}\left( B_{n}\right) $\ with $B_{n}\in 
\mathcal{B}_{i}^{n}$. We show that $\lim \inf_{m}\mu _{i,m}\left(
O\left
\vert B\right. \right) \geq \mu _{i}\left( O\left \vert B\right.
\right) $\ for every open set $O\subseteq \Theta _{i}$; with this, the
Portmanteau Theorem (Aliprantis and Border 2006, Theorem 15.3) entails that $%
\mu _{i,m}\left( \cdot \left \vert B\right. \right) $\ converges weakly to $%
\mu _{i}\left( \cdot \left \vert B\right. \right) $\ as $m\rightarrow
+\infty $.

Towards this end, we first record some facts that will be used in the proof.
By Engelking (1989, Proposition 2.5.5), a base for the topology on $\Theta
_{i}$\ is the family $\mathcal{C}$ of sets of the form $\left( \rho
_{i}^{n}\right) ^{-1}\left( O_{n}\right) $, where $O_{n}$ is a basic open
subset of $\Theta _{i}^{n}$, for some $n\geq 0$. This base\ can be taken
countable, because each $\Theta _{i}^{n}$ is a metrizable Souslin
space---hence, second countable.

Consider now any open set $U\subseteq \Theta _{i}$. Since every open set is
the union of elements of the (countable) basis, there exists a sequence $%
\left( G_{k}\right) _{k\geq 0}$ of sets such that $G_{k}\in \mathcal{C}$\
for all $k\geq 0$ and $\cup _{k\geq 0}G_{k}=U$. We derive from this sequence
a new sequence satisfying some desirable properties.\footnote{%
This argument is standard, and it can be generalized to nets whenever the
coordinate spaces are not metrizable (cf. Heifetz 1991, p. 337).} Define,
for every $n\geq 0$, the following sets:%
\begin{eqnarray*}
V_{n} &:&=\tbigcup_{k\geq 0}\left \{ U_{k}\subseteq \Theta _{i}^{n}:U_{k}%
\text{ is open and }G_{k}=\left( \rho _{i}^{n}\right) ^{-1}\left(
U_{k}\right) \right \} \text{,} \\
V_{n}^{\ast } &:&=\left( \rho _{i}^{n}\right) ^{-1}\left( V_{n}\right) \text{%
.}
\end{eqnarray*}%
It can be checked that $\left( V_{n}^{\ast }\right) _{n\geq 0}$ is an
increasing sequence such that $V_{n}^{\ast }\in \mathcal{C}$\ for all $n\geq
0$ and $\cup _{n\geq 0}V_{n}^{\ast }=U$.

We now prove the claim. Fix any open set $O\subseteq \Theta _{i}$. By the
above argument, there is an increasing sequence $\left( G_{k}\right) _{k\in 
\mathbb{N\cup }\left \{ 0\right \} }$ of open subsets of $\Theta _{i}$\ such
that: (a) $\cup _{k\geq 0}G_{k}=O$, and (b) for each $G_{k}$, there exist $%
\ell _{k}\geq 0$\ and an open set $U_{\ell _{k}}\subseteq \Theta _{i}^{\ell
_{k}}$\ such that $G_{k}=\left( \rho _{i}^{\ell _{k}}\right) ^{-1}\left(
U_{\ell _{k}}\right) $. Thus, for each $k,m\in \mathbb{N\cup }\left \{
0\right \} $,%
\begin{equation}
\mu _{i,m}\left( G_{k}\left \vert B\right. \right) =\mu _{i,m}\left( \left(
\rho _{i}^{\ell _{k}}\right) ^{-1}\left( U_{\ell _{k}}\right) \left \vert
\left( \rho _{i}^{\ell _{k}}\right) ^{-1}\left( B_{\ell _{k}}\right) \right.
\right) =\mu _{i,m}^{\ell _{k}}\left( \left( U_{\ell _{k}}\right) \left
\vert B_{\ell _{k}}\right. \right) \text{,}
\label{first equality for portmanteu}
\end{equation}%
and, for each $k\in \mathbb{N\cup }\left \{ 0\right \} $,%
\begin{equation}
\mu _{i}\left( G_{k}\left \vert B\right. \right) =\mu _{i}\left( \left( \rho
_{i}^{\ell _{k}}\right) ^{-1}\left( U_{\ell _{k}}\right) \left \vert \left(
\rho _{i}^{\ell _{k}}\right) ^{-1}\left( B_{\ell _{k}}\right) \right.
\right) =\mu _{i}^{\ell _{k}}\left( \left( U_{\ell _{k}}\right) \left \vert
B_{\ell _{k}}\right. \right) \text{.}  \label{second equality for portmanteu}
\end{equation}%
Since $\mu _{i,m}^{n}\rightarrow \mu _{i}^{n}$\ for all $n\in \mathbb{N}$,
it follows that $\mu _{i,m}^{\ell _{k}}\left( \cdot \left \vert B_{\ell
_{k}}\right. \right) $\ converges weakly to $\mu _{i}^{\ell _{k}}\left(
\cdot \left \vert B_{\ell _{k}}\right. \right) $. Hence, by the Portmanteau
Theorem,%
\begin{equation*}
\lim \inf_{m}\mu _{i,m}^{\ell _{k}}\left( \left( U_{\ell _{k}}\right) \left
\vert B_{\ell _{k}}\right. \right) \geq \mu _{i}^{\ell _{k}}\left( \left(
U_{\ell _{k}}\right) \left \vert B_{\ell _{k}}\right. \right) \text{.}
\end{equation*}%
By (\ref{first equality for portmanteu}) and (\ref{second equality for
portmanteu}), it follows that, for each $k,m\in \mathbb{N\cup }\left \{
0\right \} $,%
\begin{equation}
\lim \inf_{m}\mu _{i,m}\left( G_{k}\left \vert B\right. \right) \geq \mu
_{i}\left( G_{k}\left \vert B\right. \right) \text{.}
\label{Inequality portmanteau}
\end{equation}%
The left expression in (\ref{Inequality portmanteau}) is non-decreasing in $%
k\geq 0$ and bounded above by $\lim \inf_{m}\mu _{i,m}\left( O\left \vert
B\right. \right) $, while the right one converges to $\mu _{i}\left(
O\left
\vert B\right. \right) $ as $k\rightarrow +\infty $. Thus, $\mu
_{i,m}\left( O\left \vert B\right. \right) $\ converges weakly to $\mu
_{i}\left( O\left
\vert B\right. \right) $. Since $O\subseteq \Theta _{i}$\
and $B\in \mathcal{B}_{\Theta _{i}}$\ are arbitrary, the conclusion
follows.\hfill $\blacksquare $

\section*{Appendix C. Proofs for Section \protect\ref{Section: terminal type
structures}}

Here we prove Lemma \ref{Lemma: ontoness lubinlike}. To this end, we begin
by reviewing the required mathematical background\ and results, then we
briefly discuss the underlying idea of the proof, and finally we move on to
the formal analysis.

Fix topological spaces $X$ and $Y$. Given a Borel measure $\mu \in \Delta
\left( X\right) $, a function $f:X\rightarrow Y$ is $\mu $\textbf{-measurable%
} if $f^{-1}\left( E\right) \in \Sigma _{X}^{\mu }$ for each $E\in \Sigma
_{Y}$,\ where $\Sigma _{X}^{\mu }$\ stands for the completion of the Borel $%
\sigma $-algebra $\Sigma _{X}$\ with respect to $\mu $. Sets in $\Sigma
_{X}^{\mu }\backslash \Sigma _{X}$\ are called $\mu $-\textbf{null}. A
function $f:X\rightarrow Y$ is called \textbf{universally measurable} if it
is $\mu $-measurable\ for every $\mu \in \Delta \left( X\right) $,
or---equivalently---if $f^{-1}\left( E\right) \in \Sigma _{X}^{\ast }$ for
every $E\in \Sigma _{Y}$, where $\Sigma _{X}^{\ast }:=\cap _{\mu \in \Delta
\left( X\right) }\Sigma _{X}^{\mu }$. Sets in $\Sigma _{X}^{\ast }$\ are
called \textbf{universally measurable sets}, and $\Sigma _{X}^{\ast }$\ is
called the universally measurable $\sigma $-algebra. Clearly, $\Sigma
_{X}\subseteq \Sigma _{X}^{\ast }$.

\bigskip

\noindent \textbf{Fact C.1} \textit{Fix Souslin spaces }$X_{1}$\textit{, }$%
X_{2}$\textit{, }$Y_{1}$\textit{, and }$Y_{2}$\textit{. Fix functions }$%
f_{1}:X_{1}\rightarrow Y_{1}$\textit{\ and }$f_{2}:X_{2}\rightarrow Y_{2}$%
\textit{, and define }$f:X_{1}\times X_{2}\rightarrow Y_{1}\times Y_{2}$%
\textit{\ as }$f:=\left( f_{1},f_{2}\right) $\textit{. If }$f_{1}$\textit{\
and }$f_{2}$\textit{\ are universally measurable, so is }$f$\textit{.}

\bigskip

Fact C.1 follows from Lemma 9.1 in Friedenberg and Meier (2017) because
Souslin spaces are separable.

To understand the idea of the proof of Lemma \ref{Lemma: ontoness lubinlike}%
, fix topological spaces $X$ and $Y$, and a Borel measurable function $%
f:X\rightarrow Y$. Consider the following problem pertaining to the
pushforward-measure map induced by $f$.

\bigskip

\noindent \textbf{Problem C.1} \textit{If }$f:X\rightarrow Y$\textit{\ is
surjective, is the induced function }$\mathcal{L}_{f}:\Delta \left( X\right)
\rightarrow \Delta \left( Y\right) $\textit{\ surjective?}

\bigskip

Problem C.1 can be restated as follows: Given $\nu \in \Delta \left(
Y\right) $, does there exist a solution $\mu \in \Delta \left( X\right) $ to
the \textquotedblleft equation\textquotedblright \ $\mathcal{L}_{f}\left(
\mu \right) =\nu $? The answer is negative, as simple examples show. In
particular, there could be no solution even if $f$\ is a bijection---cf. the
discussion below Theorem \ref{Theorem terminality implies completeness} in
Section \ref{Section second result}\ of the main text. The following example
illustrates this issue.

\bigskip

\noindent \textbf{Example C.1} \textit{Suppose that }$X$\textit{\ is the
unit interval }$\left[ 0,1\right] $\textit{\ endowed with the discrete
topology, and }$Y$\textit{\ is the unit interval }$\left[ 0,1\right] $%
\textit{\ endowed with the Euclidean topology. Let }$f:X\rightarrow Y$%
\textit{\ be the identity map. Thus, }$f$\textit{\ is a continuous
bijection, but it is neither a homeomorphism nor a Borel isomorphism. The
induced function }$\mathcal{L}_{f}:\Delta \left( X\right) \rightarrow \Delta
\left( Y\right) $\textit{\ is not surjective because }$\Delta \left(
X\right) $\textit{\ does not contain the Lebesgue measure, which is an
element of }$\Delta \left( Y\right) $\textit{.}

\bigskip

To provide a positive answer, one has to impose topological restrictions on
the relevant spaces. Problem C.1 can be reframed as an instance of
\textquotedblleft measurable selection problem.\textquotedblright \ Recall
that if a map $f:X\rightarrow Y$ is surjective, it admits (by the Axiom of
Choice) a \textbf{right inverse}, i.e., an injective function $%
g:Y\rightarrow X$\ such that $f\circ g=\mathrm{Id}_{Y}$. Put differently, $g$%
\ is a selection of the correspondence $f^{-1}:Y\rightarrow 2^{X}$. The
following result, which is a version of \textbf{Von Neumann Selection Theorem%
}, gives sufficient conditions for the existence of a right inverse $%
g:Y\rightarrow X$ satisfying a \textquotedblleft nice\textquotedblright \
measurability condition. (For a proof, see Bogachev 2007, Theorem 9.1.3 or
Schwartz, 1973, Theorem 13, pp. 127-128.)

\bigskip

\noindent \textbf{Theorem C.1}\textit{\ Fix Souslin spaces }$X$\textit{\ and 
}$Y$\textit{. If }$f:X\rightarrow Y$\textit{\ is a surjective Borel
measurable function, then there exists a universally measurable function }$%
g:Y\rightarrow X$\textit{\ which is a right inverse of }$f$\textit{.}

\bigskip

Given a Borel measurable function $f:X\rightarrow Y$ between Souslin spaces\
and a measure $\nu \in \Delta \left( Y\right) $, a solution $\mu \in \Delta
\left( X\right) $\ which solves the \textquotedblleft
equation\textquotedblright \ $\mathcal{L}_{f}\left( \mu \right) =\nu $\ can
be found by means of the following proof scheme:

1) Invoke Theorem C.1 to assert the existence of a universally measurable
function $g:Y\rightarrow X$\textit{\ }which is a right inverse of $f$\textit{%
.}

2) Extend $\nu $\ to $\left( Y,\Sigma _{Y}^{\ast }\right) $: this extension,
denoted by $\bar{\nu}$, exists and it is unique (Corbae et al. 2009,
Corollary 7.9.6).

3) Use $g$ to push $\bar{\nu}$\ forward: this gives a measure $\mathcal{L}%
_{g}\left( \bar{\nu}\right) $ on $\left( X,\Sigma _{X}\right) $\ such that $%
\mathcal{L}_{f}\left( \mathcal{L}_{g}\left( \bar{\nu}\right) \right) =\nu $.

A solution to Problem C.1 was first given by Varadarajan (1963, p. 194), and
the proof follows the above scheme.\footnote{%
It is worth noting that the function $f$ in Example C.1 does not have a(n
universally) measurable right inverse.} We adopt an analogous strategy to
prove Lemma \ref{Lemma: ontoness lubinlike}.

\bigskip

\noindent \textbf{Proof of Lemma \ref{Lemma: ontoness lubinlike}}. To show
part (i), first note that$\ X$, $Y$ and $Z$ are Souslin spaces, so they are
separable. It follows from standard results (e.g., Bogachev 2007, Lemma
6.4.2) that the Borel $\sigma $-algebras generated by the product topology
on $X\times Y$\ and $X\times Z$ coincide with the product $\sigma $%
-algebras. The conclusion that\ $f_{2}:=\left( \mathrm{Id}_{X},f_{1}\right)
:X\times Y\rightarrow X\times Z$ is measurable follows from Aliprantis and
Border (2006, Lemma 4.49). Next note that every $C\in \mathcal{B}_{X\times
Y} $\ (resp. $C\in \mathcal{B}_{X\times Z}$) is such that $C=B\times Y$\
(resp. $C=B\times Z$) for some $B\in \mathcal{B}$. Hence, it is the case
that $\mathcal{B}_{X\times Y}=\left( f_{2}\right) ^{-1}\left( \mathcal{B}%
_{X\times Z}\right) $. By Lemma \ref{Lemma souslin space of CPSs}.(i), the
map $\overline{\mathcal{L}}_{f_{2}}:\Delta ^{\mathcal{B}}\left( X\times
Y\right) \rightarrow \Delta ^{\mathcal{B}}\left( X\times Z\right) $ is
well-defined.

To show part (ii), let $v$\ be a CPS on $\left( X\times Z,\Sigma _{X\times
Z},\mathcal{B}_{X\times Z}\right) $. We have to find $\mu \in \Delta ^{%
\mathcal{B}}\left( X\times Y\right) $\ such that $\overline{\mathcal{L}}%
_{f_{2}}\left( \mu \right) =v$. To this end, first note that, since $%
f_{1}:Y\rightarrow Z$\ is surjective, $f_{2}:=\left( \mathrm{Id}%
_{X},f_{1}\right) :X\times Y\rightarrow X\times Z$\ is a Borel measurable
surjection. By Theorem C.1, there exists a universally measurable right
inverse of $f_{1}$, that is, an injective function $g_{1}:Z\rightarrow Y$
such that $f_{1}\circ g_{1}=\mathrm{Id}_{Z}$ and $g_{1}^{-1}\left( E\right)
\in \Sigma _{Z}^{\ast }$ for every $E\in \Sigma _{Y}$.

Let $g_{2}:X\times Z\rightarrow X\times Y$\ be the function defined as $%
g_{2}:=\left( \mathrm{Id}_{X},g_{1}\right) $. Clearly, $g_{2}$\ satisfies $%
f_{2}\circ g_{2}=\mathrm{Id}_{X\times Z}$, and, by Fact C.1, $g_{2}$\ is a
universally measurable right inverse of $f_{2}$. Furthermore, $g_{2}$\ is
such that $\mathcal{B}_{X\times Z}=g_{2}^{-1}\left( \mathcal{B}_{X\times
Y}\right) $. Hence, by Lemma \ref{Lemma on general measurability of image
CPS}.(i), the map $\overline{\mathcal{L}}_{g_{2}}$ from the set of CPSs on $%
\left( X\times Z,\Sigma _{X\times Z}^{\ast },\mathcal{B}_{X\times Z}\right) $%
\ to the set of CPSs on $\left( X\times Y,\Sigma _{X\times Y},\mathcal{B}%
_{X\times Y}\right) $\ is well-defined.

Next, for each $B\in \mathcal{B}_{X\times Z}$, let $\bar{v}\left( \cdot
|B\right) $ denote the unique extension of $v\left( \cdot |B\right) $\ to $%
\Sigma _{X\times Z}^{\ast }$. We claim that the array $\bar{v}:=\left( \bar{v%
}\left( \cdot |B\right) \right) _{B\in \mathcal{B}_{X\times Z}}$ is a CPS on 
$\left( X\times Z,\Sigma _{X\times Z}^{\ast },\mathcal{B}_{X\times Z}\right) 
$. To see this, it is enough to verify that the chain rule holds. Pick any $%
A\in \Sigma _{X\times Z}^{\ast }$ and $B,C\in \mathcal{B}_{X\times Z}$ such
that $A\subseteq B\subseteq C$. Notice that, since $B\in \Sigma _{X\times Z}$%
, it is the case that $\bar{v}\left( B|C\right) =v\left( B|C\right) $.
Suppose first that $A\in \Sigma _{X\times Z}$. Then the chain rule is
satisfied because $\bar{v}\left( A|C\right) =v\left( A|C\right) $\ and $\bar{%
v}\left( A|B\right) =v\left( A|B\right) $. If instead $A\notin \Sigma
_{X\times Z}$, then the chain rule trivially holds as $0=0$: we have $\bar{v}%
\left( A|B\right) =\bar{v}\left( A|C\right) =0$\ because $A$ is a set that
is both $v\left( \cdot |B\right) $-null\ and $v\left( \cdot |C\right) $-null.

By Lemma \ref{Lemma on general measurability of image CPS}.(i), $\mu :=%
\overline{\mathcal{L}}_{g_{2}}\left( \bar{v}\right) $\ is a CPS on $\left(
X\times Y,\Sigma _{X\times Y},\mathcal{B}_{X\times Y}\right) $. Since the
equality $g_{2}^{-1}\left( f_{2}^{-1}\left( A\right) \right) =A$ holds for
every set $A\subseteq X\times Z$, we see that, for all $E\in \Sigma
_{X\times Z}$\ and $B\in \mathcal{B}_{X\times Z}$,%
\begin{eqnarray*}
\overline{\mathcal{L}}_{f_{2}}\left( \mu \right) \left( E|B\right) &=&\mu
\left( f_{2}^{-1}\left( E\right) |f_{2}^{-1}\left( B\right) \right) \\
&=&\bar{v}\left( g_{2}^{-1}\left( f_{2}^{-1}\left( E\right) \right)
|g_{2}^{-1}\left( f_{2}^{-1}\left( B\right) \right) \right) \\
&=&\bar{v}\left( E|B\right) \\
&=&v\left( E|B\right) \text{,}
\end{eqnarray*}%
where the last equality holds because $E\in \Sigma _{X\times Z}$. Thus, $\mu 
$\ is the desired CPS that satisfies $\overline{\mathcal{L}}_{f_{2}}\left(
\mu \right) =v$.\hfill $\blacksquare $

\end{document}


\title{Complete Conditional Type Structures\\
(Supplementary Appendix)}
\author{Nicodemo De Vito \\
{\small Department of Decision Sciences - Bocconi University}\\
{\small nicodemo.devito@unibocconi.it}}
\date{November 2023}
\maketitle

\begin{abstract}
This appendix expands on some technical results mentioned in De Vito (2023).
Section D.1 proves Fact A.3 in Appendix A by adapting some proofs in
Srivastava (1988). Section D.2 shows that the results in the paper remain
true if every topological space is assumed to be a perfectly normal
Hausdorff space. Section D.3 provides the proofs of some claims (remarks
and/or propositions) in the paper. Finally, Section E shows how some proofs
in the paper must be amended if beliefs are represented by consistent
conditional probability systems. We use the same notation as in De Vito
(2023), unless otherwise stated.
\end{abstract}

\section*{D.1 On Fact A.3 in Appendix A}

We show how the result in Fact A.3 (Appendix A of the paper) holds by
adapting some proofs in Srivastava (1998). The exposition follows Section
3.2 in Srivastava (1998), where the results are stated under the assumption
that all the relevant spaces are Polish. Here we show that analogous results
hold under the weaker assumption of Souslin and Lusin spaces, as claimed in
the paper. We first consider the case when Souslin and Lusin spaces are
metrizable; then we show how the hypothesis of metrizability can be weakened
to complete regularity.

For any set $X$, we let $\left( X,\tau \right) $\ denote the corresponding
topological space, where $\tau $\ is a topology. We let $\sigma \left( \tau
\right) $\ denote the Borel $\sigma $-algebra generated by $\tau $. If $\tau
\subseteq \tau ^{\prime }$, we say that $\tau ^{\prime }$\ is finer than $%
\tau $.

The following result is the analogue of Theorem 3.2.4 in Srivastava (1998).

\bigskip

\noindent \textbf{Theorem D.1 }\textit{Suppose }$\left( X,\tau \right) $%
\textit{\ is a metrizable Souslin space. Then for every Borel subset }$%
B\subseteq X$\textit{\ there is a finer, metrizable and Souslin topology }$%
\tau _{B}$\textit{\ on }$X$\textit{\ such that }$B$\textit{\ is clopen with
respect to }$\tau _{B}$\textit{\ and }$\sigma \left( \tau \right) =\sigma
\left( \tau _{B}\right) $\textit{.}

\bigskip

As in Srivastava (1998), the following observations are useful.

\bigskip

\noindent \textbf{Observation D.1}. Let $F$\ be a \textit{closed} set in a
metrizable Souslin space $\left( X,\tau \right) $. Let $\tau |_{F}$\ and $%
\tau |_{X\backslash F}$\ denote the relative topologies on $F$\ and $%
X\backslash F$, respectively. Then $\left( F,\tau |_{F}\right) $\ and $%
\left( X\backslash F,\tau |_{X\backslash F}\right) $\ are metrizable Souslin
spaces (Bogachev 2007, Lemma 6.6.5). The topological space $\left( X,\tau
^{\prime }\right) $, where $\tau ^{\prime }$\ is the topological sum%
\footnote{%
Suppose we are given a family of pairwise disjoint topological spaces $%
\left( X_{\lambda },\tau _{\lambda }\right) $, $\lambda \in \Lambda $\
(here, $\Lambda $ is an index set). The \textit{topological} (or \textit{%
direct}) \textit{sum} $\tau $ of the topologies $\tau _{\lambda }$ ($\lambda
\in \Lambda $) is the family of all sets $O\subseteq X:=\cup _{\lambda \in
\Lambda }X_{\lambda }$ such that $O\cap X_{\lambda }\in \tau _{\lambda }$\
for every $\lambda \in \Lambda $. See Engelking (1989, Section 2.2).} of $%
\tau |_{F}$ and $\tau |_{X\backslash F}$, is a metrizable Souslin space.
This follows from the fact that a \textit{countable} topological sum of
metrizable Souslin topologies is metrizable Souslin (Schwartz, 1973, Theorem
3, p. 96,\ and Engelking, 1989, Theorem 4.2.1). Obviously, $\tau ^{\prime }$%
\ generates the same $\sigma $-algebra as $\tau $, and $F$ in clopen\ with
respect to $\tau ^{\prime }$.\hfill $\blacksquare $

\bigskip

\noindent \textbf{Observation D.2}. Let $\left( \tau _{n}\right) _{n\in 
\mathbb{N}}$\ be a sequence of metrizable Souslin topologies on $X$ such
that for any two distinct elements $x,y\in X$, there exist disjoint sets $%
U,V\in \cap _{n\in \mathbb{N}}\tau _{n}$\ such that $x\in U$\ and $y\in V$.
The topology $\tau _{\infty }$\ generated by $\cup _{n\in \mathbb{N}}\tau
_{n}$\ is Souslin and metrizable. This can be seen as follows.

Define $f:X\rightarrow X^{\mathbb{N}}$\ by%
\begin{equation*}
f\left( x\right) =\left( x,x,...\right) \text{, }x\in X\text{.}
\end{equation*}%
Thus, $f$\ is an embedding of $\left( X,\tau _{\infty }\right) $\ in $\tprod
\limits_{n\in \mathbb{N}}\left( X,\tau _{n}\right) $. Since $\tprod
\limits_{n\in \mathbb{N}}\left( X,\tau _{n}\right) $\ is a metrizable
Souslin space (Bogachev 2007, Lemma 6.6.5.(iii)),\ the conclusion follows
because $f\left( X\right) $\ is closed in $\tprod \limits_{n\in \mathbb{N}%
}\left( X,\tau _{n}\right) $. To see this, we show that $X^{\mathbb{N}%
}\backslash f\left( X\right) $\ is open. Let $\left( x_{k}\right) _{k\in 
\mathbb{N}}\in X^{\mathbb{N}}\backslash f\left( X\right) $. Take $m$, $n$\
such that $x_{m}\neq x_{n}$. By assumption, there are disjoint sets $%
U_{n},U_{m}\in \cap _{k\in \mathbb{N}}\tau _{k}$\ such that $x_{n}\in U_{n}$%
\ and $x_{m}\in U_{m}$. Then%
\begin{equation*}
\left( x_{k}\right) _{k\in \mathbb{N}}\in \pi _{n}^{-1}\left( U_{n}\right)
\cap \pi _{m}^{-1}\left( U_{m}\right) \subseteq X^{\mathbb{N}}\backslash
f\left( X\right) \text{,}
\end{equation*}%
where $\pi _{n}$\ and $\pi _{m}$\ are the coordinate projection from $\tprod
\limits_{n\in \mathbb{N}}\left( X,\tau _{n}\right) $\ onto, respectively, $%
\left( X,\tau _{n}\right) $\ and $\left( X,\tau _{m}\right) $.\hfill $%
\blacksquare $

\bigskip

\noindent \textbf{Proof of Theorem D.1}. Given Observations D.1-D.2, it is
enough to replace the word \textquotedblleft Polish\textquotedblright \ with
\textquotedblleft metrizable Souslin\textquotedblright \ in the proof of
Theorem 3.2.4 in Srivastava (1998).\hfill $\blacksquare $

\bigskip

Fact A.3 in Appendix A follows as a corollary of Theorem D.1.

\bigskip

\noindent \textbf{Corollary D.1}. \textit{Suppose }$\left( X,\tau \right) $%
\textit{\ is a metrizable Souslin space. Then for every sequence\footnote{%
Note the statement of Fact A.3 pertains to a \textit{countable family} of
Borel subsets of $X$, not necessarily a sequence. This difference is
immaterial for the result to hold.} }$\left( B_{n}\right) _{n\in \mathbb{N}}$%
\textit{\ of Borel subsets of }$X$\textit{\ there exists a finer metrizable
Souslin topology }$\tau ^{\prime }$\textit{\ on }$X$\textit{\ generating the
same Borel }$\sigma $\textit{-algebra and making each }$B_{n}$\textit{\
clopen.}

\bigskip

We now consider that case of a Lusin space $\left( X,\tau \right) $. We
argue that this case is, in a sense, \textquotedblleft
implicitly\textquotedblright \ covered by Theorem 3.2.4 in Srivastava
(1998); that is, the assumption of $\left( X,\tau \right) $\ being Polish is
without loss of generality. To understand what this really means, consider
first the following characterization of Lusin spaces.

\bigskip

\noindent \textbf{Remark D.1 }\textit{Fix a topological space }$\left(
X,\tau \right) $\textit{. Then }$\left( X,\tau \right) $\textit{\ is a Lusin
space if and only if there exists a topology }$\tau ^{\prime }$\textit{\ (on 
}$X$\textit{) finer than }$\tau $\textit{\ such that }$\left( X,\tau
^{\prime }\right) $\textit{\ is a Polish space.}

\bigskip

The characterization result in Remark D.1 is well-known (see, e.g., Schwartz
1973, p. 94), and it is often taken as a primary definition of Lusin spaces.
We provide a proof for convenience.

\bigskip

\noindent \textbf{Proof of Remark D.1}. Let $\left( Y,\tau \right) $\ be a
Lusin space. Then there exists a Polish space $\left( X,\tau _{X}\right) $\
and a continuous injective function $f:X\rightarrow Y$\ such that $f\left(
X\right) =Y$. Define a topology $\tau ^{\prime }$\ on $Y$ as follows. A set $%
O\subseteq Y$\ is such that $O\in \tau ^{\prime }$\ if only if $f^{-1}\left(
O\right) \in \tau _{X}$.\footnote{%
In the standard terminology, $\tau ^{\prime }$\ is called the \textit{final
topology} on $Y$ induced by $f$.} By construction, $\tau \subseteq \tau
^{\prime }$. To see this, note that if $O\in \tau _{X}$, then $f\left(
O\right) \in \tau ^{\prime }$\ because $f^{-1}\left( f\left( O\right)
\right) =O$\ by injectivity of $f$. Thus, $\left( Y,\tau ^{\prime }\right) $%
\ is a Polish space because the map $f:X\rightarrow Y$\ turns out to be a
homeomorphism between $\left( X,\tau _{X}\right) $\ and $\left( Y,\tau
^{\prime }\right) $. It follows that $\tau ^{\prime }$\ satisfies the
required properties.

Conversely, fix a topological space $\left( X,\tau \right) $\ so that there
is a topology $\tau ^{\prime }$ (on $X$) finer than $\tau $\ and for which $%
\left( X,\tau ^{\prime }\right) $ is a Polish space. It is immediate to see
that the identity map from $\left( X,\tau ^{\prime }\right) $\ onto $\left(
X,\tau \right) $\ is a continuous bijection. Hence, $\left( X,\tau \right) $%
\ is a Lusin space.\hfill $\blacksquare $

\bigskip

Next, note the following result (Schwartz 1973, Corollary 2, p. 101).

\bigskip

\noindent \textbf{Remark D.2}\textit{\ If }$\tau $\textit{\ and }$\tau
^{\prime }$\textit{\ are Souslin comparable topologies on }$X$\textit{\
(that is, }$\tau \subseteq \tau ^{\prime }$\textit{\ or }$\tau ^{\prime
}\subseteq \tau $\textit{), then they generate the same Borel }$\sigma $%
\textit{-algebra}.

\bigskip

The \textquotedblleft Lusin version\textquotedblright \ of Theorem D.1\
works as follows. Suppose that $\left( X,\tau \right) $\ is a Lusin space.
By Remark D.1, replace $\left( X,\tau \right) $\ with the corresponding
Polish space $\left( X,\tau ^{\prime }\right) $. Both $\left( X,\tau \right) 
$\ and $\left( X,\tau ^{\prime }\right) $\ generate the same Borel $\sigma $%
-algebra by Remark D.2. With this, apply Theorem 3.2.4 in Srivastava (1998)
to the topological Polish space $\left( X,\tau ^{\prime }\right) $. The
result of Theorem D.1 follows immediately.

Finally, we clarify our claim according to which the result in Fact A.3
still holds if metrizability is replaced by complete regularity.\footnote{%
As we shall see in Section D.2, a Souslin space is completely regular if and
only if it is perfectly normal.} A (Hausdorff) topological space $\left(
X,\tau \right) $\ is completely regular if for any closed set $F\subseteq X$%
\ and any point $x\in X\backslash F$\ there is a continuous function $%
f:X\rightarrow \mathbb{R}$ such that $f\left( x\right) =1$\ and $f\left(
y\right) =0$\ for every $y\in F$. A metrizable space is completely regular.

Let us recall the following property of completely regular Souslin spaces
(see Castaing et al. 2004, p. 17).

\bigskip

\noindent \textbf{Remark D.3} \textit{If }$\left( X,\tau \right) $\textit{\
is a completely regular Souslin space, then there exists on }$X$\textit{\ a
metrizable Souslin topology }$\tau ^{\prime }$\textit{\ which is finer than }%
$\tau $\textit{.}

\bigskip

With this, we obtain the following extension of Fact A.3 to completely
regular Souslin spaces.

\bigskip

\noindent \textbf{Fact A.3.bis }\textit{Suppose that }$\left( X,\tau \right) 
$\textit{\ is a completely regular Souslin space. Then for every sequence }$%
\left( B_{n}\right) _{n\in \mathbb{N}}$\textit{\ of Borel subsets of }$X$%
\textit{\ there exists a finer, completely regular Souslin topology }$\tau
^{\prime }$\textit{\ on }$X$\textit{\ generating the same Borel }$\sigma $%
\textit{-algebra and making each }$B_{n}$\textit{\ clopen.}

\bigskip

\noindent \textbf{Proof}. Let $\left( X,\tau \right) $\ be a completely
regular Souslin space. By Remarks D.3, there is metrizable Souslin topology $%
\tau ^{\prime }$ which is finer than $\tau $. By Remark D.2, both $\left(
X,\tau \right) $\ and $\left( X,\tau ^{\prime }\right) $\ generate the same
Borel $\sigma $-algebra. Hence, apply Corollary D.1 to the topological space 
$\left( X,\tau ^{\prime }\right) $. The result follows.\hfill $\blacksquare $

\section*{D.2 The non-metrizable case}

We now show that the results in the paper remain true if every topological
space is assumed to be a perfectly normal Hausdorff space. We use the same
notation as in the paper; the only difference is that we make explicit the
reference to a topology $\tau $ on a set $X$.

Recall that a topological space $\left( X,\tau \right) $ is normal if, given
any disjoint closed sets $E$\ and $F$, there are disjoint sets $%
O_{E},O_{F}\in \tau $\ such that $E\subseteq O_{E}$\ and $F\subseteq O_{F}$.
A set $E\subseteq X$\ is a $G_{\delta }$-set\ if there is a sequence $\left(
O_{n}\right) _{n\in \mathbb{N}}$ of subsets of $X$ such that $O_{n}\in \tau $%
\ for every $n\in \mathbb{N}$, and $E=\cap _{n\in \mathbb{N}}O_{n}$.

\bigskip

\noindent \textbf{Definition D.1} \textit{A topological Hausdorff space\ is
perfectly normal if it is normal and every closed set is a }$G_{\delta }$%
\textit{-set.}

\bigskip

Throughout this section, we assume that every topological space is
Hausdorff. Therefore, we drop the qualifier \textquotedblleft
Hausdorff\textquotedblright \ in the statements and proofs of the results.
Let us first record some useful properties of perfectly normal (Hausdorff)
spaces:

\begin{description}
\item[P1] Every metrizable space is a perfectly normal space (Engelking
1989, Corollary 4.1.13).

\item[P2] Every perfectly normal space is normal, and so completely regular
(Bogachev 2007, Chapter 6).

\item[P3] If $\left( X,\tau \right) $\ is a completely regular Souslin
space, then $\left( X,\tau \right) $\ is perfectly normal (see Castaing et
al. 2004, p. 4, and Bogachev 2007, Theorem 6.7.7). Consequently, $\left(
X,\tau \right) $\ is a completely regular Souslin space if and only if $%
\left( X,\tau \right) $\ is perfectly normal Souslin space.

\item[P4] If $\left( X,\tau \right) $\ is a a perfectly normal space, then
the Baire $\sigma $-algebra\footnote{%
The Baire $\sigma $-algebra on a topological space $\left( X,\tau \right) $\
is the $\sigma $-algebra generated by the family of real-valued, continuous
functions on $\left( X,\tau \right) $. The Baire $\sigma $-algebra is
coarser than the Borel $\sigma $-algebra.} coincides with the Borel $\sigma $%
-algebra (Bogachev 2007, Proposition 6.3.4).
\end{description}

We let $\left( \Delta \left( X\right) ,\tau _{w}\right) $\ denote the
topological space of Borel probability measures on a perfectly normal space $%
\left( X,\tau \right) $, where $\tau _{w}$\ stands for the topology of weak
convergence.

The required modifications in the main text of the paper pertain only to
Lemma 1, Lemma 2, and the statement of part (ii) of Theorem 1. Precisely,
both lemmas must be replaced by the following \textquotedblleft
weaker\textquotedblright \ versions.

\bigskip

\noindent \textbf{Lemma 1.bis} \textit{Fix a perfectly normal space }$\left(
X,\tau \right) $\textit{\ and a countable family }$\mathcal{B}\subseteq
\Sigma _{X}$ \textit{of conditioning events.}\newline
\textit{\ \ (i) If }$\left( X,\tau \right) $\textit{\ is Souslin or Lusin,
so is }$\left( \Delta \left( X\right) ,\tau _{w}\right) $\textit{. \newline
\ \ (ii) Suppose that }$\mathcal{B}$\textit{\ is clopen and }$\left( X,\tau
\right) $\textit{\ is a Souslin space. Then }$\Delta ^{\mathcal{B}}\left(
X\right) $\textit{\ is compact if and only if }$\left( X,\tau \right) $%
\textit{\ is compact.}

\bigskip

\noindent \textbf{Lemma 2.bis} \textit{Fix measurable spaces }$\left(
X,\Sigma _{X}\right) $\textit{\ and }$\left( Y,\Sigma _{Y}\right) $\textit{,
and a family }$\mathcal{B}\subseteq \Sigma _{X}$\textit{\ of conditioning
events. Suppose that }$f:X\rightarrow Y$\textit{\ is a measurable function
such that }$f^{-1}\left( \mathcal{B}_{Y}\right) =\mathcal{B}_{X}$\textit{.
The following statements hold.\newline
\ \ (i) The map }$\overline{\mathcal{L}}_{f}:\Delta ^{\mathcal{B}}\left(
X\right) \rightarrow \Delta ^{\mathcal{B}_{f}}\left( Y\right) $\textit{\ is
well-defined.\newline
\ \ (ii) Suppose that }$\mathcal{B}$\textit{\ is countable, }$\left( X,\tau
\right) $\textit{\ is a perfectly normal space and }$\left( Y,\tau ^{\prime
}\right) $\textit{\ is a completely regular Souslin space. If }$f$\textit{\
is Borel measurable (resp. continuous), then }$\overline{\mathcal{L}}_{f}$%
\textit{\ is Borel measurable (resp. continuous).}

\bigskip

To prove both lemmas, we show how metrizability can be replaced by perfect
normality in the results of Appendix A. Then we will also discuss the
appropriate modifications of the results and proofs in Appendix B and
Appendix C.

As far as Theorem 1.(ii) is concerned, the following amendment is needed.
The statement must be:%
\begin{equation*}
\text{\textquotedblleft If }\mathcal{T}\text{\ is Souslin, complete, compact
and continuous, then }\mathcal{T}\text{ is terminal.\textquotedblright }
\end{equation*}

\subsection*{D.2.1 Appendix A revisited}

Fact A.1 in Appendix A must be replaced by the following

\bigskip

\noindent \textbf{Fact A.1.bis }\textit{Fix a perfectly normal space }$%
\left( X,\tau \right) $. \textit{If }$\left( X,\tau \right) $\textit{\ is a
Souslin (resp. Lusin, compact) space, then }$\left( \Delta \left( X\right)
,\tau _{w}\right) $\textit{\ is a Souslin (resp. Lusin, compact) space.}

\bigskip

The results in Fact A.1.bis are well-known; see Bogachev (2007, Chapter
8.9). Notice that every compact Souslin space is metrizable.

Next, consider the following version of Fact A.2.

\bigskip

\noindent \textbf{Fact A.2.bis }\textit{Fix a perfectly normal space }$%
\left( X,\tau \right) $\textit{\ and a countable family }$\mathcal{B}%
\subseteq \Sigma _{X}$\textit{\ of conditioning events. If }$\mathcal{B}$%
\textit{\ is clopen, then }$\Delta ^{\mathcal{B}}\left( X\right) $\textit{\
is a closed subset of }$\Delta \left( X\right) ^{\mathcal{B}}$.

\bigskip

\noindent \textbf{Proof}. (Sketch) Let $\left( X,\tau \right) $\ be
perfectly normal. We can use the same proof as in Battigalli and Siniscalchi
(1999, Lemma 1) by weakening the hypothesis of metrizability on $\left(
X,\tau \right) $ to perfect normality. Indeed, the proof in Battigalli and
Siniscalchi (1999) relies on two relevant properties implied by perfect
normality of $\left( X,\tau \right) $. First, every Borel probability
measure is both outer regular (w.r.t. open sets) and inner regular (w.r.t.
closed sets). This follows from two facts:

(1) Every Borel probability measure on a perfectly normal space is inner
regular (Bogachev 2007, Corollary 7.1.9).

(2) A Borel probability measure is outer regular if and only if it is inner
regular; this fact is a consequence of Lemma 12.3 in Aliprantis and Border
(2006).\footnote{%
Alternatively, the result can be proved using arguments similar to those in
the proof of Theorem 12.5 in Aliprantis and Border (2006). To see this, fix
a Borel probability measure $\mu $. Let $\mathcal{F}$ be the family of Borel
sets $E\subseteq X$ such that%
\begin{equation*}
\mu \left( E\right) =\sup \left \{ \mu \left( F\right) :F\subseteq E\text{
is closed}\right \} =\inf \left \{ \mu \left( G\right) :G\supseteq E\text{
is open}\right \} \text{.}
\end{equation*}%
Since each closed set $F$ is a $G_{\delta }$-set, we have $F=\cap _{n\in 
\mathbb{N}}O_{n}$ for some sequence $\left( O_{n}\right) _{n\in \mathbb{N}}$
of open sets; hence, $\mathcal{F}$ contains the closed sets. It can be
checked that $\mathcal{F}$ is closed under complementation and countable
unions. Therefore, $\mathcal{F}$ is the Borel $\sigma $-algebra.}

Second, every perfectly normal space is a normal space. Such fact yields the
existence of a Urysohn function satisfying the required properties for some
steps of Battigalli and Siniscalchi's (1999) proof. Finally, the assumption
that $\mathcal{B}$\ is clopen allows to apply the Portmanteau Theorem for
continuity sets (Bogachev 2007, Corollary 8.2.10). This theorem is used to
prove that $\Delta ^{\mathcal{B}}\left( X\right) $\ is a closed subset (in
the product topology) of $\Delta \left( X\right) ^{\mathcal{B}}$.\hfill $%
\blacksquare $

\bigskip

Fact A.3 in Appendix A is true under the assumption of perfect normality
because, as shown in Section D.1 of this Supplementary Appendix, the result
still holds if $\left( X,\tau \right) $ is completely regular Souslin space
(cf. Fact A.3.bis); as seen, complete regularity is equivalent to perfect
normality for Souslin spaces. With these modifications, the proof of Lemma
1.bis is identical to the proof of Lemma 1 in the paper.

Next, to prove Lemma 2.bis, we need a preliminary result (Fact D.1 below).
In the proof that follows, we use the following notation concerning a
special map. Fix a measurable space $\left( X,\Sigma _{X}\right) $. For
every $E\in \Sigma _{X}$, the evaluation map $\mathbf{e}_{E}:\Delta \left(
X\right) \rightarrow \left[ 0,1\right] $ is defined by $\mathbf{e}_{E}\left(
\mu \right) :=\int \boldsymbol{1}_{E}d\mu $,\ where, as customary, $%
\boldsymbol{1}_{E}$\ denotes the indicator function on the set $E$.

\bigskip

\noindent \textbf{Fact D.1} \textit{Let }$\left( X,\tau \right) $\textit{\
be a perfectly normal space. Then}%
\begin{equation*}
\mathcal{F}_{\Delta \left( X\right) }\subseteq \Sigma _{\Delta \left(
X\right) }\text{.}
\end{equation*}%
\textit{Furthermore, if }$\left( X,\tau \right) $\textit{\ is Souslin, then}%
\begin{equation*}
\mathcal{F}_{\Delta \left( X\right) }=\Sigma _{\Delta \left( X\right) }\text{%
.}
\end{equation*}%
\textit{\ }

\bigskip

\noindent \textbf{Proof}. Recall that, according to the definition given in
Appendix A, $\mathcal{F}_{\Delta \left( X\right) }$\ is the $\sigma $%
-algebra\ on $\Delta \left( X\right) $ generated by all sets of the form%
\begin{equation*}
b^{p}\left( E\right) :=\left \{ \mu \in \Delta \left( X\right) :\mu \left(
E\right) \geq p\right \} \text{,}
\end{equation*}%
where $E\in \Sigma _{X}$ and $p\in \mathbb{Q\cap }\left[ 0,1\right] $. It is
enough to show that $b^{p}\left( E\right) \in \Sigma _{\Delta \left(
X\right) }$ for every $E\in \Sigma _{X}$. We first show that $b^{p}\left(
E\right) \in \Sigma _{\Delta \left( X\right) }$\ whenever $E$ is closed,
then we will make use of the $\pi $-$\lambda $\ Theorem to prove the claim.

It is well-known (Topsoe 1970, p. 40, and Schwartz 1973, Appendix) that if $%
\left( X,\tau \right) $ is a perfectly normal (so completely regular) space,
then $\tau _{w}$ is such that every\ evaluation map $\mathbf{e}_{F}:\Delta
\left( X\right) \rightarrow \left[ 0,1\right] $, where $F\subseteq X$ is 
\textit{closed}, is upper semicontinuous and $\mu \mapsto \mathbf{e}%
_{X}\left( \mu \right) =1$\ is continuous.\footnote{%
This is a property of the so-called \textquotedblleft weak
topology\textquotedblright \ (see Topsoe 1970, p. 40). If $\left( X,\tau
\right) $\ is a perfectly normal space, then the \textquotedblleft weak
topology\textquotedblright \ coincides with the topology of weak
convergence. The \textquotedblleft weak topology\textquotedblright \ is
called narrow topology and $A$-topology in Schwartz (1973) and Bogachev
(2007), respectively.} With this, we can claim that each set%
\begin{equation*}
b^{p}\left( E\right) :=\left \{ \mu \in \Delta \left( X\right) :\mu \left(
E\right) \geq p\right \} \text{,}
\end{equation*}%
with $E\subseteq X$ closed and $p\in \mathbb{Q\cap }\left[ 0,1\right] $, is
closed. Indeed, by definition (see Aliprantis and Border 2006, p. 43), a
function $g:\Delta \left( X\right) \rightarrow \mathbb{R}$\ is upper
semicontinuous if, for each $c\in \mathbb{R}$, the set $\left \{ \mu \in
\Delta \left( X\right) :g\left( \mu \right) \geq c\right \} $\ is closed.
Therefore, by writing $b^{p}\left( E\right) =\left \{ \mu \in \Delta \left(
X\right) :\mathbf{e}_{E}\left( \mu \right) \geq p\right \} $, the claim
follows. Thus, $b^{p}\left( E\right) \in \Sigma _{\Delta \left( X\right) }$
because $\Sigma _{\Delta \left( X\right) }$\ is the smallest $\sigma $%
-algebra containing the closed sets.

Define now the following classes of subsets of $X$:%
\begin{eqnarray*}
\mathcal{E} &:&=\left \{ F\in \Sigma _{X}:F\text{ is closed}\right \} \text{,%
} \\
\mathcal{D} &:&=\left \{ E\in \Sigma _{X}:b^{p}\left( E\right) \in \Sigma
_{\Delta \left( X\right) }\right \} \text{.}
\end{eqnarray*}%
Clearly, $\mathcal{E}$\ is stable under finite intersections, and $\mathcal{E%
}\subseteq \mathcal{D}$. We show that $\mathcal{D}$\ is a $\lambda $-system.
With this,%
\begin{equation*}
\Sigma _{\Delta \left( X\right) }=\sigma \left( \mathcal{E}\right) \subseteq 
\mathcal{D}\text{,}
\end{equation*}%
which follows from the $\pi $-$\lambda $\ Theorem.

(i) $X\in \mathcal{D}$. Immediate.

(ii) If $E_{1},E_{2}\in \mathcal{D}$\ and $E_{1}\supseteq E_{2}$, then $%
E_{1}\backslash E_{2}\in \mathcal{D}$. To see this, observe that $\mu \left(
E_{1}\backslash E_{2}\right) +\mu \left( E_{2}\right) =\mu \left(
E_{1}\right) $. Hence%
\begin{eqnarray*}
b^{p}\left( E_{1}\backslash E_{2}\right) &=&\left \{ \mu \in \Delta \left(
X\right) :\mu \left( E_{1}\backslash E_{2}\right) \geq p\right \} \\
&=&\left \{ \mu \in \Delta \left( X\right) :\mu \left( E_{2}\right) \leq \mu
\left( E_{1}\right) -p\right \} \\
&=&\Delta \left( X\right) \backslash \left \{ \mu \in \Delta \left( X\right)
:\mu \left( E_{2}\right) >\mu \left( E_{1}\right) -p\right \} \text{.}
\end{eqnarray*}%
Let $\alpha :=\mu \left( E_{1}\right) $. Note that the set%
\begin{eqnarray*}
\left \{ \mu \in \Delta \left( X\right) :\mu \left( E_{2}\right) >\alpha
-p\right \} &=&\tbigcup_{q=1}^{\infty }\left \{ \mu \in \Delta \left(
X\right) :\mu \left( E_{2}\right) \geq \alpha -p+\frac{1}{q}\right \} \\
&=&\tbigcup_{q=1}^{\infty }b^{\alpha -p+\frac{1}{q}}\left( E_{2}\right)
\end{eqnarray*}%
belongs to $\Sigma _{\Delta \left( X\right) }$. Thus, $b^{p}\left(
E_{1}\backslash E_{2}\right) \in \Sigma _{\Delta \left( X\right) }$.

(iii) If $\left( E_{n}\right) _{n\in \mathbb{N}}\in \mathcal{D}^{\mathbb{N}}$%
\ is such that $E_{n}\subseteq E_{n+1}$ for every $n\in \mathbb{N}$, then $%
\cup _{n\in \mathbb{N}}E_{n}\in \mathcal{D}$. To see this, first note that
if $\cup _{n\in \mathbb{N}}E_{n}=X$ then the result follows from part (i).
Therefore, assume $\cup _{n\in \mathbb{N}}E_{n}\neq X$, and observe that%
\begin{equation*}
\mu \left( \tbigcup_{n\in \mathbb{N}}E_{n}\right) +\mu \left( X\backslash
\left( \tbigcup_{n\in \mathbb{N}}E_{n}\right) \right) =\mu \left(
\tbigcup_{n\in \mathbb{N}}E_{n}\right) +\mu \left( \tbigcap_{n\in \mathbb{N}%
}\left( X\backslash E_{n}\right) \right) =1\text{.}
\end{equation*}%
The sequence $\left( X\backslash E_{n}\right) _{n\in \mathbb{N}}$\ is
non-increasing and $E:=\cap _{n\in \mathbb{N}}\left( X\backslash
E_{n}\right) \neq \emptyset $. Thus,%
\begin{eqnarray*}
b^{p}\left( \tbigcup_{n\in \mathbb{N}}E_{n}\right) &=&\left \{ \mu \in
\Delta \left( X\right) :\mu \left( \tbigcup_{n\in \mathbb{N}}E_{n}\right)
\geq p\right \} \\
&=&\left \{ \mu \in \Delta \left( X\right) :\mu \left( E\right) \leq
1-p\right \} \\
&=&\Delta \left( X\right) \backslash \left \{ \mu \in \Delta \left( X\right)
:\mu \left( E\right) >1-p\right \} \\
&=&\Delta \left( X\right) \backslash \left( \tbigcup_{q=1}^{\infty }b^{1-p+%
\frac{1}{q}}\left( E\right) \right) \text{,}
\end{eqnarray*}%
where the forth equality follows from the same argument as the one in step
(ii). Thus, $b^{p}\left( \cup _{n\in \mathbb{N}}E_{n}\right) \in \Sigma
_{\Delta \left( X\right) }$.

Summing up, $\mathcal{D}$\ is a $\lambda $-system, hence $\sigma \left( 
\mathcal{E}\right) \subseteq \mathcal{D}$. This means that $b^{p}\left(
E\right) \in \Sigma _{\Delta \left( X\right) }$ for every $E\in \Sigma _{X}$%
. Hence, $\mathcal{F}_{\Delta \left( X\right) }\subseteq \Sigma _{\Delta
\left( X\right) }$.

If $\left( X,\tau \right) $\ is Souslin, then the equality $\mathcal{F}%
_{\Delta \left( X\right) }=\Sigma _{\Delta \left( X\right) }$ is shown in
Bogachev (2018, Corollary 5.1.9). An alternative, short proof---which uses
the properties of Souslin topologies---is as follows. By P3, $\left( X,\tau
\right) $\ is a completely regular Souslin\ space. By Remarks D2-D3, there
exists (on $X$) a metrizable Souslin topology $\tau ^{\prime }$\ which is
finer than $\tau $ and such that $\sigma \left( \tau \right) =\sigma \left(
\tau ^{\prime }\right) =\Sigma _{\Delta \left( X\right) }$. Since $\tau
^{\prime }$\ is separable, Theorem 2.3 in Gaudard and Handwin (1989) yields $%
\sigma \left( \tau ^{\prime }\right) =\mathcal{F}_{\Delta \left( X\right) }$%
.\hfill $\blacksquare $

\bigskip

We are now ready to state and prove the following analogue of Fact A.4 in
Appendix A. For the definitions of the families $\mathcal{F}_{\Delta \left(
X\right) ^{\mathcal{B}}}$, $\mathcal{G}_{\Delta \left( X\right) ^{\mathcal{B}%
}}$\ and $\Sigma _{\Delta \left( X\right) ^{\mathcal{B}}}$, see Appendix A.

\bigskip

\noindent \textbf{Fact A.4.bis }\textit{Fix\ a perfectly normal space }$%
\left( X,\tau \right) $\textit{\ and a countable family }$\mathcal{B}%
\subseteq \Sigma _{X}$\textit{\ of conditioning events. The following
statements hold.}\newline
\textit{\ \ (i)} $\mathcal{F}_{\Delta \left( X\right) ^{\mathcal{B}}}$\ 
\textit{is generated by all sets of the form}%
\begin{equation*}
\left \{ \mu \in \Delta \left( X\right) ^{\mathcal{B}}:\mu \left( E|B\right)
\geq p\right \} \text{,}
\end{equation*}%
\textit{where }$E\in \Sigma _{Y}$\textit{, }$B\in B$\textit{\ and }$p\in
Q\cap \left[ 0,1\right] $\textit{.}\newline
\textit{\ \ (ii) }$\mathcal{F}_{\Delta \left( X\right) ^{\mathcal{B}%
}}\subseteq \mathcal{G}_{\Delta \left( X\right) ^{\mathcal{B}}}\subseteq
\Sigma _{\Delta \left( X\right) ^{\mathcal{B}}}$\textit{.}\newline
\textit{\ \ (iii) If }$\left( X,\tau \right) $\textit{\ is a Souslin space,
then }$\mathcal{F}_{\Delta \left( X\right) ^{\mathcal{B}}}=\mathcal{G}%
_{\Delta \left( X\right) ^{\mathcal{B}}}=\Sigma _{\Delta \left( X\right) ^{%
\mathcal{B}}}$\textit{.}

\bigskip

\noindent \textbf{Proof}. Part (i) follows by inspection of the definitions
of $\mathcal{F}_{\Delta \left( X\right) }$ and $\mathcal{F}_{\Delta \left(
X\right) ^{\mathcal{B}}}$. To show part (ii), note that, by Fact D.1, $%
\mathcal{F}_{\Delta \left( X\right) }\subseteq \Sigma _{\Delta \left(
X\right) }$. This entails $\mathcal{F}_{\Delta \left( X\right) ^{\mathcal{B}%
}}\subseteq \mathcal{G}_{\Delta \left( X\right) ^{\mathcal{B}}}$. Moreover,
by Lemma 6.4.1 in Bogachev (2007), $\mathcal{G}_{\Delta \left( X\right) ^{%
\mathcal{B}}}\subseteq \Sigma _{\Delta \left( X\right) ^{\mathcal{B}}}$. As
for part (iii), assume that $\left( X,\tau \right) $\ is Souslin. By Fact
D.1, $\mathcal{F}_{\Delta \left( X\right) }=\Sigma _{\Delta \left( X\right)
} $. This yields $\mathcal{F}_{\Delta \left( X\right) ^{\mathcal{B}}}=%
\mathcal{G}_{\Delta \left( X\right) ^{\mathcal{B}}}$. Furthermore, if $%
\left( X,\tau \right) $ is Souslin then $\left( \Delta \left( X\right) ,\tau
_{w}\right) $ is also Souslin, so that it is a separable space. Thus, by
Lemma 6.4.2 in Bogachev (2007), $\mathcal{G}_{\Delta \left( X\right) ^{%
\mathcal{B}}}=\Sigma _{\Delta \left( X\right) ^{\mathcal{B}}}$.\hfill $%
\blacksquare $

\bigskip

Finally:

\bigskip

\noindent \textbf{Proof of Lemma 2.bis}. The proof of part (i) is in the
paper. As for the proof of part (ii), the expression \textquotedblleft Fact
A.4\textquotedblright \ must be replaced by \textquotedblleft Fact
A.4.bis,\textquotedblright \ and nets (instead of sequences) must be used
for the proof concerning continuity of the pushforward-CPS map.\hfill $%
\blacksquare $

\subsection*{D.2.2 Appendix B revisited}

Here, the required modifications are easy to identify. Specifically, in the
proofs of Remark 1 and Proposition B.2 sequences must be replaced by nets.
The rest remains unaltered. The proof of Theorem B.1 does not depend on
metrizability of Souslin spaces. As far as the proof of Proposition B.1 is
concerned, the changes to be made are:

(1) use nets in place of sequences to prove continuity of the function $%
f_{i} $;

(2) as indicated in the footnote, the argument pertaining to the existence
of an increasing sequence covering the open set $U$ can be extended to nets.

\subsection*{D.2.3 Appendix C revisited}

Fact C.1 holds without the metrizability assumption because Souslin spaces
are separable. In the proof of Lemma 9.1 in Friedenberg and Meier (2017),
separability is used to conclude the following fact: If $\left( X,\tau
\right) $ and $\left( Y,\tau ^{\prime }\right) $ are separable, then the
Borel $\sigma $-algebra generated by the product topology on $X\times Y$\
coincides with the product of the Borel $\sigma $-algebras generated by $%
\tau $\ and $\tau ^{\prime }$. This property holds for Souslin spaces. The
remainder of the appendix applies verbatim.

\section*{D.3 Omitted Proofs}

We prove \textbf{Remark 5}\ in the main text, which is reported here for
convenience.

\bigskip

\noindent \textbf{Remark 5\ }\textit{The map }$h:=\left( h_{i}\right) _{i\in
I}:\tprod_{i\in I}T_{i}\rightarrow \tprod_{i\in I}H_{i}$\textit{\ is a type
morphism from }$\mathcal{T}$\textit{\ to }$\mathcal{T}^{\mathrm{c}}$\textit{%
. That is, for each }$i\in I$\textit{\ and for each }$t_{i}\in T_{i}$\textit{%
,}%
\begin{equation*}
\beta _{i}^{\mathrm{c}}\left( h_{i}\left( t_{i}\right) \right) =\overline{%
\mathcal{L}}_{\left( \mathrm{Id}_{S},h_{j}\right) }\left( \beta _{i}\left(
t_{i}\right) \right) \text{\textit{.}}
\end{equation*}

\bigskip

\noindent \textbf{Proof}. Fix a player $i\in I$\ and $t_{i}\in T_{i}$. We
prove the following claim: for every $E\in \cup _{n\in \mathbb{N}}\left(
\rho _{i}^{n-1}\right) ^{-1}\left( \Sigma _{\Theta _{i}^{n-1}}\right) $\ and
every $B\in \mathcal{B}_{i}$,%
\begin{equation}
\beta _{i}^{\mathrm{c}}\left( h_{i}\left( t_{i}\right) \right) \left( E\left
\vert B\times H_{j}\right. \right) =\overline{\mathcal{L}}_{\left( \mathrm{Id%
}_{S},h_{j}\right) }\left( \beta _{i}\left( t_{i}\right) \right) \left(
E\left \vert B\times H_{j}\right. \right) \text{.}  \tag{D.1}
\label{equality type morphism}
\end{equation}%
That is, we show that the equality holds for every element of the cylinder
algebra on $\Theta _{i}=S\times H_{j}$. With this, the same monotone class
argument as in the proof of Theorem B.1 shows that the equality holds for
every Borel subset of $\Theta _{i}$.

Let us first record some properties of the involved maps: for all $n\in 
\mathbb{N}$,%
\begin{equation}
T_{j}=\left( h_{j}\right) ^{-1}\left( H_{j}\right) =\left( h_{j}^{n}\right)
^{-1}\left( H_{j}^{n}\right) \text{,}  \tag{D.2}  \label{property h_j}
\end{equation}%
and%
\begin{equation}
h_{-i}^{n}=\rho _{i}^{n}\circ h_{-i}\text{.}  \tag{D.3}
\label{property maps coherence limit}
\end{equation}%
Specifically, (\ref{property h_j})\ follows from Claim 1 and Remark 4. To
see why (\ref{property maps coherence limit}) holds, note that $%
h_{j}^{n}=\pi _{j}^{n}\circ h_{j}$, and the result follows by definitions of 
$h_{-i}^{n}=\left( \mathrm{Id}_{\Theta _{i}^{0}},h_{j}^{n}\right) $, $\rho
_{i}^{n}=\left( \mathrm{Id}_{\Theta _{i}^{0}},\pi _{j}^{n}\right) $ and $%
h_{-i}=\left( \mathrm{Id}_{\Theta _{i}^{0}},h_{j}\right) $.

Now, pick any $E\in \cup _{n\in \mathbb{N}}\left( \rho _{i}^{n-1}\right)
^{-1}\left( \Sigma _{\Theta _{i}^{n-1}}\right) $ and $B\in \mathcal{B}_{i}$.
Then there are $n\in \mathbb{N}$\ and $E_{n-1}\in \Sigma _{\Theta
_{i}^{n-1}} $\ such that $E=\left( \rho _{i}^{n-1}\right) ^{-1}\left(
E_{n-1}\right) $. Since $\rho _{i}^{n-1}=\rho _{i}^{n-1,n}\circ \rho
_{i}^{n} $, we have $E=\left( \rho _{i}^{n}\right) ^{-1}\left( E_{n}\right) $
where $E_{n}:=\left( \rho _{i}^{n-1,n}\right) ^{-1}\left( E_{n-1}\right) \in
\Sigma _{\Theta _{i}^{n}}$. By (\ref{property maps coherence limit}),%
\begin{equation*}
\left( h_{-i}^{n}\right) ^{-1}\left( E_{n}\right) =\left( h_{-i}\right)
^{-1}\left( \left( \rho _{i}^{n}\right) ^{-1}\left( E_{n}\right) \right) 
\text{.}
\end{equation*}%
With this, using (\ref{property h_j}), the right-hand side of (\ref{equality
type morphism}) is equal to%
\begin{eqnarray*}
\overline{\mathcal{L}}_{\left( \mathrm{Id}_{S},h_{j}\right) }\left( \beta
_{i}\left( t_{i}\right) \right) \left( E\left \vert B\times H_{j}\right.
\right) &=&\beta _{i}\left( t_{i}\right) \left( \left( h_{-i}\right)
^{-1}\left( E\right) \left \vert B\times \left( h_{j}\right) ^{-1}\left(
H_{j}\right) \right. \right) \\
&=&\beta _{i}\left( t_{i}\right) \left( \left( h_{-i}\right) ^{-1}\left(
\left( \rho _{i}^{n}\right) ^{-1}\left( E_{n}\right) \right) \left \vert
B\times \left( h_{j}^{n}\right) ^{-1}\left( H_{j}^{n}\right) \right. \right)
\\
&=&\beta _{i}\left( t_{i}\right) \left( \left( h_{-i}^{n}\right) ^{-1}\left(
E_{n}\right) \left \vert \left( h_{-i}^{n}\right) ^{-1}\left( B\times
H_{j}^{n}\right) \right. \right) \\
&=&\overline{\mathcal{L}}_{h_{-i}^{n}}\left( \beta _{i}\left( t_{i}\right)
\right) \left( E_{n}\left \vert B\times H_{j}^{n}\right. \right) \text{.}
\end{eqnarray*}%
For convenience, let $\left( \mu _{i}^{1},\mu _{i}^{2},...\right)
:=h_{i}\left( t_{i}\right) $ so that $\mu _{i}^{m+1}=\overline{\mathcal{L}}%
_{h_{-i}^{m}}\left( \beta _{i}\left( t_{i}\right) \right) $\ for every $%
m\geq 0$. Also, let $\mu _{i}:=\beta _{i}^{\mathrm{c}}\left( h_{i}\left(
t_{i}\right) \right) $. The left-hand side of (\ref{equality type morphism})
is equal to%
\begin{eqnarray*}
\beta _{i}^{\mathrm{c}}\left( h_{i}\left( t_{i}\right) \right) \left( E\left
\vert B\times H_{j}\right. \right) &=&\beta _{i}^{\mathrm{c}}\left(
h_{i}\left( t_{i}\right) \right) \left( \left( \rho _{i}^{n}\right)
^{-1}\left( E_{n}\right) \left \vert B\times \left( \pi _{j}^{n}\right)
^{-1}\left( H_{j}^{n}\right) \right. \right) \\
&=&\mu _{i}\left( \left( \rho _{i}^{n}\right) ^{-1}\left( E_{n}\right) \left
\vert \left( \rho _{i}^{n}\right) ^{-1}\left( B\times H_{j}^{n}\right)
\right. \right) \\
&=&\mu _{i}^{n+1}\left( E_{n}\left \vert B\times H_{j}^{n}\right. \right) \\
&=&\overline{\mathcal{L}}_{h_{-i}^{n}}\left( \beta _{i}\left( t_{i}\right)
\right) \left( E_{n}\left \vert B\times H_{j}^{n}\right. \right) \text{.}
\end{eqnarray*}%
This shows that (\ref{equality type morphism})\ holds, as desired.\hfill $%
\blacksquare $

\bigskip

We now prove \textbf{Remark 6} in the main text.

\bigskip

\noindent \textbf{Remark 6\ }\textit{An }$\left( S,\left( \mathcal{B}%
_{i}\right) _{i\in I}\right) $\textit{-based type structure }$\mathcal{T}$%
\textit{\ is finitely terminal if and only if, for each }$\left( S,\left( 
\mathcal{B}_{i}\right) _{i\in I}\right) $\textit{-based type structure }$%
\mathcal{T}^{\ast }$\textit{, for each }$i\in I$\textit{\ and each }$n\in N$%
\textit{,}%
\begin{equation*}
h_{i}^{\ast ,n}\left( T_{i}^{\ast }\right) \subseteq h_{i}^{n}\left(
T_{i}\right) \text{.}
\end{equation*}%
\textit{An }$\left( S,\left( \mathcal{B}_{i}\right) _{i\in I}\right) $%
\textit{-based type structure }$\mathcal{T}$\textit{\ is terminal if and
only if, for each }$\left( S,\left( \mathcal{B}_{i}\right) _{i\in I}\right) $%
\textit{-based type structure }$\mathcal{T}^{\ast }$\textit{, and for each }$%
i\in I$\textit{,}%
\begin{equation*}
h_{i}^{\ast }\left( T_{i}^{\ast }\right) \subseteq h_{i}\left( T_{i}\right) 
\text{.}
\end{equation*}

\bigskip

\noindent \textbf{Proof}. We present only the proof for the case of a
finitely terminal structure $\mathcal{T}$, since the other case is
analogous. Suppose that $\mathcal{T}$\ is finitely terminal. Fix any
structure $\mathcal{T}^{\ast }$. Fix also $i\in I$\ and $n\in \mathbb{N}$.
We show that $h_{i}^{\ast ,n}\left( T_{i}^{\ast }\right) \subseteq
h_{i}^{n}\left( T_{i}\right) $. To this end, let $\left( \mu
_{i}^{1},...,\mu _{i}^{n}\right) \in h_{i}^{\ast ,n}\left( T_{i}^{\ast
}\right) $. (We know from Claim 1 that $h_{i}^{\ast ,n}\left( T_{i}^{\ast
}\right) \subseteq H_{i}^{n}$, so each element is a finite sequence of order 
$n$ of CPSs.) We need to show that $\left( \mu _{i}^{1},...,\mu
_{i}^{n}\right) \in h_{i}^{n}\left( T_{i}\right) $, that is, by definition
of image of a function, we need to show the existence of $t_{i}\in T_{i}$\
such that $h_{i}^{n}\left( t_{i}\right) =\left( \mu _{i}^{1},...,\mu
_{i}^{n}\right) $. Note that, by definition of $h_{i}^{\ast ,n}\left(
T_{i}^{\ast }\right) $, there is some $t_{i}^{\ast }\in T_{i}^{\ast }$\ such
that $h_{i}^{\ast ,n}\left( t_{i}^{\ast }\right) =\left( \mu
_{i}^{1},...,\mu _{i}^{n}\right) $.\ Since $\mathcal{T}$\ is finitely
terminal, by definition there exists $t_{i}\in T_{i}$\ such that $%
h_{i}^{\ast ,n}\left( t_{i}^{\ast }\right) =h_{i}^{n}\left( t_{i}\right) $.
Hence, we have found $t_{i}\in T_{i}$\ such that $h_{i}^{n}\left(
t_{i}\right) =\left( \mu _{i}^{1},...,\mu _{i}^{n}\right) $, as desired.

Conversely, consider any type structure $\mathcal{T}^{\ast }$\ such that $%
h_{i}^{\ast ,n}\left( T_{i}^{\ast }\right) \subseteq h_{i}^{n}\left(
T_{i}\right) $\ for each $i\in I$\ and each $n\in \mathbb{N}$. Fix $i\in I$\
and $n\in \mathbb{N}$. Pick any $t_{i}^{\ast }\in T_{i}^{\ast }$. We have to
show the existence of some $t_{i}\in T_{i}$\ such that $h_{i}^{\ast
,n}\left( t_{i}^{\ast }\right) =h_{i}^{n}\left( t_{i}\right) $. Since $i\in
I $, $n\in \mathbb{N}$ and $t_{i}^{\ast }\in T_{i}^{\ast }$\ are arbitrary,
this will imply that $\mathcal{T}$\ is finitely terminal. Certainly, we have 
$h_{i}^{\ast ,n}\left( t_{i}^{\ast }\right) \in h_{i}^{\ast ,n}\left(
T_{i}^{\ast }\right) $. By assumption, $h_{i}^{\ast ,n}\left( t_{i}^{\ast
}\right) \in h_{i}^{n}\left( T_{i}\right) $. Hence, by definition of image
of a function, there is $t_{i}\in T_{i}$\ such that $h_{i}^{\ast ,n}\left(
t_{i}^{\ast }\right) =h_{i}^{n}\left( t_{i}\right) $.\hfill $\blacksquare $

\bigskip

For completeness, we also provide the proof of part (i) of \textbf{%
Proposition 2}.

\bigskip

\noindent \textbf{Proof of Proposition 2.(i)}. Suppose that $\mathcal{T}$\
is finitely terminal, and fix a player $i\in I$ and $n\in \mathbb{N}$.
Since\ $h_{i}^{n}\left( T_{i}\right) \subseteq H_{i}^{n}$ holds true (as
shown in Claim 1), we only need to show that\ $H_{i}^{n}\subseteq
h_{i}^{n}\left( T_{i}\right) $. We make use of the characterization of the
notion of finite terminality given in Remark 6. Since $\mathcal{T}$\ is
finitely terminal, by considering structure $\mathcal{T}^{\mathrm{c}}$\ we
get%
\begin{equation*}
h_{i}^{\mathrm{c},n}\left( H_{i}^{n}\right) =H_{i}^{n}\subseteq
h_{i}^{n}\left( T_{i}\right) \text{,}
\end{equation*}%
where the first equality holds because the hierarchy map $h_{i}^{\mathrm{c}%
}:T_{i}^{\mathrm{c}}\rightarrow H_{i}$ is the identity. Conversely, suppose
that $\mathcal{T}$\ satisfies $h_{i}^{n}\left( T_{i}\right) =H_{i}^{n}$ for
each $i\in I$ and each $n\in \mathbb{N}$. For any structure $\mathcal{T}%
^{\ast }$, we have $h_{i}^{\ast ,n}\left( T_{i}^{\ast }\right) \subseteq
H_{i}^{n}=h_{i}^{n}\left( T_{i}\right) $ for each $i\in I$ and each $n\in 
\mathbb{N}$, where the inclusion follows again from Claim 1. Thus, $\mathcal{%
T}$\ is finitely terminal.\hfill $\blacksquare $

\bigskip

We now prove one step of the proof of \textbf{Theorem 1} which was taken for
granted. Specifically, it is that part of the proof (inductive step of part
(i)) where it is claimed that, \textquotedblleft ...by coherence of $\mu
_{i}^{n+1}$ with respect to $\mu _{i}^{m}$ ($m\leq n$), it follows that%
\begin{equation*}
\mu _{i}^{m}=\overline{\mathcal{L}}_{h_{-i}^{m-1}}\left( \beta _{i}\left(
t_{i}\right) \right)
\end{equation*}%
for all $m$\ such that $1\leq m\leq n$.\textquotedblright

To show why this is the case, pick any $m$\ such that $1\leq m\leq n$.
Recall the map $\overline{\mathcal{L}}_{h_{-i}^{m-1}}\circ \beta
_{i}:T_{i}\rightarrow \Delta ^{\mathcal{B}_{i}}\left( \Theta
_{i}^{m-1}\right) $, where $\Delta ^{\mathcal{B}_{i}}\left( \Theta
_{i}^{m-1}\right) =\Delta ^{\mathcal{B}_{i}^{m-1}}\left( \Theta
_{i}^{m-1}\right) $. Pick any event $E\subseteq \Theta _{i}^{m-1}$\ and any $%
B\in \mathcal{B}_{i}^{m-1}$. We have%
\begin{eqnarray*}
\mu _{i}^{m}\left( E\left \vert B\right. \right) &=&\overline{\mathcal{L}}%
_{\rho _{i}^{m-1,n}}\left( \mu _{i}^{n+1}\right) \left( E\left \vert
B\right. \right) \\
&=&\overline{\mathcal{L}}_{\rho _{i}^{m-1,n}}\left( \overline{\mathcal{L}}%
_{h_{-i}^{n}}\left( \beta _{i}\left( t_{i}\right) \right) \right) \left(
E\left \vert B\right. \right) \\
&=&\overline{\mathcal{L}}_{h_{-i}^{n}}\left( \beta _{i}\left( t_{i}\right)
\right) \left( \left( \rho _{i}^{m-1,n}\right) ^{-1}\left( E\right) \left
\vert \left( \rho _{i}^{m-1,n}\right) ^{-1}\left( B\right) \right. \right) \\
&=&\beta _{i}\left( t_{i}\right) \left( \left( \rho _{i}^{m-1,n}\circ
h_{-i}^{n}\right) ^{-1}\left( E\right) \left \vert \left( \rho
_{i}^{m-1,n}\circ h_{-i}^{n}\right) ^{-1}\left( B\right) \right. \right) \\
&=&\overline{\mathcal{L}}_{h_{-i}^{m-1}}\left( \beta _{i}\left( t_{i}\right)
\right) \left( E\left \vert B\right. \right)
\end{eqnarray*}%
where the first equality follows from coherence of $\mu _{i}^{n+1}$\ with
respect to $\mu _{i}^{m}$, the following three equalities are obvious, and
the last equality follows from condition (4.2) in the text (actually, such
condition is stated as $\rho _{i}^{n-1,n}\circ h_{-i}^{n}=h_{-i}^{n-1}$, but
this is irrelevant, given the recursive definition of the maps).

\bigskip

Next, we prove \textbf{Proposition B.1} in Appendix B.

\bigskip

\noindent \textbf{Proposition B.1 }\textit{Consider the projective sequences 
}$\left( H_{i}^{n},\pi _{i}^{n,n+1}\right) _{n\in \mathbb{N}}$\textit{\ and }%
$\left( \Theta _{i}^{n-1},\rho _{i}^{n-1,n}\right) _{n\in \mathbb{N}}$%
\textit{. Then, for each }$i\in I$\textit{\ and each }$n\in \mathbb{N}$%
\textit{, the bonding maps }$\pi _{i}^{n,n+1}:H_{i}^{n+1}\rightarrow
H_{i}^{n}$\textit{\ and }$\rho _{i}^{n-1,n}:\Theta _{i}^{n}\rightarrow
\Theta _{i}^{n-1}$\textit{\ are surjective and open.}

\bigskip

\noindent \textbf{Proof}. For every $\left( \mu _{i}^{1},...,\mu
_{i}^{n}\right) \in H_{i}^{n}$\ we have to find a $\mu _{i}^{n+1}\in \Delta
^{\mathcal{B}_{i}^{n}}\left( \Theta _{i}^{n}\right) $ such that $\left( \mu
_{i}^{1},...,\mu _{i}^{n},\mu _{i}^{n+1}\right) \in H_{i}^{n+1}$, i.e., $%
\overline{\mathcal{L}}_{\rho _{i}^{n-1,n}}\left( \mu _{i}^{n+1}\right) =\mu
_{i}^{n}$. That is, any coherent $n$-level hierarchy can be extended to a
coherent $\left( n+1\right) $-level hierarchy. We will prove, by induction
on $n\in \mathbb{N}$, the following claim: For each $i\in I$, there exists a
continuos function $\psi _{i}^{n}:\Theta _{i}^{n-1}\rightarrow \Theta
_{i}^{n}$ such that $\rho _{i}^{n-1,n}\circ \psi _{i}^{n}=\mathrm{Id}%
_{\Theta _{i}^{n-1}}$\ and $\overline{\mathcal{L}}_{\rho _{i}^{n-1,n}}\left( 
\overline{\mathcal{L}}_{\psi _{i}^{n}}\left( \mu _{i}^{n}\right) \right)
=\mu _{i}^{n}$ for every $\mu _{i}^{n}\in \Delta ^{\mathcal{B}%
_{i}^{n-1}}\left( \Theta _{i}^{n-1}\right) $.

For $n=1$, fix a player $i\in I$. Fix also a CPS $\mu _{j}^{1}\in \Delta ^{%
\mathcal{B}_{j}^{0}}\left( \Theta _{j}^{0}\right) $\ and define the function 
$\psi _{i}^{1}:\Theta _{i}^{0}\rightarrow \Theta _{i}^{1}$($:=\Theta
_{i}^{0}\times \Delta ^{\mathcal{B}_{j}^{0}}(\Theta _{j}^{0})$) by%
\begin{equation*}
\psi _{i}^{1}\left( s\right) :=\left( \mathrm{Id}_{\Theta _{i}^{0}}\left(
s\right) ,\mu _{j}^{1}\right)
\end{equation*}%
for all $s\in S$. Clearly, $\psi _{i}^{1}$\ is continuous and satisfies $%
\rho _{i}^{0,1}\circ \psi _{i}^{1}=\mathrm{Id}_{\Theta _{i}^{0}}$. Note that
the map $\overline{\mathcal{L}}_{\psi _{i}^{1}}:\Delta ^{\mathcal{B}%
_{i}^{0}}\left( \Theta _{i}^{0}\right) \rightarrow \Delta ^{\mathcal{B}%
_{i}^{1}}\left( \Theta _{i}^{1}\right) $ is well-defined because the
hypothesis of Lemma 2\ are satisfied: by construction, $\mathcal{B}%
_{i}^{0}=\left( \psi _{i}^{1}\right) ^{-1}\left( \mathcal{B}_{i}^{1}\right) $%
. Moreover, $\overline{\mathcal{L}}_{\psi _{i}^{1}}$\ is continuous. For
each $\mu _{i}^{1}\in \Delta ^{\mathcal{B}_{i}^{0}}\left( \Theta
_{i}^{0}\right) $, the image CPS $\overline{\mathcal{L}}_{\psi
_{i}^{1}}\left( \mu _{i}^{1}\right) $ is a coherent extension of $\mu
_{i}^{1}$ because, for all $E\in \Sigma _{\Theta _{i}^{0}}$\ and $B\in 
\mathcal{B}_{i}^{0}$,%
\begin{eqnarray*}
\overline{\mathcal{L}}_{\rho _{i}^{0,1}}\left( \overline{\mathcal{L}}_{\psi
_{i}^{1}}\left( \mu _{i}^{1}\right) \right) \left( E\left \vert B\right.
\right) &=&\overline{\mathcal{L}}_{\psi _{i}^{1}}\left( \mu _{i}^{1}\right)
\left( \left( \rho _{i}^{0,1}\right) ^{-1}\left( E\right) \left \vert \left(
\rho _{i}^{0,1}\right) ^{-1}\left( B\right) \right. \right) \\
&=&\mu _{i}^{1}\left( \left( \psi _{i}^{1}\right) ^{-1}\left( \left( \rho
_{i}^{0,1}\right) ^{-1}\left( E\right) \right) \left \vert \left( \psi
_{i}^{1}\right) ^{-1}\left( \left( \rho _{i}^{0,1}\right) ^{-1}\left(
B\right) \right) \right. \right) \\
&=&\mu _{i}^{1}\left( \left( \rho _{i}^{0,1}\circ \psi _{i}^{1}\right)
^{-1}\left( E\right) \left \vert \left( \rho _{i}^{0,1}\circ \psi
_{i}^{1}\right) ^{-1}\left( B\right) \right. \right) \\
&=&\mu _{i}^{1}\left( E\left \vert B\right. \right) \text{,}
\end{eqnarray*}%
as required.

Suppose that we have proved the result for $n\geq 1$. Fix a player $i\in I$%
.\ Define $\psi _{i}^{n+1}:\Theta _{i}^{n}\rightarrow \Theta _{i}^{n+1}$ by%
\begin{equation*}
\psi _{i}^{n+1}\left( s,\mu _{j}^{1},...,\mu _{j}^{n}\right) :=\left( \left(
s,\mu _{j}^{1},...,\mu _{j}^{n}\right) ,\overline{\mathcal{L}}_{\psi
_{j}^{n}}(\mu _{j}^{n})\right)
\end{equation*}%
for every $(s,\mu _{j}^{1},...,\mu _{j}^{n})\in \Theta _{i}^{n}:=\Theta
_{i}^{0}\times H_{j}^{n}$. Note that $\psi _{i}^{n+1}$ is continuous: by the
inductive hypothesis, $\psi _{j}^{n}$\ is continuous, so that, by Lemma 2, $%
\overline{\mathcal{L}}_{\psi _{j}^{n}}$\ is continuous as well. By
construction, $\rho _{i}^{n,n+1}\circ \psi _{i}^{n+1}=\mathrm{Id}_{\Theta
_{i}^{n}}$ and $\mathcal{B}_{i}^{n}=\left( \psi _{i}^{n+1}\right)
^{-1}\left( \mathcal{B}_{i}^{n+1}\right) $. Thus, the map $\overline{%
\mathcal{L}}_{\psi _{i}^{n+1}}:\Delta ^{\mathcal{B}_{i}^{n}}\left( \Theta
_{i}^{n}\right) \rightarrow \Delta ^{\mathcal{B}_{i}^{n+1}}\left( \Theta
_{i}^{n+1}\right) $\ is well-defined and continuous by Lemma 2. For every $%
\mu _{i}^{n+1}\in \Delta ^{\mathcal{B}_{i}^{n}}\left( \Theta _{i}^{n}\right) 
$, the image CPS $\overline{\mathcal{L}}_{\psi _{i}^{n+1}}\left( \mu
_{i}^{n+1}\right) \in \Delta ^{\mathcal{B}_{i}^{n+1}}\left( \Theta
_{i}^{n+1}\right) $ is a coherent extension of $\mu _{i}^{n+1}$ because, for
all $E\in \Sigma _{\Theta _{i}^{n}}$\ and $B\in \mathcal{B}_{i}^{n}$,%
\begin{eqnarray*}
&&\overline{\mathcal{L}}_{\rho _{i}^{n,n+1}}\left( \overline{\mathcal{L}}%
_{\psi _{i}^{n+1}}\left( \mu _{i}^{n+1}\right) \right) \left( E\left \vert
B\right. \right) \\
&=&\overline{\mathcal{L}}_{\psi _{i}^{n+1}}\left( \mu _{i}^{n+1}\right)
\left( \left( \rho _{i}^{n,n+1}\right) ^{-1}\left( E\right) \left \vert
\left( \rho _{i}^{n,n+1}\right) ^{-1}\left( B\right) \right. \right) \\
&=&\mu _{i}^{n+1}\left( \left( \psi _{i}^{n+1}\right) ^{-1}\left( \left(
\rho _{i}^{n,n+1}\right) ^{-1}\left( E\right) \right) \left \vert \left(
\psi _{i}^{n+1}\right) ^{-1}\left( \left( \rho _{i}^{n,n+1}\right)
^{-1}\left( B\right) \right) \right. \right) \\
&=&\mu _{i}^{n+1}\left( \left( \rho _{i}^{n,n+1}\circ \psi _{i}^{n+1}\right)
^{-1}\left( E\right) \left \vert \left( \rho _{i}^{n,n+1}\circ \psi
_{i}^{n+1}\right) ^{-1}\left( B\right) \right. \right) \\
&=&\mu _{i}^{n+1}\left( E\left \vert B\right. \right) \text{,}
\end{eqnarray*}%
where the last equality follows from the fact that $\rho _{i}^{n,n+1}\circ
\psi _{i}^{n+1}=\mathrm{Id}_{\Theta _{i}^{n}}$. This concludes the proof of
the inductive step.

Therefore, for every $n\in \mathbb{N}$ and $i\in I$, the map$\ \pi
_{i}^{n,n+1}$\ is surjective. By construction, $H_{i}^{n+1}=H_{i}^{n}\times
\Delta _{C}^{\mathcal{B}_{i}}\left( \Theta _{i}^{n}\right) $\ for each $n\in 
\mathbb{N}$, where $\Delta _{C}^{\mathcal{B}_{i}}\left( \Theta
_{i}^{n}\right) $\ is the subset of $\Delta ^{\mathcal{B}_{i}}\left( \Theta
_{i}^{n}\right) $\ consisting of all $\mu _{i}^{n+1}$\ which are coherent
extensions of $\left( \mu _{i}^{1},...,\mu _{i}^{n}\right) \in H_{i}^{n}$.
Since $H_{i}^{n+1}$\ is endowed with the product topology, the projection $%
\pi _{i}^{n,n+1}$ is an open map. Next note that, for each $i\in I$, the
projection map $\rho _{i}^{0,1}:\Theta _{i}^{0}\times \Delta ^{\mathcal{B}%
_{i}}(\Theta _{j}^{0})\rightarrow \Theta _{i}^{0}$ is surjective and open.
Since $\rho _{i}^{n-1,n}=\left( \mathrm{Id}_{\Theta _{i}^{0}},\pi
_{j}^{n-1,n}\right) $ for all $n\geq 2$,\ the result follows from the
properties of $\pi _{j}^{n-1,n}$.\hfill $\blacksquare $

\bigskip

Finally, we prove \textbf{Proposition B.2} in Appendix B.

\bigskip

\noindent \textbf{Proposition B.2 }\textit{For each }$i\in I$\textit{, }$%
H_{i}$\textit{\ and }$\Theta _{i}$\textit{\ are homeomorphic to }$%
\underleftarrow{\lim }H_{i}^{n}$\textit{\ and }$\underleftarrow{\lim }\Theta
_{i}^{n-1}$\textit{, respectively.}

\bigskip

\noindent \textbf{Proof.} Fix a player $i\in I$. Recall that (by definition
of $\underleftarrow{\lim }H_{i}^{n}$) for all $m,n\in \mathbb{N}$ such that $%
m\leq n$,\ the functions $\overline{\pi }_{i}^{m}$\ and $\overline{\pi }%
_{i}^{n}$\ satisfy the equality $\overline{\pi }_{i}^{m}=\pi _{i}^{m,n}\circ 
\overline{\pi }_{i}^{n}$. We first construct a homeomorphism $\gamma
_{i}:H_{i}\rightarrow \underleftarrow{\lim }H_{i}^{n}$\ such that $\pi
_{i}^{n}=\overline{\pi }_{i}^{n}\circ \gamma _{i}$ for each $n\in \mathbb{N}$%
. Let $\gamma _{i}:H_{i}\rightarrow \underleftarrow{\lim }H_{i}^{n}$\ be the
function defined by $\gamma _{i}\left( \cdot \right) :=\left( \pi
_{i}^{1}\left( \cdot \right) ,\pi _{i}^{2}\left( \cdot \right) ,...\right) $%
.\ The map $\gamma _{i}$\ is injective. To see this, consider any $h^{\prime
},h^{\prime \prime }\in H_{i}$\ such that $\gamma _{i}\left( h^{\prime
}\right) =\gamma _{i}\left( h^{\prime \prime }\right) $; thus, $h^{\prime
}=h^{\prime \prime }$\ because all the coordinates of $h^{\prime }$\ are
equal to all those of $h^{\prime \prime }$. To show that $\gamma _{i}$ is
surjective, pick any $\left( \left( \mu _{i}^{1}\right) ,\left( \mu
_{i}^{1},\mu _{i}^{2}\right) ,...\right) \in \underleftarrow{\lim }H_{i}^{n}$%
. The element $\bar{h}:=\left( \mu _{i}^{1},\mu _{i}^{2},...\right) \in
H_{i} $ is such that%
\begin{equation*}
\gamma _{i}\left( \bar{h}\right) =\left( \pi _{i}^{1}\left( \bar{h}\right)
,\pi _{i}^{2}\left( \bar{h}\right) ,...\right) =\left( \left( \mu
_{i}^{1}\right) ,\left( \mu _{i}^{1},\mu _{i}^{2}\right) ,...\right) \text{,}
\end{equation*}%
as desired. Hence, $\gamma _{i}:H_{i}\rightarrow \underleftarrow{\lim }%
H_{i}^{n}$\ is a bijection. Since each projection $\pi _{i}^{n}\left( \cdot
\right) $ ($n\in \mathbb{N}$) is a continuous function, so is $\gamma _{i}$;
see Engelking (1989, Proposition 2.3.6). The inverse of $\gamma _{i}$ is
also continuous: consider a sequence $\left( \hat{h}_{m}\right) _{m\in 
\mathbb{N}}:=\left( \left( \mu _{i,m}^{1}\right) ,\left( \mu _{i,m}^{1},\mu
_{i,m}^{2}\right) ,...\right) _{m\in \mathbb{N}}$ in $\underleftarrow{\lim }%
H_{i}^{n}$\ converging to $\hat{h}:=\left( \left( \mu _{i}^{1}\right)
,\left( \mu _{i}^{1},\mu _{i}^{2}\right) ,...\right) $. By definition of
product topology (i.e., the topology of pointwise convergence), it is the
case that, for all $n\in \mathbb{N}$, $\mu _{i,m}^{n}\rightarrow \mu
_{i}^{n} $ as $m\rightarrow \infty $. Hence, $\gamma _{i}^{-1}\left( \hat{h}%
_{m}\right) =\left( \mu _{i,m}^{1},\mu _{i,m}^{2},...\right) $\ converges to 
$\gamma _{i}^{-1}\left( \hat{h}\right) =\left( \mu _{i}^{1},\mu
_{i}^{2},...\right) $.

To show the result for the spaces $\Theta _{i}$\ and $\underleftarrow{\lim }%
\Theta _{i}^{n-1}$, let $\mathbb{N}_{0}:=\mathbb{N}\cup \left \{ 0\right \} $
be the set of natural numbers including $0$. Denote by $\theta _{i}:=\left(
\theta _{i}^{0},\theta _{i}^{1},...\right) $\ each element of the product
set $\tprod_{n\in \mathbb{N}_{0}}\Theta _{i}^{n}$. Note that, by definition,%
\begin{equation*}
\underleftarrow{\lim }\Theta _{i}^{n-1}=\left \{ \theta _{i}\in \tprod_{n\in 
\mathbb{N}_{0}}\Theta _{i}^{n}:%
\begin{tabular}{l}
1) $\rho _{i}^{0,1}\left( \theta _{i}^{1}\right) =\theta _{i}^{0}$, \\ 
2) $\forall n\in \mathbb{N},\rho _{i}^{n-1,n}\left( \theta _{i}^{n}\right)
=\theta _{i}^{n-1}$%
\end{tabular}%
\right \} \text{,}
\end{equation*}%
where each $\theta _{i}^{n}$\ ($n\in \mathbb{N}$) can be written as $\theta
_{i}^{n}=\left( s,\left( \mu _{j}^{m}\right) _{m=1}^{n}\right) $. That is,
each element $\theta _{i}\in \underleftarrow{\lim }\Theta _{i}^{n-1}$\ has
the form $\theta _{i}=\left( \left( s\right) ,\left( s,\left( \mu
_{j}^{1}\right) \right) ,\left( s,\left( \mu _{j}^{1},\mu _{j}^{2}\right)
\right) ,...\right) $. Therefore, $\underleftarrow{\lim }\Theta _{i}^{n}=$%
\textrm{D}$\left( S^{\mathbb{N}_{0}}\right) \times \underleftarrow{\lim }%
H_{j}^{n}$, where \textrm{D}$\left( S^{\mathbb{N}_{0}}\right) $\ is the set
of all constant sequences on $S$ (i.e., \textrm{D}$\left( S^{\mathbb{N}%
_{0}}\right) $ is the \textquotedblleft diagonal\textquotedblright \ of $S^{%
\mathbb{N}_{0}}$). The map $\psi _{i}:S\rightarrow S^{\mathbb{N}_{0}}$\
defined by%
\begin{equation*}
\psi _{i}\left( s\right) =\left( s,s,...\right)
\end{equation*}%
is a homeomorphism from $S$\ onto \textrm{D}$\left( S^{\mathbb{N}%
_{0}}\right) $. Consequently, the map $\varphi _{i}:=\left( \psi _{i},\gamma
_{j}\right) :\Theta _{i}\rightarrow \underleftarrow{\lim }\Theta _{i}^{n}$\
is a homeomorphism such that $\rho _{i}^{n-1}=\overline{\rho }%
_{i}^{n-1}\circ \varphi _{i}$\ for every $n\in \mathbb{N}$.\hfill $%
\blacksquare $

\section*{E Adapting the results for consistent CPSs}

This section shows how the results in the paper remain true if conditional
beliefs are represented by \textbf{consistent conditional probability systems%
} (CCPSs), a subfamily of CPSs used by Siniscachi (2022) to formalize the
notion of structural rationality (a refinement of weak sequential
rationality).

The following definition of CCPS, adapted to our notation, is taken from
Siniscalchi (2020, Definition 1).

\bigskip

\noindent \textbf{Definition E.1 }\textit{Let }$\left( X,\Sigma _{X}\right) $%
\textit{\ be a measurable space and }$\mathcal{B}\subseteq \Sigma _{X}$%
\textit{\ be a family\ of conditioning events. A consistent conditional
probability system (CCPS) on }$\left( X,\Sigma _{X},\mathcal{B}\right) $%
\textit{\ is an array of probability measures }$\mu :=\left( \mu \left(
\cdot |B\right) \right) _{B\in \mathcal{B}}$\textit{\ such that:\newline
\ \ (i) for every }$B\in \mathcal{B}$\textit{, }$\mu \left( B|B\right) =1$%
\textit{;\newline
\ \ (ii) for every ordered list }$B_{1},...,B_{L}\in \mathcal{B}$\textit{\
and every }$E\in \Sigma _{X}$\textit{\ such that }$E\subseteq B_{1}\cap
B_{L} $\textit{,}%
\begin{equation*}
\mu \left( E|B_{1}\right) \tprod_{\ell =1}^{L-1}\mu \left( B_{\ell }\cap
B_{\ell +1}|B_{\ell +1}\right) =\mu \left( E|B_{L}\right) \tprod_{\ell
=1}^{L-1}\mu \left( B_{\ell }\cap B_{\ell +1}|B_{\ell }\right) \text{.}
\end{equation*}

\bigskip

As explained in Siniscalchi (2020), a CCPS is a CPS satisfying a requirement
that is stronger than the chain rule, as it imposes a more stringent
consistency condition on beliefs across different conditioning events.
Consider the case of $L=2$\ and $B_{1}\subseteq B_{L}$\ in Condition (ii) of
Definition E.1. In this case, it can be checked that the conditioning
requirement corresponds to the chain rule. See Siniscalchi (2020) for a
detailed analysis and examples.

Here we show that all the results in the paper still hold if CPSs are
replaced by CCPSs. The changes to be made are easy to identify, and they
mainly pertain to some proofs in Appendices A-C.

As for Appendix A, first note that there is an analogue of Fact A.2 for
CCPSs: the proof is due to Siniscalchi (2023).\footnote{%
An alternative proof, due to the present author, is available on request.}
Second, the proof of Lemma 2.(i) must be adapted to take into account the
conditioning requirement of a CCPS. Namely, Lemma 2.(i) must be replaced by
the following version.

\bigskip

\noindent \textbf{Lemma E.1 }\textit{Fix measurable spaces }$\left( X,\Sigma
_{X}\right) $\textit{\ and }$\left( Y,\Sigma _{Y}\right) $\textit{, and
families }$\mathcal{B}_{X}\subseteq \Sigma _{X}$\textit{\ and }$\mathcal{B}%
_{Y}\subseteq \Sigma _{Y}$\textit{\ of conditioning events. Suppose that }$%
f:X\rightarrow Y$\textit{\ is a measurable function such that }$f^{-1}\left( 
\mathcal{B}_{Y}\right) =\mathcal{B}_{X}$\textit{. The following statements
hold.\newline
\ \ (i) The map }$\overline{\mathcal{L}}_{f}:\Delta ^{\mathcal{B}_{X}}\left(
X\right) \rightarrow \Delta ^{\mathcal{B}_{Y}}\left( Y\right) $\textit{\ is
well-defined.\newline
\ \ (ii) Suppose that }$\mathcal{B}_{X}$\textit{\ and }$\mathcal{B}_{Y}$%
\textit{\ are countable, }$X$\textit{\ is a metrizable space and }$Y$\textit{%
\ is a Souslin space. If }$f$\textit{\ is Borel measurable (resp.
continuous), then }$\overline{\mathcal{L}}_{f}$\textit{\ is Borel measurable
(resp. continuous).}

\bigskip

\noindent \textbf{Proof}. The proof of part (ii) is unaltered. To prove part
(i), fix some CCPS $\mu \in \Delta ^{\mathcal{B}_{X}}\left( X\right) $. We
show that $\overline{\mathcal{L}}_{f}\left( \mu \right) $\ is a CCPS on $%
\left( Y,\Sigma _{Y},\mathcal{B}_{Y}\right) $. Pick any $B\in \mathcal{B}%
_{Y} $. Since $f^{-1}\left( \mathcal{B}_{Y}\right) =\mathcal{B}_{X}$, we
have $f^{-1}\left( B\right) \in \mathcal{B}_{X}$ and%
\begin{equation*}
\overline{\mathcal{L}}_{f}\left( \mu \right) \left( B|B\right) =\mu \left(
f^{-1}\left( B\right) |f^{-1}\left( B\right) \right) =1\text{,}
\end{equation*}%
because $\mu $\ is a CCPS. Thus, Condition (i) in Definition E.1 holds. To
show that Condition (ii) is satisfied, consider any $E\in \Sigma _{Y}$\ and $%
B_{1},...,B_{L}\in \mathcal{B}_{Y}$\ such that $E\subseteq B_{1}\cap B_{L}$.
We have $f^{-1}\left( B_{1}\right) $,..., $f^{-1}\left( B_{L}\right) \in 
\mathcal{B}_{X}$. Furthermore, $f^{-1}\left( E\right) \in \Sigma _{X}$\ and $%
f^{-1}\left( E\right) \subseteq f^{-1}\left( B_{1}\right) \cap f^{-1}\left(
B_{L}\right) $. Using the fact that $\mu $\ is a CCPS, we get%
\begin{eqnarray*}
&&\overline{\mathcal{L}}_{f}\left( \mu \right) \left( E|B_{1}\right)
\tprod_{\ell =1}^{L-1}\overline{\mathcal{L}}_{f}\left( \mu \right) \left(
B_{\ell }\cap B_{\ell +1}|B_{\ell +1}\right) \\
&=&\mu \left( f^{-1}\left( E\right) |f^{-1}\left( B_{1}\right) \right)
\tprod_{\ell =1}^{L-1}\mu \left( f^{-1}\left( B_{\ell }\cap B_{\ell
+1}\right) |f^{-1}\left( B_{\ell +1}\right) \right) \\
&=&\mu \left( f^{-1}\left( E\right) |f^{-1}\left( B_{L}\right) \right)
\tprod_{\ell =1}^{L-1}\mu \left( f^{-1}\left( B_{\ell }\cap B_{\ell
+1}\right) |f^{-1}\left( B_{\ell }\right) \right) \\
&=&\overline{\mathcal{L}}_{f}\left( \mu \right) \left( E|B_{L}\right)
\tprod_{\ell =1}^{L-1}\overline{\mathcal{L}}_{f}\left( \mu \right) \left(
B_{\ell }\cap B_{\ell +1}|B_{\ell }\right) \text{,}
\end{eqnarray*}%
as required.\hfill $\blacksquare $

\bigskip

As for Appendix B, we just need to substitute Lemma 2 with Lemma E.1 to
provide an analogue of Proposition B.1 for CCPSs. The other change pertains
to the proof of Theorem B.1, specifically the part which shows that the
array $\left( \mu \left( \cdot |B\right) \right) _{B\in \mathcal{B}_{Y}}$\
is the limit CCPS.

\bigskip

\noindent \textbf{Modified proof of Theorem B.1}. Each CCPS $\mu _{n}$\ on $%
\left( Y_{n},\Sigma _{Y_{n}},\mathcal{B}_{n}\right) $\ is an array%
\begin{equation*}
\mu _{n}=\left( \mu _{n}\left( \cdot |B\right) \right) _{B\in \mathcal{B}%
_{n}}
\end{equation*}%
such that, for all $B\in \mathcal{B}_{n}$, $\mu _{n}\left( \cdot |B\right) $%
\ is a Radon probability measure on $\left( Y_{n},\Sigma _{Y_{n}}\right) $,
i.e., for each $E\in \Sigma _{Y_{n}}$ and each $\varepsilon >0$, there
exists a compact set $K\subseteq E$\ such that $\mu \left( E|B_{n}\right)
-\mu \left( K|B\right) \leq \varepsilon $. This follows from the fact that
every Borel probability measure on a Souslin space is Radon (Bogachev 2007,
Theorem 7.4.3). With this, we can apply Prokhorov's Theorem for Radon
probability measures (Schwartz 1974, Theorem 21 and Corollary, p. 81) to
claim the existence of a unique array $\left( \mu \left( \cdot |B\right)
\right) _{B\in \mathcal{B}_{Y}}$\ of (Radon) probability measures on the
Borel $\sigma $-algebra $\Sigma _{Y}$ of the Souslin space $Y$\ (see Lemma
B1.(ii)) such that%
\begin{equation*}
\mu \left( \overline{f}_{n}^{-1}\left( \cdot \right) \left \vert \overline{f}%
_{n}^{-1}\left( B_{n}\right) \right. \right) =\mu _{n}\left( \cdot
|B_{n}\right)
\end{equation*}%
for all $n\in \mathbb{N}$ and $B_{n}\in \mathcal{B}_{n}$. It remains to
check that the array $\left( \mu \left( \cdot |B\right) \right) _{B\in 
\mathcal{B}_{Y}}$\ is a CCPS. Condition (i) of Definition E.1 clearly holds.

To show that Condition (ii) of Definition E.1 is satisfied, pick any $%
B_{1},...,B_{L}\in \mathcal{B}_{Y}$. We show that Condition (ii) is
satisfied for every $A\in \Sigma _{Y}$\ such that $A\subseteq B_{1}\cap
B_{L} $. To this end, we first consider the case when $A$\ belongs to the
algebra $\mathcal{A}_{Y}:=\cup _{n\in \mathbb{N}}\overline{f}_{n}^{-1}\left(
\Sigma _{n}\right) $, which generates the Borel $\sigma $-algebra $\Sigma
_{Y}$ (Lemma B.1.(ii)). If $A\in \mathcal{A}_{Y}$, then there exist $n\in 
\mathbb{N}$\ and $A_{n}\in \Sigma _{Y_{n}}$\ such that $A=\overline{f}%
_{n}^{-1}\left( A_{n}\right) $. Since $\mathcal{B}_{Y}=\overline{f}%
_{n}^{-1}\left( \mathcal{B}_{n}\right) $, there are $B_{1,n},...,B_{L,n}\in 
\mathcal{B}_{n}$\ such that $B_{\ell }=\overline{f}_{n}^{-1}\left( B_{\ell
,n}\right) $ for all $\ell =1,...L$. By surjectivity of $\overline{f}_{n}$
(Lemma B.1.(i)), $A_{n}\subseteq B_{1,n}\cap B_{L,n}$. Using the fact that $%
\mu _{n}$\ is a CCPS, we obtain that Condition (ii) holds:%
\begin{eqnarray*}
&&\mu \left( A|B_{1}\right) \tprod_{\ell =1}^{L-1}\mu \left( B_{\ell }\cap
B_{\ell +1}|B_{\ell +1}\right) \\
&=&\mu \left( \overline{f}_{n}^{-1}\left( A_{n}\right) |\overline{f}%
_{n}^{-1}\left( B_{1,n}\right) \right) \tprod_{\ell =1}^{L-1}\mu \left( 
\overline{f}_{n}^{-1}\left( B_{\ell ,n}\right) \cap \overline{f}%
_{n}^{-1}\left( B_{\ell +1,n}\right) |\overline{f}_{n}^{-1}\left( B_{\ell
+1,n}\right) \right) \\
&=&\mu \left( \overline{f}_{n}^{-1}\left( A_{n}\right) |\overline{f}%
_{n}^{-1}\left( B_{1,n}\right) \right) \tprod_{\ell =1}^{L-1}\mu \left( 
\overline{f}_{n}^{-1}\left( B_{\ell ,n}\cap B_{\ell +1,n}\right) |\overline{f%
}_{n}^{-1}\left( B_{\ell +1,n}\right) \right) \\
&=&\mu _{n}\left( A_{n}|B_{1,n}\right) \tprod_{\ell =1}^{L-1}\mu _{n}\left(
B_{\ell ,n}\cap B_{\ell +1,n}|B_{\ell +1,n}\right) \\
&=&\mu _{n}\left( A_{n}|B_{L,n}\right) \tprod_{\ell =1}^{L-1}\mu _{n}\left(
B_{\ell ,n}\cap B_{\ell +1,n}|B_{\ell ,n}\right) \\
&=&\mu \left( \overline{f}_{n}^{-1}\left( A_{n}\right) |\overline{f}%
_{n}^{-1}\left( B_{L,n}\right) \right) \tprod_{\ell =1}^{L-1}\mu \left( 
\overline{f}_{n}^{-1}\left( B_{\ell ,n}\cap B_{\ell +1,n}\right) |\overline{f%
}_{n}^{-1}\left( B_{\ell ,n}\right) \right) \\
&=&\mu \left( A|B_{L}\right) \tprod_{\ell =1}^{L-1}\mu \left( \overline{f}%
_{n}^{-1}\left( B_{\ell ,n}\right) \cap \overline{f}_{n}^{-1}\left( B_{\ell
+1,n}\right) |\overline{f}_{n}^{-1}\left( B_{\ell ,n}\right) \right) \\
&=&\mu \left( A|B_{L}\right) \tprod_{\ell =1}^{L-1}\mu \left( B_{\ell }\cap
B_{\ell +1}|B_{\ell }\right) \text{.}
\end{eqnarray*}%
Next, let $\mathcal{M}\left( B_{1},...,B_{L}\right) \subseteq \Sigma _{Y}$\
be the collection of all Borel subsets of $Y$ such that Condition (ii) of
Definition E.1 holds for those fixed $B_{1},...,B_{L}\in \mathcal{B}_{Y}$.
By $\sigma $-additivity of probability measures, $\mathcal{M}\left(
B_{1},...,B_{L}\right) $\ is a monotone class. As seen, $\mathcal{A}%
_{Y}\subseteq \mathcal{M}\left( B_{1},...,B_{L}\right) $. It follows that $%
\Sigma _{Y}\subseteq \mathcal{M}\left( B_{1},...,B_{L}\right) $ because, by
the Monotone Class Theorem, the smallest monotone class containing $\mathcal{%
A}_{Y}$ is $\Sigma _{Y}$. Thus, $\mathcal{M}\left( B_{1},...,B_{L}\right)
=\Sigma _{Y}$. This shows that Condition (ii) of Definition E.1 is
satisfied, and concludes the proof.\hfill $\blacksquare $

\bigskip

As for Appendix C, we need to show (proof of Lemma 3.(ii)) that the
extension of a CCPS to the universally measurable $\sigma $-algebra is well
defined.

\bigskip

\noindent \textbf{Modified proof of Lemma 3.(ii)}. Let $v$\ be a CCPS on $%
\left( X\times Z,\Sigma _{X\times Z},\mathcal{B}_{X\times Z}\right) $. We
have to find $\mu \in \Delta ^{\mathcal{B}}\left( X\times Y\right) $\ such
that $\overline{\mathcal{L}}_{f_{2}}\left( \mu \right) =v$. To this end,
first note that, since $f_{1}:Y\rightarrow Z$\ is surjective, $f_{2}:=\left( 
\mathrm{Id}_{X},f_{1}\right) :X\times Y\rightarrow X\times Z$\ is a Borel
measurable surjection. By Theorem C.1, there exists a universally measurable
right inverse of $f_{1}$, that is, an injective function $g_{1}:Z\rightarrow
Y$ such that $f_{1}\circ g_{1}=\mathrm{Id}_{Z}$ and $g_{1}^{-1}\left(
E\right) \in \Sigma _{Z}^{\ast }$ for every $E\in \Sigma _{Y}$.

Let $g_{2}:X\times Z\rightarrow X\times Y$\ be the function defined as $%
g_{2}:=\left( \mathrm{Id}_{X},g_{1}\right) $. Clearly, $g_{2}$\ satisfies $%
f_{2}\circ g_{2}=\mathrm{Id}_{X\times Z}$, and, by Fact C.1, $g_{2}$\ is a
universally measurable right inverse of $f_{2}$. Furthermore, $g_{2}$\ is
such that $\mathcal{B}_{X\times Z}=g_{2}^{-1}\left( \mathcal{B}_{X\times
Y}\right) $. Hence, by Lemma E.1.(i), the map $\overline{\mathcal{L}}%
_{g_{2}} $ from the set of CCPSs on $\left( X\times Z,\Sigma _{X\times
Z}^{\ast },\mathcal{B}_{X\times Z}\right) $\ to the set of CCPSs on $\left(
X\times Y,\Sigma _{X\times Y},\mathcal{B}_{X\times Y}\right) $\ is
well-defined.

Next, for each $B\in \mathcal{B}_{X\times Z}$, let $\bar{v}\left( \cdot
|B\right) $ denote the unique extension of $v\left( \cdot |B\right) $\ to $%
\Sigma _{X\times Z}^{\ast }$. We claim that the array $\bar{v}:=\left( \bar{v%
}\left( \cdot |B\right) \right) _{B\in \mathcal{B}_{X\times Z}}$ is a CCPS
on $\left( X\times Z,\Sigma _{X\times Z}^{\ast },\mathcal{B}_{X\times
Z}\right) $. To see this, it is enough to verify that Condition (ii) of
Definition E.1 is satisfied. Pick any $E\in \Sigma _{X\times Z}^{\ast }$ and 
$B_{1},...,B_{L}\in \mathcal{B}_{X\times Z}$ such that $E\subseteq B_{1}\cap
B_{L}$. Notice that, since $B_{\ell }\in \Sigma _{X\times Z}$ for each $\ell
=1$, it is the case that%
\begin{eqnarray*}
\tprod_{\ell =1}^{L-1}\bar{v}\left( B_{\ell }\cap B_{\ell +1}|B_{\ell
+1}\right) &=&\tprod_{\ell =1}^{L-1}v\left( B_{\ell }\cap B_{\ell
+1}|B_{\ell +1}\right) \text{,} \\
\tprod_{\ell =1}^{L-1}\bar{v}\left( B_{\ell }\cap B_{\ell +1}|B_{\ell
}\right) &=&\tprod_{\ell =1}^{L-1}v\left( B_{\ell }\cap B_{\ell +1}|B_{\ell
}\right) \text{.}
\end{eqnarray*}%
Suppose first that $E\in \Sigma _{X\times Z}$. Thus, $\bar{v}\left(
E|B_{1}\right) =v\left( E|B_{1}\right) $\ and $\bar{v}\left( E|B_{L}\right)
=v\left( E|B_{L}\right) $, and we have%
\begin{equation*}
\bar{v}\left( E|B_{1}\right) \tprod_{\ell =1}^{L-1}\bar{v}\left( B_{\ell
}\cap B_{\ell +1}|B_{\ell +1}\right) =\bar{v}\left( E|B_{L}\right)
\tprod_{\ell =1}^{L-1}\bar{v}\left( B_{\ell }\cap B_{\ell +1}|B_{\ell
}\right) \text{.}
\end{equation*}%
If $E\notin \Sigma _{X\times Z}$, then Condition (ii) of Definition E.1
trivially holds as $0=0$: we have $\bar{v}\left( E|B_{1}\right) =\bar{v}%
\left( E|B_{L}\right) =0$\ because $E$ is a set that is both $v\left( \cdot
|B_{1}\right) $-null\ and $v\left( \cdot |B_{L}\right) $-null.

By Lemma E.1.(i), $\mu :=\overline{\mathcal{L}}_{g_{2}}\left( \bar{v}\right) 
$\ is a CCPS on $\left( X\times Y,\Sigma _{X\times Y},\mathcal{B}_{X\times
Y}\right) $. Since the equality $g_{2}^{-1}\left( f_{2}^{-1}\left( A\right)
\right) =A$ holds for every set $A\subseteq X\times Z$, we see that, for all 
$E\in \Sigma _{X\times Z}$\ and $B\in \mathcal{B}_{X\times Z}$,%
\begin{eqnarray*}
\overline{\mathcal{L}}_{f_{2}}\left( \mu \right) \left( E|B\right) &=&\mu
\left( f_{2}^{-1}\left( E\right) |f_{2}^{-1}\left( B\right) \right) \\
&=&\bar{v}\left( g_{2}^{-1}\left( f_{2}^{-1}\left( E\right) \right)
|g_{2}^{-1}\left( f_{2}^{-1}\left( B\right) \right) \right) \\
&=&\bar{v}\left( E|B\right) \\
&=&v\left( E|B\right) \text{,}
\end{eqnarray*}%
where the last equality holds because $E\in \Sigma _{X\times Z}$. Thus, $\mu 
$\ is the desired CCPS that satisfies $\overline{\mathcal{L}}_{f_{2}}\left(
\mu \right) =v$.\hfill $\blacksquare $

\bigskip

With these modifications, all the results\ and proofs in the paper still
hold if \textquotedblleft CPS\textquotedblright \ is replaced by
\textquotedblleft CCPS.\textquotedblright